\pdfoutput=1
\documentclass[11pt,twoside,a4paper,cmspaper,final,collab]{cms-tdr}
\def\svnVersion{52532ec-D}\def\svnDate{2021/09/02}\def\cmsCernNoTag{CERN-EP-2021-033}\def\cmsCernDate{\today}\def\cmsMessage{"Published in the European Physical Journal C as \href{http://dx.doi.org/10.1140/epjc/s10052-021-09538-2}{\doi{10.1140/epjc/s10052-021-09538-2}.}"}
\begin{document}\cmsNoteHeader{LUM-17-003}

\ifthenelse{\boolean{cms@external}}{\providecommand{\cmsLeft}{upper\xspace}}{\providecommand{\cmsLeft}{left\xspace}} 
\ifthenelse{\boolean{cms@external}}{\providecommand{\cmsRight}{lower\xspace}}{\providecommand{\cmsRight}{right\xspace}}
\ifthenelse{\boolean{cms@external}}{\providecommand{\cmsTable}[1]{#1}}{\providecommand{\cmsTable}[1]{\resizebox{\textwidth}{!}{#1}}}
\newlength\cmsTabSkip\setlength{\cmsTabSkip}{1ex}

\xspaceaddexceptions{\%\fbinv}
\newcommand{\pp}{\ensuremath{\Pp\Pp}\xspace}
\newcommand{\sigmaVis}{\ensuremath{\sigma_{\text{vis}}}\xspace}
\newcommand{\muVis}{\ensuremath{\mu_{\text{vis}}}\xspace}
\newcommand{\sigmaVisPCC}{\ensuremath{\sigma_{\text{vis}}^{\text{PCC}}}\xspace}
\newcommand{\freq}{\ensuremath{\nu_r}\xspace}
\newcommand{\vdM}{vdM\xspace}
\newcommand{\Lb}{\ensuremath{\lumi_\text{b}}\xspace}
\newcommand{\sigmab}{\ensuremath{\sigma_{\text{b}}}\xspace}
\newcommand{\Aeff}{\ensuremath{A_{\text{eff}}}\xspace}
\newcommand{\Weff}{\ensuremath{W_{\text{eff}}}\xspace}
\newcommand{\Heff}{\ensuremath{H_{\text{eff}}}\xspace}
\newcommand{\betastar}{\ensuremath{\beta^{*}}\xspace}
\newcommand{\nb}{\ensuremath{n_\text{b}}\xspace}
\newcommand{\highEnergy}{\ensuremath{\sqrt{s}=13\TeV}\xspace}
\newcommand{\FillNumberI}{4266\xspace}
\newcommand{\NBunchCollI}{30\xspace}
\newcommand{\DateAbvI}{Aug. 2015\xspace}
\newcommand{\FillNumberII}{4954\xspace}
\newcommand{\FillNumberIII}{4945\xspace}
\newcommand{\NBunchCollII}{32\xspace}
\newcommand{\DateAbvII}{May 2016\xspace}
\newcommand{\PCCINoUnit}{\ensuremath{9.166\pm0.056\stat}\xspace}
\newcommand{\PCCII}{\ensuremath{8.429\pm0.029\stat\unit{barns}}\xspace}
\newcommand{\BeamCurUnc}{0.2\xspace}
\newcommand{\BIEffI}{1.3\xspace}
\newcommand{\BIffII}{0.9\xspace}
\newcommand{\BIUnc}{0.5\xspace}
\newcommand{\BIAndLRLowEffI}{0.8\xspace}
\newcommand{\BIAndLRUpEffI}{1.3\xspace}
\newcommand{\BIAndLREffII}{0.6\xspace}
\newcommand{\BIAndLRUnc}{0.5\xspace}
\newcommand{\DeadTimeUncI}{0.5\xspace}
\newcommand{\DeadTimeUncII}{0.1\xspace}
\newcommand{\DeflEffI}{2.0\xspace}
\newcommand{\DeflEffII}{1.6\xspace}
\newcommand{\DefAndFocEffI}{0.6\xspace}
\newcommand{\DefAndFocEffII}{0.4\xspace}
\newcommand{\DefAndFocUnc}{0.5\xspace}
\newcommand{\FocEffI}{1.7\xspace}
\newcommand{\FocEffII}{1.4\xspace}
\newcommand{\LinearityUncI}{0.5\xspace}
\newcommand{\LinearityUncII}{0.3\xspace}
\newcommand{\LRLowEffI}{0.6\xspace}
\newcommand{\LRUpEffI}{1.1\xspace}
\newcommand{\LREffII}{0.2\xspace}
\newcommand{\LRUnc}{0.6\xspace}
\newcommand{\LSCEffI}{0.4\xspace}
\newcommand{\LSCEffII}{1.3\xspace}
\newcommand{\LSCUncI}{0.2\xspace}
\newcommand{\LSCUncII}{0.3\xspace}
\newcommand{\ODSystLowEffI}{\ensuremath{-}0.6\xspace}
\newcommand{\ODSystUpEffI}{\ensuremath{+}0.4\xspace}
\newcommand{\ODSystLowEffII}{\ensuremath{-}0.5\xspace}
\newcommand{\ODSystUpEffII}{\ensuremath{-}0.2\xspace}
\newcommand{\ODSystEffI}{\ensuremath{-}0.1\xspace}
\newcommand{\ODSystEffII}{\ensuremath{-}0.3\xspace}
\newcommand{\ODSystUncI}{0.8\xspace}
\newcommand{\ODSystUncII}{0.5\xspace}
\newcommand{\ODRandomLowEffI}{0.6\xspace}
\newcommand{\ODRandomUpEffI}{1.0\xspace}
\newcommand{\ODRandomLowEffII}{0.2\xspace}
\newcommand{\ODRandomUpEffII}{1.0\xspace}
\newcommand{\ODRandomUncI}{0.2\xspace}
\newcommand{\ODRandomUncII}{0.1\xspace}
\newcommand{\OtherVarUncI}{0.6\xspace}
\newcommand{\OtherVarUncII}{0.3\xspace}
\newcommand{\SpuriousEffI}{0.2\xspace}
\newcommand{\SpuriousEffII}{0.3\xspace}
\newcommand{\SpuriousUnc}{0.1\xspace}
\newcommand{\StabUncI}{0.6\xspace}
\newcommand{\StabUncII}{0.5\xspace}
\newcommand{\TypeIUncI}{0.3\xspace}
\newcommand{\TypeIUncII}{0.3\xspace}
\newcommand{\TypeIIUncI}{0.1\xspace}
\newcommand{\TypeIIUncII}{0.3\xspace}
\newcommand{\VdmUncI}{1.3\xspace}
\newcommand{\VdmUncII}{1.0\xspace}
\newcommand{\IntUncI}{1.0\xspace}
\newcommand{\IntUncII}{0.7\xspace}
\newcommand{\TotUncI}{1.6\xspace}
\newcommand{\TotUncII}{1.2\xspace}
\newcommand{\LumiI}{2.27\xspace}
\newcommand{\LumiII}{36.3\xspace}
\hyphenation{mar-gin-al-ized}
\hyphenation{bun-ches}
\hyphenation{lu-mi-nos-i-ty}

\cmsNoteHeader{LUM-17-003}
\title{Precision luminosity measurement in proton-proton collisions at \texorpdfstring{\highEnergy}{sqrt(s) = 13 TeV} in 2015 and 2016 at CMS}

\date{\today}
\abstract{
The measurement of the luminosity recorded by the CMS detector installed at LHC interaction point 5, using proton-proton collisions at \highEnergy in 2015 and 2016, is reported. The absolute luminosity scale is measured for individual bunch crossings using beam-separation scans (the van der Meer method), with a relative precision of \VdmUncI\% and \VdmUncII\% in 2015 and 2016, respectively. The dominant sources of uncertainty are related to residual differences between the measured beam positions and the ones provided by the operational settings of the LHC magnets, the factorizability of the proton bunch spatial density functions in the coordinates transverse to the beam direction, and the modeling of the effect of electromagnetic interactions among protons in the colliding bunches. When applying the van der Meer calibration to the entire run periods, the integrated luminosities when CMS was fully operational are \LumiI and \LumiII\fbinv in 2015 and 2016, with a relative precision of \TotUncI\% and \TotUncII\%, respectively. These are among the most precise luminosity measurements at bunched-beam hadron colliders.
}
\hypersetup{pdfauthor={CMS Collaboration},
pdftitle={Precision luminosity measurement at CMS in proton-proton collisions at sqrt(s) = 13 TeV in 2015 and 2016},
pdfsubject={CMS},
pdfkeywords={CMS, luminosity, van der Meer}}

\maketitle

\section{Introduction}

Luminosity, \lumi, is a key parameter at particle colliders. Along with the energy available in the 
center-of-mass system, it is one of the two main figures of merit that quantify the potential for delivering 
large data samples and producing novel massive particles.  The instantaneous luminosity $\lumi(t)$ is the 
process-independent ratio of the rate $R(t)$ of events produced per unit of time $\rd t$ to the cross 
section $\sigma$ for a given process.   The fundamental limitations on precise predictions for these cross sections (\eg, from 
quantum chromodynamics) motivate the techniques used for luminosity measurements at various types of colliders. The 
precise determination of the integrated luminosity,  $\int\lumi(t)\rd t$, has proven particularly 
challenging at hadron colliders, with an achieved precision typically ranging from 1 to 
15\%~\cite{Grafstrom:2015foa}.  The ``precision frontier'' target of 1\%~\cite{HLHE-LHC} does not reflect a 
fundamental limitation, but rather results from a variety of uncorrelated sources of systematic uncertainty with typical 
magnitudes of 0.1--0.5\%.  In this paper, we report the precise determination of the absolute luminosity at the CERN LHC 
interaction point (IP) 5 with the CMS detector~\cite{CMS}, using data from proton-proton (\pp{}) collisions at \highEnergy collected in 2015 
and 2016.

A central component of the physics program at the LHC consists of measurements that can 
precisely test the validity of standard model (SM) predictions, \eg, 
cross sections for the production of electroweak gauge bosons~\cite{SMP-16-005,SMP-17-001} or top quark pairs~\cite{TOP-16-006,TOP-17-001}.  
A good understanding of the luminosity is critical to minimize the systematic uncertainty in these measurements.  
The uncertainty in the luminosity measurement is often the dominant systematic uncertainty~\cite{SMP-17-001,TOP-16-006,TOP-17-001},
motivating an effort to improve its precision.

Stable luminosity information is also crucial to the efforts of the LHC operators to optimize the performance 
of the accelerator~\cite{emittance_scans,Antoniou:2293678}.  In this context, it is important to provide luminosity 
information in real time at a high enough frequency to facilitate rapid optimization.
The ability to measure the luminosity of individual bunch crossings (bunch-by-bunch luminosity) is also necessary 
so that the distribution of number of collisions per crossings is known to the experiments.  
This information is important when preparing simulations as well as optimization of thresholds 
to keep event-recording rates near data acquisition design targets.
 
An absolute luminosity scale is obtained with good accuracy using the direct method of van der Meer (\vdM)
scans~\cite{VdM:1968,VdM:1977,Aaij:2014ida,Aaboud2016}. In these scans, the transverse separation of the two beams is varied over time 
and the resulting rate of some physical observables 
(\eg, number of charged particles passing through a silicon detector or energy deposited in a calorimeter) as a 
function of separation is used to extract the effective beam size.  The absolute luminosity at one point in 
time can then be calculated from measurable beam parameters---namely, the transverse spatial widths of the 
overlap of the beams and the number of protons in each beam. To achieve the desired accuracy in the absolute 
luminosity calibration, the \vdM scans are typically performed under carefully tailored conditions and with beam 
parameters optimized for that purpose~\cite{Grafstrom:2015foa}, in 
conjunction with processing the input from accelerator instrumentation and multiple detector systems.  A
relative normalization method is then needed to transfer the absolute luminosity calibration to the complete
data-taking period.  To this end, for a given subdetector, the cross section \sigmaVis in the ``visible'' phase space region, defined by 
its acceptance, is measured for several observables. The integrated luminosity is obtained 
from the \sigmaVis-calibrated counts accumulated for a given period of data taking. Changes in the detector
response over time can result in variations in \sigmaVis, which could appear as nonlinearity and/or long-term instability
in the measured luminosity.

To address these challenges, CMS employs a multifaceted approach, in which measurements from
various individual subsystems are used to produce a final luminosity value with high precision, good linearity, and stability.
Several methods and independent detectors are used to provide redundancy and to minimize 
any bias originating from detector effects. 

The LHC orbit is divided into a total of 3564 time windows 25\unit{ns} long (bunch crossing slots), 
each of which can potentially contain a colliding bunch.  However, the total number of filled bunch 
crossings is limited by design to a maximum of 2808 by the choice of 
the beam production scheme in the injectors and constraints from the rise times of 
injection and extraction kicker magnets in the various accelerators involved~\cite{Evans:2008zzb}. 
Furthermore, the length of the injections in 2015 and 2016 was limited by the maximal tolerable heat
load in the arcs due to electron clouds (2015) and safety considerations in the LHC injection system 
with very luminous beams (2016)~\cite{Salvachua:2750272}.
The bunch crossings are numbered with an identification number (BCID) in the range 1--3564.
The specific pattern of filled and empty bunch crossings used in a single fill is known as the ``filling scheme''; a typical filling scheme is composed of
long strings of consecutive bunches, up to 72 bunches long, called a ``train'', with the individual trains
separated by gaps of varying lengths. Generally, filling schemes also include some number of noncolliding
bunch crossings, where one beam is filled but the other remains empty; these can be used to study effects from
beam-induced background. The two LHC beams are designated ``beam 1'' and ``beam 2'', where beam 1 (beam 2) circulates in
the clockwise (counterclockwise) direction, as viewed from above~\cite{Evans:2008zzb}.

For Run 2 of the LHC, the period from 2015 to 2018 featuring \pp collisions at \highEnergy, the CMS luminosity
systems were significantly upgraded and expanded. We report the results for the first two years~\cite{evian_2017}, in which
the operational conditions feature a wide range in the number of colliding bunches \nb and 
instantaneous luminosity, reaching a maximum of 2232 and 2208, and $0.5\ten{34}$ and 
$1.5\ten{34}\percms$ in 2015 and 2016, respectively.  In the majority of \pp LHC fills in Run 2,
the bunches are spaced 25\unit{ns} apart.  The initial Run 2 data set delivered with a bunch spacing of 50\unit{ns} is 
negligibly small~\cite{SMP-15-007}, and hence not included in this paper.
In this paper, ``pileup'' refers to the total number of \pp interactions in a single bunch crossing, and ``out-of-time pileup'' refers to 
additional \pp collisions in nearby bunches. For a total inelastic \pp cross section of 80\unit{mb}~\cite{Skands:2014pea,Sjostrand:2014zea},
the pileup during nominal physics data-taking conditions in 2015 (2016) extended from 5 to 35 (10 to 50) with an 
expected average ($\mu$) of about 14 (27) \pp interactions.

This paper is structured as follows. In Section~\ref{SEC:Detector} the CMS detector is described 
with special emphasis on the subdetectors used to derive observables for luminosity estimation, 
and in Section~\ref{SEC:lumi_algos} we review the methods to obtain the luminosity information. 
Section~\ref{SEC:Calibration} describes the \vdM scan calibration method and the associated systematic 
uncertainty. Sections~\ref{SEC:RateCorrections} and~\ref{SEC:Performance} outline the corrections applied to 
the luminosity algorithms and their resulting performance, respectively.  Finally, Section~\ref{SEC:SysErr} 
outlines the sources of corrections and the associated systematic uncertainties, and presents the main results.  A summary 
is given in Section~\ref{SEC:Summary}.

\section{The CMS detector}
\label{SEC:Detector}

The CMS detector is a multipurpose apparatus designed to study high-\pt physics processes in \pp
collisions, as well as a broad range of phenomena in heavy ion collisions. The central element of 
CMS is a 3.8\unit{T} superconducting solenoid, 13\unit{m} in length and 6\unit{m} in diameter. 
Within the solenoid volume are---in order of increasing radius from the beam pipe---a silicon 
pixel and strip tracker of high granularity for measuring charged particles up to pseudorapidity 
($\eta$) of $\pm2.5$; a lead tungstate crystal electromagnetic calorimeter for measurements of the 
energy of photons, electrons, and the electromagnetic component of hadronic showers (``jets''); and a brass and scintillator hadron calorimeter,
each composed of a barrel and two endcap sections, for jet energy measurements.
The forward hadron (HF) calorimeter uses steel as an absorber and quartz fibers as the sensitive
material. The two halves of the HF are located 11.2\unit{m} from the interaction region, one on each end, and
together they provide coverage in the range $3.0 < \abs{\eta} < 5.2$, hence extending the pseudorapidity coverage provided by the barrel and endcap detectors.
Outside the magnet, and within the range $\abs{\eta} < 2.4$, is the muon system~\cite{MUO-16-001}, which is embedded in the 
iron flux-return yoke.  It is composed of detection planes 
made using three technologies: drift tubes (DTs) in the barrel, cathode strip chambers (CSCs) in the endcaps, 
and resistive plate chambers (RPCs) both in the barrel and in the endcaps. 

Events of interest for physics are selected using a two-tiered trigger system~\cite{Khachatryan:2016bia}. The 
first-level trigger, composed of custom hardware processors, uses information from the calorimeters 
and muon detectors to select events at a rate of around 100\unit{kHz}.  The second level, known 
as the high-level trigger, consists of a farm of processors running a version of the full 
event reconstruction software optimized for fast processing, and reduces the event rate to around 1\unit{kHz} before data storage.

Several subdetectors, although not part of the main CMS data acquisition (DAQ) system, provide 
additional inputs (\eg, binary logic signals) to the triggering system. 
The two beam monitors closest to the IP for each LHC experiment, the 
Beam Pick-up Timing for eXperiments (BPTX) detectors~\cite{bptx}, are reserved for timing measurements. They are
located on either side of IP 5 at a distance of approximately 175\unit{m}. The BPTX system can be used to
provide a set of zero-bias events (\ie, events from nominally colliding bunch crossings but without a
requirement for any specific activity in the event) by requiring a coincidence between the two BPTX sides.
To suppress noise in triggers with high background, the presence of this coincidence is
typically required~\cite{Khachatryan:2016bia}.

The knowledge of the integrated luminosity requires stability over long periods of time, 
and hence benefits greatly from redundant measurements whose combination can lead to an 
improved precision.
 To that end, several upgrades were completed during the first LHC long shutdown (LS1), the 
transition period between LHC Run 1 (2009--2012) and Run 2. The main luminosity subdetectors 
(luminometers) in Run 1 were the silicon pixel detector and the HF. 
The HF back-end electronics, which were upgraded during LS1, consist of two independent readout 
systems:  a primary readout over optical
links for physics data taking, and a secondary readout using Ethernet links,
explicitly reserved for luminosity data.  In addition, 
two other luminometers were designed, constructed, and commissioned: the Pixel Luminosity 
Telescope (PLT)~\cite{Lujan:2017kvh} and the Fast Beam Conditions Monitor (BCM1F)~\cite{Hempel:2017nvn}.
Finally, a separate DAQ system was developed that is independent of the central DAQ 
system~\cite{Khachatryan:2016bia,Tapper:1556311}, so that HF, PLT, and BCM1F data, as well as LHC 
beam-related data, are collected and stored in a time- rather than event-based manner.

\begin{figure*}[!tbh]
  \centering
  \includegraphics[width=0.99\textwidth]{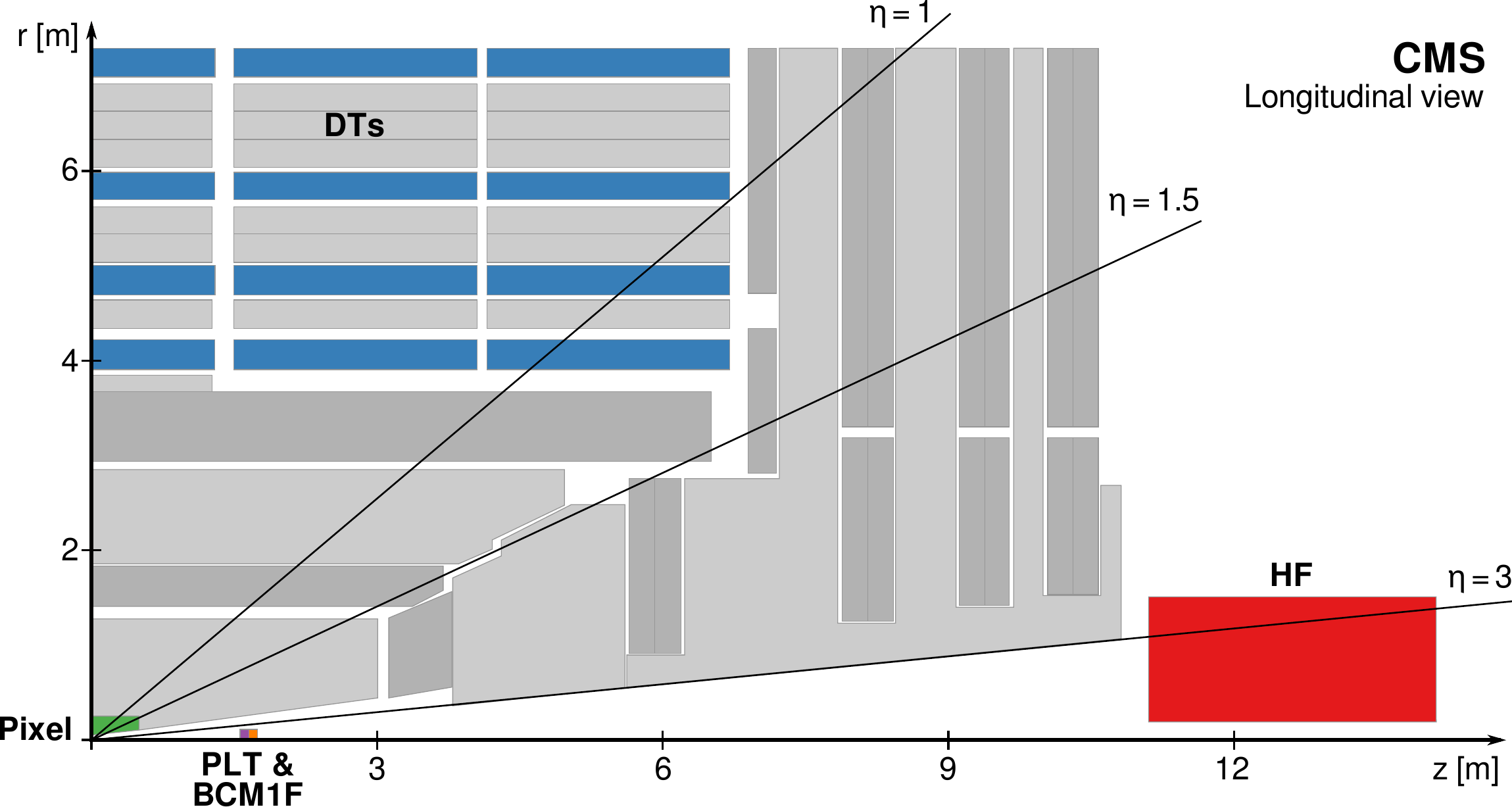}
  \caption{\label{fig:luminometers}
    Schematic cross section through the CMS detector in the $r$-$z$ plane. The main luminometers in Run 2, 
    as described in the text, are highlighted, showing the silicon pixel detector, PLT, BCM1F, DTs, and HF. 
    The two RAMSES monitors used as a luminometer in Run 2 are located directly behind HF. 
    In this view, the detector is symmetric about the horizontal and vertical axes, so only one quarter is shown here. 
    The center of the detector, corresponding to the approximate position of the \pp collision point, is located at the origin. 
    Solid lines represent distinct $\eta$ values.
  }
\end{figure*}

The luminometers, along with the accompanying algorithms used to estimate the instantaneous luminosity 
in Run 2, are briefly described in the following. Figure~\ref{fig:luminometers} shows an overview of
the position of these luminometers within CMS.
A more detailed description of the rest of the CMS detector, together with a definition of the 
coordinate system used and the relevant kinematic variables, is reported in Ref.~\cite{CMS}.

\subsection{Silicon pixel cluster counting} 

\label{sec:PCC}

\begin{figure}[htp]
\centering
\includegraphics[width=0.495\textwidth]{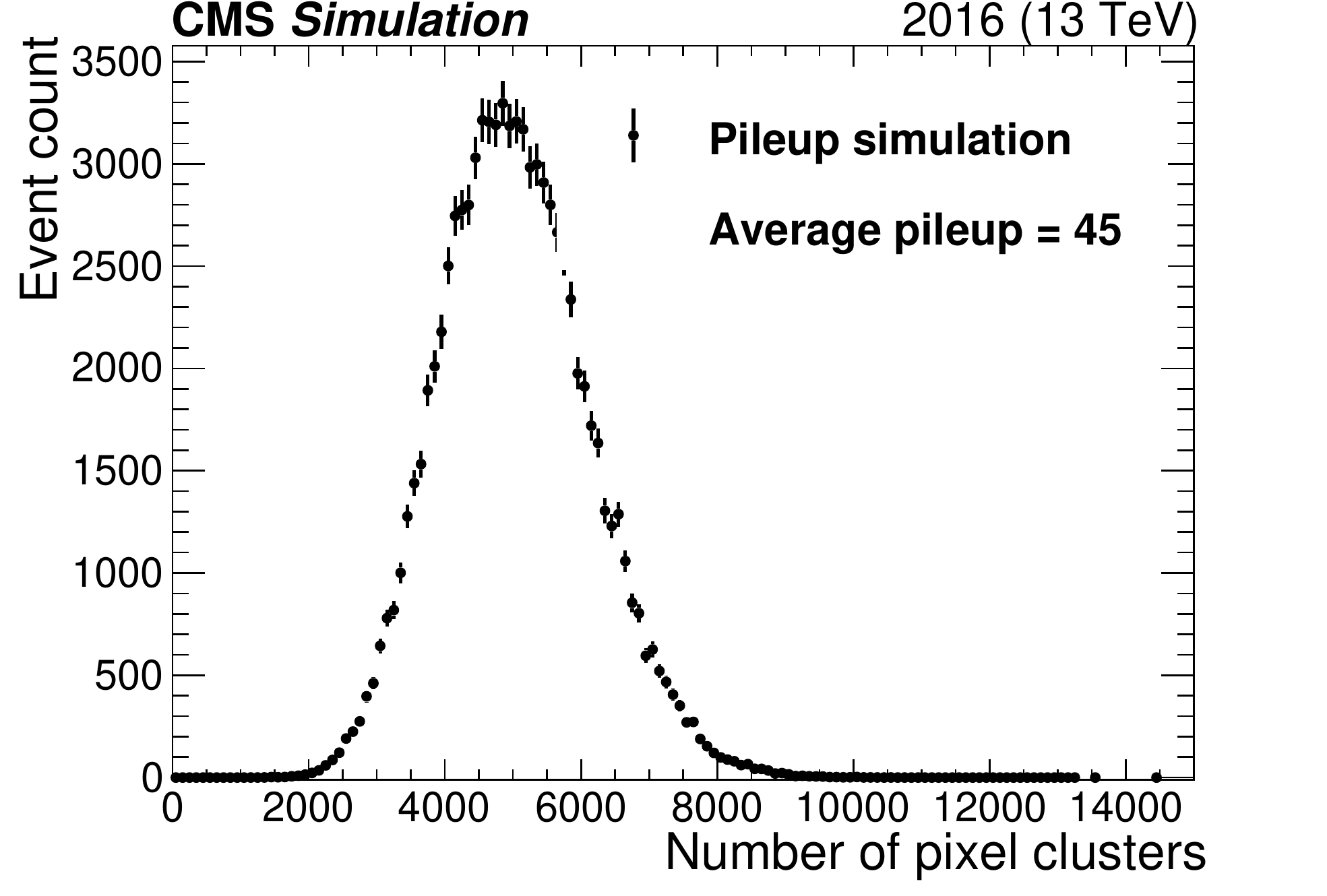}
\includegraphics[width=0.495\textwidth]{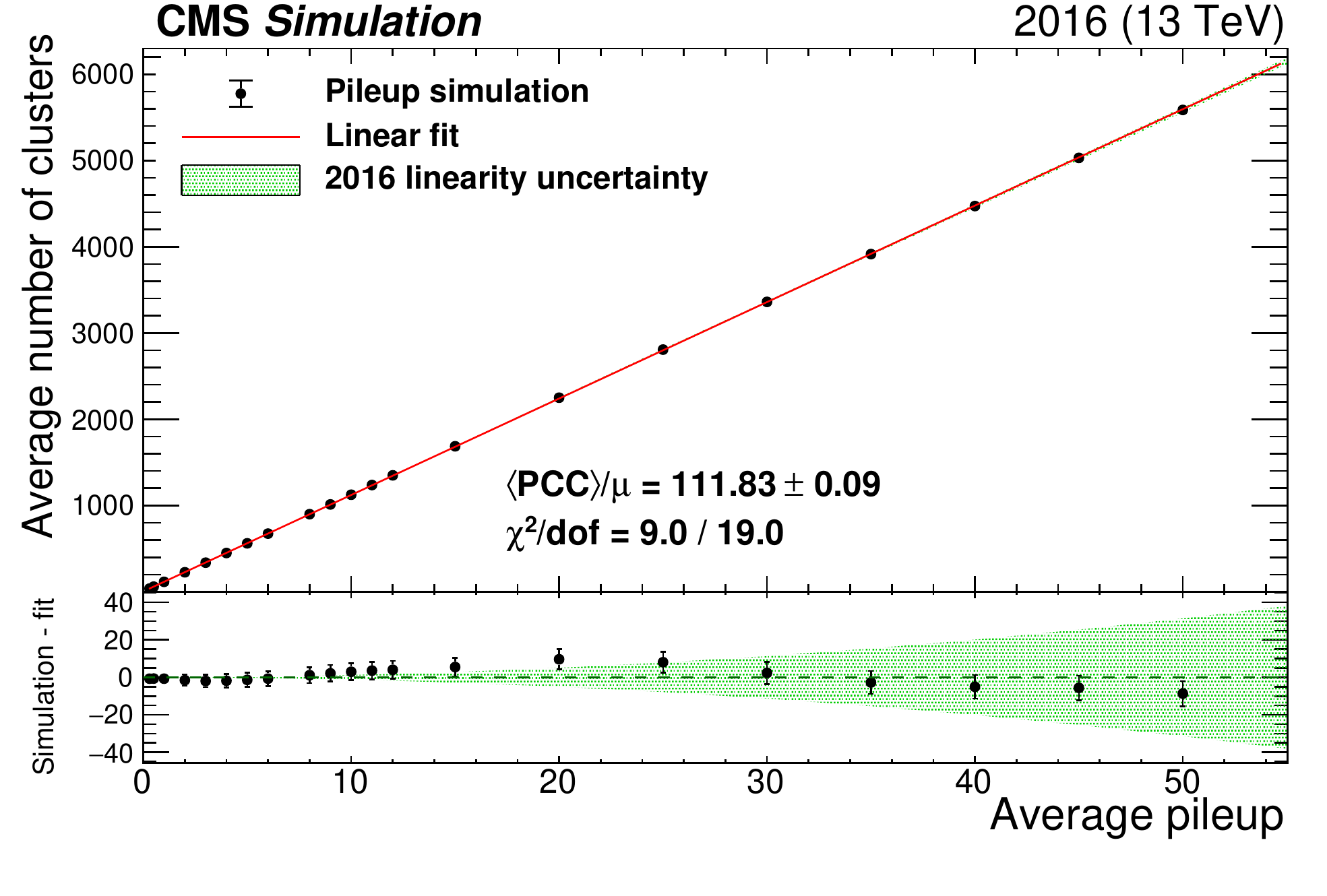}
\caption{The \cmsLeft plot shows the number of pixel clusters and their statistical uncertainty from simulation 
of pileup following a Poisson distribution with a mean of 45. 
The \cmsRight plot shows the mean number of pixel clusters from simulation as a function of mean
pileup.  The red curve is a first-order polynomial fit with slope and $\chi^2/\text{dof}$ values 
shown in the legend. Only pixel modules considered for the PCC measurement in data are included.
The lower panel of the \cmsRight plot shows the difference between the simulation and the linear fit in black points.
The green band is the final linearity uncertainty for the 2016 data set.
\label{fig:pcc_linearity_mc}}
\end{figure}

The pixel cluster counting (PCC) method, which uses the mean number of pixel clusters in the silicon 
pixel detector, exploits the very large number of pixels in the inner part of the CMS tracking 
system.  The number of pixels in 2015--2016 was about $7\ten{7}$, which means that the probability of a 
given pixel being hit by two different charged particles from the same bunch crossing is 
exceedingly small.  The mean number of pixel clusters in simulated zero-bias events is 
of the order of 100 per \pp collision, although the precise 
mean depends on the fraction of the detector used for a given data set.  Assuming each pixel cluster 
comprises five pixels and using a typical pileup for the 2016 running of $\mu= 27$, 
the fraction of pixels hit in a typical bunch crossing is roughly: 
\begin{linenomath}
\begin{equation}
f = \frac{N^{\text{hit}}_{\text{pixel}} }{ N^{\text{total}}_{\text{pixel}}} \simeq \frac{ 100 \times 5 \times 27}{ 7 \times 10^7} = 0.02\%.
\label{eq:fraction} 
\end{equation}
\end{linenomath}
The probability of accidental overlap between pixel clusters is correspondingly small, and, as a 
consequence, the number of pixel clusters per bunch crossing is linearly dependent 
on pileup, and therefore an accurate measure of instantaneous luminosity. {\tolerance=800 Simulated \pp
collision events that contain only in-time pileup and detector noise are generated using
\PYTHIA version 8.223~\cite{Sjostrand:2014zea} with the CUETP8M1~\cite{Khachatryan:2015pea,Skands:2014pea}
tune. The simulated events include a full simulation of the CMS detector response based on
\GEANTfour~\cite{geant}.  For the sake of simplicity, the number of pileup interactions present in
each simulated event is randomly generated from a Poisson distribution with $\mu$ up to 50.
Figure~\ref{fig:pcc_linearity_mc} shows a representative PCC distribution at $\mu=45$ and the average
PCC as a function of $\mu$.  The latter distribution is fitted with a first-order polynomial, assuming
no correlations among different values of $\mu$.  Good agreement is seen based on the estimated
goodness-of-fit $\chi^2$ per degree of freedom (dof) value of about 0.5~\cite{Verkerke:2003ir},
indicating linearity under simulated conditions.\par}

Only the components (modules) of the pixel subdetector that are stable for the entire period of data 
taking are used for the PCC rate measurements, 
excluding pixel modules known to be defective or significantly affected by the limited size of the 
readout buffer~\cite{TRK-11-001}. The measured \sigmaVis for PCC, \sigmaVisPCC, therefore depends on the 
data-taking period (\ie, one calibration per year).

\subsection{Primary vertex counting}
The primary vertex counting (PVC) method uses the vertices that have been reconstructed using the tracks in
the CMS detector. For this method, a good primary vertex is defined to be one with 11 or more tracks.  This
requirement is sufficient to suppress spurious vertices~\cite{TRK-11-001}, and results in better vertex resolution.

The PVC method is simple and robust, but suffers from mild nonlinearity effects when there are many collisions
in a single bunch crossing.  There are two competing effects.  In one effect, primary vertices from two collisions
occurring close to one another in space are merged, leading to an undercounting of vertices.
In the other effect, the very large numbers of tracks associated with numerous collisions can produce spurious
vertices, leading to overcounting. The precision with which these effects are understood falls short of the
$\approx$1\% level needed for luminosity studies. However, during \vdM scans these effects are minimal 
because of the very low pileup, and so PVC is very useful as a validation tool for the \vdM analysis in
the measurement of beam-dependent parameters.

\subsection{Forward hadron calorimeter}
\label{SEC:HF}

The HF luminosity measurement uses the separate readout described above, so the measurement
can be performed at the full 40\unit{MHz} bunch crossing rate.
The back-end electronics upgrade during LS1 added new electronics using field-programmable gate array (FPGA) 
technology such that several features of the readout 
were separately programmable for luminosity histogramming, \ie, identifying and counting the readout channels.
Although the whole HF is capable of being read out for luminosity use, only the two outer rings in $\eta$ are used 
to ensure uniform occupancy and minimize minor nonlinearities expected from simulation.

The computation of the HF observable is based on the occupancy method (HFOC).  In this method, 
the fraction of channel measurements above an analog-to-digital converter (ADC) threshold is
used for each bunch slot in a configurable time window.  The ADC threshold is set high enough to avoid most 
noise and as low as possible otherwise.  Both the ADC threshold and the integration time of the histograms 
between readouts are configurable, but they were fixed during data taking in 2015 and 2016.  The 
number of valid measurements is also stored, so the fraction of events with hits above threshold can be computed.

\subsection{Pixel Luminosity Telescope}
The PLT is a dedicated system for measuring luminosity using silicon pixel sensors, installed
in Run 2, at the beginning of 2015.  There are a total of 48 sensors arranged into 16 ``telescopes'',
eight at either end of CMS outside the pixel endcap.  Each telescope contains three sensor
planes which are arranged in a triplet that faces the IP.  
The sensors measure $8{\times}8 \unit{mm}^2$, divided into
80 rows and 52 columns, although only the central region of the sensors is used to reduce the
contribution from background.  The PLT measures the rate of triple coincidences, where a hit is 
observed in all three planes, typically corresponding to a track from a particle originating at the IP. The
overall mean rate for PLT is estimated using the fraction of events where no triple coincidences are 
observed (as described in Section~\ref{SEC:lumi_algos}) in order to avoid potential systematic effects from 
overlapping tracks being counted as a single hit.

\subsection{Fast Beam Conditions Monitor}
The BCM1F measures luminosity and beam-induced background separately. {\tolerance=800 It consists of
a total of 24 sensors mounted on the same carriage as the PLT.  Single-crystal diamond 
sensors are used with split-pad metallization. Each sensor has two readout channels to keep 
the overall occupancy low, given the experimental conditions in Run 2. The BCM1F features 
a fast readout with 6.25\unit{ns} time resolution. The precise time measurement allows hits from
collision products to be separated from beam-induced background hits, while the incoming background is separated in time from the outgoing collision products
due to the position of BCM1F 1.8 m from the center of CMS.\par}

\subsection{Drift tube muon detector}
{\tolerance=800 The luminosity measurement based on the DT muon detector~\cite{MUO-16-001} is based on an efficient 
trigger on a low-background physics object: muons produced in the CMS barrel.  Muon track segments 
from barrel muon DT stations are sent every bunch crossing to track finder hardware, where tracks are built and 
later used to generate first level triggers.  The number of tracks in time windows of approximately 23 seconds is read out 
and stored in a database.  These data are used to estimate luminosity.  The rate of muons in the DTs is significantly lower 
than the rate for most other observables from other luminometers.  Thus, there are not enough muon 
tracks during the \vdM scans to provide a precise measurement of \sigmaVis, and so the system must be 
calibrated to the normalized PCC luminosity measurement.  On the other hand, the muon candidate rate has 
been observed to be linear with luminosity and rather stable over time.  The luminosity data of this system are integrated over 
all bunches.\par}

\subsection{Radiation Monitoring System for the Environment and Safety}
The Radiation Monitoring System for the Environment and Safety (RAMSES) is a monitoring subsystem of the unified supervisory 
CERN system~\cite{RAMSES2005,Ledeul:PCaPAC2018-FRCC3}.  There are 10 ionization chambers filled with air at atmospheric pressure
that are used as monitors installed in the CMS experimental cavern. They are sensitive to ionizing radiation and 
can monitor the ambient dose equivalent rate.  Thus, they generate alarms and interlocks to ensure the safety 
of the personnel.  This system is maintained and calibrated by the LHC radiation protection group.

While not designed as a luminometer, the two chambers with the highest rates (designated PMIL55X14 and PMIL55X15) have been used to produce a luminosity measurement
with good linearity and stability over time. However, similarly to the DT luminosity measurement, the overall rates
are too low for bunch-by-bunch measurements or extracting an absolute calibration during \vdM scans. The
RAMSES luminosity is thus calibrated to the normalized PCC luminosity measurement and is used as an additional measurement
for assessing the luminometer stability with time.

\section{Luminosity determination algorithms}
\label{SEC:lumi_algos}

Each bunch crossing gives rise to a certain number of \pp interactions.  In a given luminometer, 
each interaction results in some number of observables (\eg, hits, tracks, or clusters).  If one 
averages over several unbiased measurements, the mean number of observables is
\begin{linenomath}
\ifthenelse{\boolean{cms@external}}
{\begin{equation}
\label{eqn:nobs}
\begin{aligned}
\left< N_{\text{observables}} \right> & = \left< N_{\text{observables/interaction}} \right> \left< N_{\text{interactions}} \right>\\
& \equiv \left< N_{\text{observables/interaction}} \right> \mu,
\end{aligned}
\end{equation}}
{\begin{equation}
\label{eqn:nobs}
\left< N_{\text{observables}} \right> = \left< N_{\text{observables/interaction}} \right> \left< N_{\text{interactions}} \right>
\equiv \left< N_{\text{observables/interaction}} \right> \mu,
\end{equation}}
\end{linenomath}
where the average number of interactions per bunch crossing is denoted by 
$\mu$, in keeping with the Poisson nature of the underlying probability distribution.  Typically these 
observables are averaged over seconds or tens of seconds.

To measure the instantaneous luminosity, we use the fact that $\mu$ is proportional to the single-bunch 
crossing instantaneous luminosity \Lb via:
\begin{linenomath}
\begin{equation}
\mu = \frac{\sigma\Lb}{\freq},
\label{eqn:mulumi}
\end{equation}
\end{linenomath}
where $\freq=11\,245.6\unit{Hz}$ is the LHC revolution frequency during collisions, and $\sigma$ 
is the total interaction cross section. At the LHC, \Lb is typically expressed in units of $\text{Hz}/\mu\text{b} \equiv 10^{30}\percms$.

{\tolerance=800 Two algorithms have been developed for extracting the instantaneous luminosity.  One method is 
rate-scaling, where the raw rate of observables is scaled with calibration constants to the 
luminosity.  Rearranging Eqs.~(\ref{eqn:nobs}) and~(\ref{eqn:mulumi}), one can estimate the 
instantaneous luminosity using the average number of observables at a given time:
\begin{linenomath}
\begin{equation}
\Lb = \frac{\left< N_{\text{observables}} \right>}{\left< N_{\text{observables/interaction}} \right>} \frac{\freq}{\sigma} 
\equiv \left< N_{\text{observables}} \right>  \frac{\freq }{\sigmaVis}.
\label{eqn:ratescaling}
\end{equation}
\end{linenomath}
Luminosity is estimated from PCC, PVC, DT, and RAMSES data using the rate-scaling algorithm.\par}

The second method (zero counting) uses the average fraction of bunch crossings where no
observables in a detector are produced.  This zero fraction is then used to infer the mean number of
observables per bunch crossing. The principal advantage of the zero-counting method is that it is not affected
by cases where two or more separate signals overlap in the detector and produce only one reconstructed observable.

Assuming that the probability of 
no observables in a single collision is $p$, then the probability of no observables seen in a 
bunch crossing with $k$ interactions is thus simply $p^k$.  Averaged over a large number of 
bunch crossings, with the number of interactions per bunch crossing distributed according to 
a Poisson distribution of mean $\mu$, the expected fraction of events with zero observables 
recorded, $\langle f_0 \rangle$, can be expressed as:
\begin{linenomath}
\begin{equation}
\langle f_0 \rangle = \sum_{k=0}^{\infty}\frac{\mathrm{e}^{-\mu}\mu^k}{k!}p^k = \mathrm{e}^{-\mu\left(1-p\right)}.
\label{eqn:zerocounting}
\end{equation}
\end{linenomath}
The logarithm of Eq.~(\ref{eqn:zerocounting}) is proportional to the the mean number of \pp
interactions per bunch crossing, and hence to the \Lb according to Eq.~(\ref{eqn:mulumi}): 
\begin{linenomath}
\begin{equation}
\Lb = \mu \frac{\freq}{\sigma} = -\ln{\langle f_0 \rangle}\frac{1}{1-p} \frac{\freq}{\sigma}
\equiv -\ln{\langle f_0 \rangle} \frac{\freq}{\sigmaVis}.
\label{eqn:zerocounting2}
\end{equation}
\end{linenomath}
The actual value of $p$ does not need to be known beforehand, since it is effectively absorbed in \sigmaVis, although it
could be extracted from the measured \sigmaVis value. The raw inputs from HFOC, PLT, and BCM1F are converted to
luminosity using the zero-counting method.

\section{Absolute luminosity calibration}
\label{SEC:Calibration}

Any luminometer requires an externally determined absolute calibration.
Approximate \sigmaVis values can be obtained using Monte Carlo (MC) simulation, but these ultimately rely on 
theory, \ie, the inelastic \pp cross section, and are not expected to be reliable at the percent level that represents the target accuracy for the CMS luminosity measurement.
At the LHC, the 
precision of theoretical predictions for SM processes is typically limited by the knowledge of the parton distribution functions in the proton.
Although methods independent of 
theoretical assumptions have been proposed at the expense of introducing correlations between low- and high-$\mu$ 
data-taking periods~\cite{Salfeld-Nebgen:2018bdp}, a more precise and purely experimental method to determine the 
luminosity is based on the \vdM scan technique, which is used in this paper.

Beam-separation scans are therefore performed to obtain calibrated \sigmaVis for the 
luminosity measurement.  These were pioneered by Simon van der Meer at the ISR~\cite{VdM:1968}, extended by 
Carlo Rubbia to the case of a collider with bunched beams~\cite{VdM:1977}, and have been extensively used by all 
four major LHC experiments~\cite{Aaij:2014ida,Aaboud2016}.  The key principle of the \vdM scan method is to 
infer the beam-overlap integral from the rates measured at different beam separations---provided the beam 
displacements are calibrated as absolute distances---as opposed to measuring the bunch density functions 
directly.  The basic formalism is described in the following.

\subsection{The van der Meer method}
\label{SEC:Scan}

The instantaneous luminosity for a single colliding bunch pair in a colliding-beam accelerator is given by:
\begin{linenomath}
\begin{equation}
\Lb = \frac{\freq n_1 n_2 }{\Aeff},
\label{EQ:VDM1}
\end{equation}
\end{linenomath}
where $n_1$ and $n_2$ are the numbers of particles in each of the two bunches, and \Aeff is the effective area of overlap 
between the bunches.  In general, each of the bunches will be distributed in the plane transverse to the 
beam direction, in which case $1/\Aeff$ can be replaced by an overlap integral of the bunch densities, \ie, 
\begin{linenomath}
\begin{equation}
\Lb = \freq n_1 n_2 \iint \rho_1(x,y) \rho_2(x,y) \rd x \rd y,
\label{EQ:VDM2}
\end{equation}
\end{linenomath}
where $x$ and $y$ represent the horizontal and vertical coordinates in the plane transverse to the 
beams, and $\rho_1$ and $\rho_2$ are the normalized two-dimensional density distributions for the 
two bunches. Here, we have integrated over time and the longitudinal coordinate $z$.

If one assumes that the bunch profiles can be factorized into terms depending only on $x$ and $y$~\cite{VdM:1968,VdM:1977}, then $\rho_i$ can be written 
as the product of one-dimensional density functions of the form $\rho_i(x,y) = f_i(x) g_i(y)$ 
($i=1,2$), and $1/\Aeff$ can be written
\begin{linenomath}
\begin{equation}
\frac{1}{\Aeff} = \int f_1(x) f_2(x) \rd x \int g_1(y) g_2(y) \rd y \equiv \frac{1}{\Weff} \frac{1}{\Heff},
\label{EQ:Aeff}
\end{equation}
\end{linenomath}
where \Weff and \Heff are the effective width and the effective height of the 
luminous region. For the ideal case of Gaussian-distributed bunches with the same width in both beams and undergoing head-on collisions, Eq.~(\ref{EQ:VDM2}) reduces to:
\begin{linenomath}
\begin{equation}
\Lb = \frac{\freq n_1 n_2 }{ 4 \pi \sigma_x \sigma_y},
\label{EQ:basiclumi}
\end{equation}
\end{linenomath}
where $\sigma_x$ and $\sigma_y$ are the root-mean-square (RMS) widths of the horizontal and vertical bunch profiles in either beam, respectively.     
In the case of round beams, $\sigma_x=\sigma_y \equiv \sigmab \equiv \sqrt{\smash[b]{\epsilon_\text{N} \betastar/\gamma}}$, where
$\epsilon_\text{N}$ is the so-called normalized emittance,  
$\gamma$ the relativistic Lorentz factor, and $\betastar$ corresponds to the value of the optical function
$\beta$ at the IP~\cite{lhcdesign}.

We designate the luminosity when the beams are displaced with respect to each other by an amount $w$ in the
$x$ direction, or an amount $h$ in the $y$ direction, as $\Lumi(w,h)$.  As shown in
Ref.~\cite{VdM:1968}, when a separation scan is performed in the $x$ direction, in which $w$ is varied in a
systematic way from $-\infty$ to $+\infty$, the effective width can be determined from:
\begin{linenomath}
 \begin{equation}
 \Weff = \frac{\iint  f_1(x) f_2(x-w) \rd x \rd w  }{  \int f_1(x) f_2(x) \rd x } = \frac{\int \Lb(w,0) \rd w  }{  \Lb(0,0)},
 \label{EQ:Weff}
 \end{equation}
\end{linenomath}
where common normalization factors have been canceled in the second step.
Similarly, if a scan is performed in the $y$ direction, the effective beam-overlap height is given by
\begin{linenomath}
 \begin{equation}
 \Heff = \frac{\iint  g_1(y) g_2(y-h) \rd y \rd h  }{  \int  g_1(y) g_2(y) \rd y } = \frac{\int \Lb(0,h) \rd h  }{  \Lb(0,0)}.
  \label{EQ:Heff}
 \end{equation} 
\end{linenomath}
For Gaussian-distributed bunches, the resulting scan curves, $\Lumi(w,0)$ and $\Lumi(0,h)$, are also Gaussian 
with RMS widths of $\Sigma_x= \Weff = \sqrt{2}\sigma_x$ and $\Sigma_y = \Heff = \sqrt{2}\sigma_y$, yielding
\begin{linenomath}
 \begin{equation}
\Lb = \frac{\freq n_1 n_2 }{ 2 \pi \Sigma_x \Sigma_y}.
\label{EQ:basiclumi2}
\end{equation}
\end{linenomath}
Equations~(\ref{EQ:Weff}) and~(\ref{EQ:Heff}) are quite general, and do not depend on the 
assumption of Gaussian-distributed bunches.  Indeed, it is frequently the case that simple Gaussians do not 
provide an adequate description of the scan-curve data.  In such cases, we use double-Gaussian functions 
of the form 
\begin{linenomath}
\ifthenelse{\boolean{cms@external}}
{\begin{multline}
\label{eq:DG}
f(x) = \frac{1}{\sqrt{2\pi}}\Bigg[\frac{\epsilon_{x}}{\sigma_{1x}}\exp{\Bigg(-\frac{x^2}{2\sigma_{1x}^2}\Bigg)}\\
+ \frac{1-\epsilon_{x}}{\sigma_{2x}}\exp{\Bigg(-\frac{x^2}{2\sigma_{2x}^2}\Bigg)}\Bigg],
\end{multline}}
{\begin{equation}
\label{eq:DG}
f(x) = \frac{1}{\sqrt{2\pi}}\Bigg[\frac{\epsilon_{x}}{\sigma_{1x}}\exp{\Bigg(-\frac{x^2}{2\sigma_{1x}^2}\Bigg)} + \frac{1-\epsilon_{x}}{\sigma_{2x}}\exp{\Bigg(-\frac{x^2}{2\sigma_{2x}^2}\Bigg)}\Bigg],
\end{equation}}
\end{linenomath}
where $\epsilon_{x}$ is the fraction of the Gaussian with width $\sigma_{1x}$.
Normally the Gaussian with the smaller width $\sigma_{1x}$ is considered the core Gaussian, while the 
Gaussian with the larger width $\sigma_{2x}$ is used to fit the tails of the scan curve.  Similar relations
apply for the $y$ coordinate.  The effective value of $\Sigma_i$ ($i=x,y$) is then given by
\begin{linenomath}
\begin{equation}
  \label{eq:EffectiveSigma}
  \Sigma_i  =   \frac{\sigma_{1i}\,\sigma_{2i}}{\epsilon_i\sigma_{2i}+ (1-\epsilon_i)\sigma_{1i}}.
\end{equation}
\end{linenomath}
To calibrate a given luminosity algorithm, the absolute luminosity computed from beam parameters via Eq.~(\ref{EQ:basiclumi2})
is used in conjunction with Eq.~(\ref{eqn:mulumi}) to obtain
\begin{linenomath}
\begin{equation}
  \sigmaVis = \muVis \frac{2 \pi \Sigma_x \Sigma_y }{ n_1 n_2},
  \label{EQ:sigmaVis}
\end{equation}
\end{linenomath}
where \muVis is the visible interaction rate.  In this analysis, \muVis is taken as the arithmetic mean of the 
peak values from $\Lumi(w,0)$ and $\Lumi(0,h)$ in scans that are performed sufficiently close in time to 
minimize the impact of varying bunch distributions over the course of a fill.
Equation~(\ref{EQ:sigmaVis}) therefore provides a direct calibration of the visible cross section for each algorithm in terms of $\Sigma_x \Sigma_y$
and $n_1 n_2$.

In the LHC, bunches typically cross at a small angle $\phi$ in the horizontal plane at IP 5.  This introduces a reduction in the luminosity relative to the 
case of head-on collisions~\cite{Grafstrom:2015foa}, given by:
\begin{linenomath}
\begin{equation}
\frac{\Lumi}{\Lumi_0} =  \left[ 1 + \left( \frac{\sigma_z} {\sigma_x} \tan \frac{\phi}{2} \right)^2 \right]^{-1/2},
\label{EQ:crossingAngle}
\end{equation}
\end{linenomath}
where $\sigma_x$ is the width of the luminous region in the crossing plane and $\sigma_z$ is the 
width in the longitudinal direction.   For typical LHC physics running conditions in 2016, 
$\phi \simeq 140\unit{$\mu$rad}$, $\sigma_x \simeq 12\mum$, and $\sigma_z \simeq 8\unit{cm}$,
and so the reduction from Eq.~(\ref{EQ:crossingAngle}) is around 10\%~\cite{Hostettler:2018}.
The \vdM scans are typically carried out under special conditions, where $\phi=0$, as described in the following.
The values of \sigmaVis do not depend on the crossing angle.

\subsection{Analysis of vdM scan data}
\label{SEC:VdMScanData}

While \sigmaVis does not depend on beam conditions, the LHC delivers beams under special conditions to
improve the precision of measurements and to reduce systematic effects.  The \vdM filling schemes are
characterized by a low number of colliding bunch pairs at IP 5 ($\nb=30$--50). The bunches are widely separated from
each other in the LHC orbit, to reduce the effect of afterglow (as discussed in
Section~\ref{SUBSEC:afterglow}). Special beam optics with $\betastar\approx 19\unit{m}$ and transverse emittance of
$\epsilon_\text{N} \approx 3.0\mum$ are implemented to produce a relatively large bunch size of approximately
$\sigmab = 100\mum$.  Large bunches reduce the impact of vertex reconstruction resolution in analyses where
vertex positions are utilized.  A crossing angle of 0 is used for collisions at IP 5 in \vdM scans.
To minimize the effect of potential nonlinear response in the luminometers, the target pileup
 is set to $\mu
\approx 0.6$, which is 1--2 orders of magnitude lower than typical physics fills. To achieve that goal, in
addition to the large beam size, the beams have relatively low intensities, which typically begin at
(8--9)\ten{10} protons per filled bunch, resulting in a total intensity of (3.5--4.0)\ten{12} per beam for
44 bunches.

The total beam intensities are measured with the DC current transformers
(DCCT)~\cite{Barschel:1425904}, and the bunch currents measured with the fast beam current transformers
(FBCT)~\cite{Belohrad:1267400}, and cross-checked with the longitudinal density monitors
(LDMs)~\cite{Jeff:2012zz,Jeff:1513180} and the beam quality monitors~\cite{bqm}.  Because of the low beam
intensity and low collision rate, the luminosity remains nearly constant over the course of time, in contrast
to typical physics fills~\cite{Antoniou:2293678}. The beam orbit is monitored using two systems, the Diode Orbit and Oscillation (DOROS) 
beam position monitors (BPMs)~\cite{Gasior:1476070} located near IP 5, and the
BPMs located in the LHC arcs adjacent to CMS (referred to as ``LHC arc BPMs'').  The latter are transformed to
a beam position at IP 5 using the LHC optics files that are centrally provided by LHC
operators~\cite{lhc_optics}.  The orbit is also tracked using the movements of the luminous region at IP 5 
based on the vertices reconstructed with the CMS tracker.

\begin{figure*}[!tbh]
  \centering
  \includegraphics[width=0.99\textwidth]{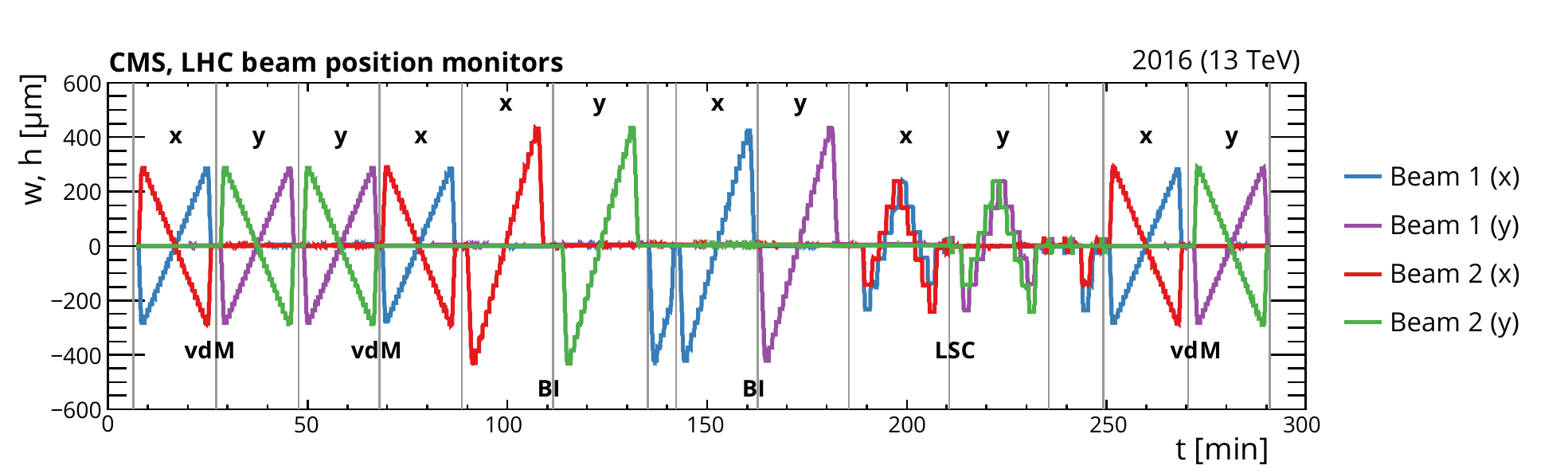}
  \caption{\label{fig:BeamPos}
    Relative change in the positions of beams 1 and 2 measured by the DOROS BPMs during fill 4954 in the horizontal ($x$) or vertical ($y$) directions, 
    as a function of the time elapsed from the beginning of the program.  
    The gray vertical lines delineate \vdM, BI, or LSC scans.
  }
\end{figure*}

The \vdM scan program at IP 5 consists of a series of $x$-$y$ scan pairs. Figure~\ref{fig:BeamPos} shows the progression
of these scans in a calibration fill, with the beam displacement measured by the DOROS BPMs~\cite{Gasior:1476070,Gasior:2313935}.

Typical scan sessions consist of at least three \vdM scan pairs, with one scan in each 
of the transverse coordinates per pair.  There are two at the start of the fill and another at the end of fill.  
In the absence of systematic effects, all scans are expected to produce compatible results. In each pair, the 
scans are typically performed first in the $x$ and then in the $y$ direction, although sometimes 
the pair is performed in the opposite order.  In the \vdM scans, the two beams are separated by up to 
6\sigmab, and scanned across one another in a sequence of 25 steps of 30\unit{s} each to 
obtain a statistically significant measurement.

Dedicated length scale calibration (LSC) scans (described in Section~\ref{sec:LS}), which are used to calibrate
the distance by which the steering magnets displace the beams, are also performed typically close in time
to the rest of the scans and using the same collision optics configuration.  The LSC implemented at IP 5 is of
a constant-separation type, in which the two beams are positioned at $-2.5$ and $-1.5\sigmab$ relative to
nominal and moved together forward in steps of 1\sigmab, maintaining the 1\sigmab separation between the
two beams, until they reach the $+2.5\sigmab$ point.  Then, their positions are swapped, and they are moved 
together backward in $-1 \sigmab$ steps back to $-2.5\sigmab$. The scan is performed once in the $x$ direction and once in the $y$
direction, with a total of 10 steps of 60\unit{s} in each direction. The LSC scans are performed with
successive forward and backward displacements for multiple measurements under slightly different conditions
in case there are compounding effects that limit precision.  The transverse position of the luminous region
is needed for this calibration and is measured using reconstructed primary vertices in CMS data. 

{\tolerance=800 To test the assumption of transversely factorizable bunch profiles in Eq.~(\ref{EQ:Aeff}), four dedicated beam-imaging
(BI) scans are performed, one for each beam and each transverse direction. One beam is kept fixed at its head-on 
position, while the other is moved and scanned in 19 steps from $-4.5$ to $+4.5\sigmab$ along $x$ or $y$
with a duration of 40\unit{s} per step. Primary vertices are reconstructed, and their 
positions are then analyzed to perform a global fit to derive the transverse bunch density distributions 
of the beams (as discussed in Section~\ref{sec:BI}).
The BI scans are also analyzed as regular beam-separation scans.
During both BI and regular \vdM scans, the transverse bunch density distributions are also determined by simultaneously
fitting the beam-separation dependence (``evolution'') on the luminosity and the luminous region position, orientation, 
and spatial width, as reflected in the reconstructed primary vertices (as discussed in Section~\ref{sec:LR}).\par}

The LHC conditions at IP 5 for the luminosity calibration fills discussed in this paper for 2015 and 2016 are summarized in Table~\ref{Tab:ScanSummary}.  

\begin{table*}[htbp]
  \centering
  \topcaption{
    Summary of the LHC conditions at IP 5 for the scan sessions in \pp collisions in 2015 and 2016.
    The column labeled $\mu$ is the average pileup corresponding to $\lumi_{\text{init}}$,
    the latter denoting the initial instantaneous luminosity.
    The columns corresponding to ``No. of scans'' indicate the total number of vdM, BI, and LSC scans that were performed in
    either transverse coordinate, counting only scans used for analysis.
  }
  \cmsTable{
    \begin{tabular}{ccccccccccc}
      \hline
      \multirow{2}{*}{Fill} & $\sqrt{s}$ & \multirow{2}{*}{Date} & \multirow{2}{*}{\nb}  & $\phi$            & $\betastar$ & \multirow{2}{*}{$\mu$} & $\lumi_{\text{init}}$ & \multicolumn{3}{c}{No. of scans}  \\
      &  [\TeVns{}]     &      &  & [$\mu$rad]        & [cm]        &       & [$\ten{30} \percms$] & \vdM  &   BI & LSC     \\ \hline
      \FillNumberI     & 13            & \DateAbvI  &  \NBunchCollI      &   0     & 1917 &       0.6   &    2.7 & 6    &  4   &     3\\
      \FillNumberIII   & 13            & \DateAbvII &  \NBunchCollII     &   0     & 1917 &       0.6   &    2.5 & \NA  &  \NA &     2\\
      \FillNumberII    & 13            & \DateAbvII &  \NBunchCollII     &   0     & 1917 &       0.6   &    2.5 & 6    &  4   &     2\\
      \hline
    \end{tabular}
  }
  \label{Tab:ScanSummary}
\end{table*}

\begin{figure*}[htbp]
  \centering
  \includegraphics[width=.495\textwidth]{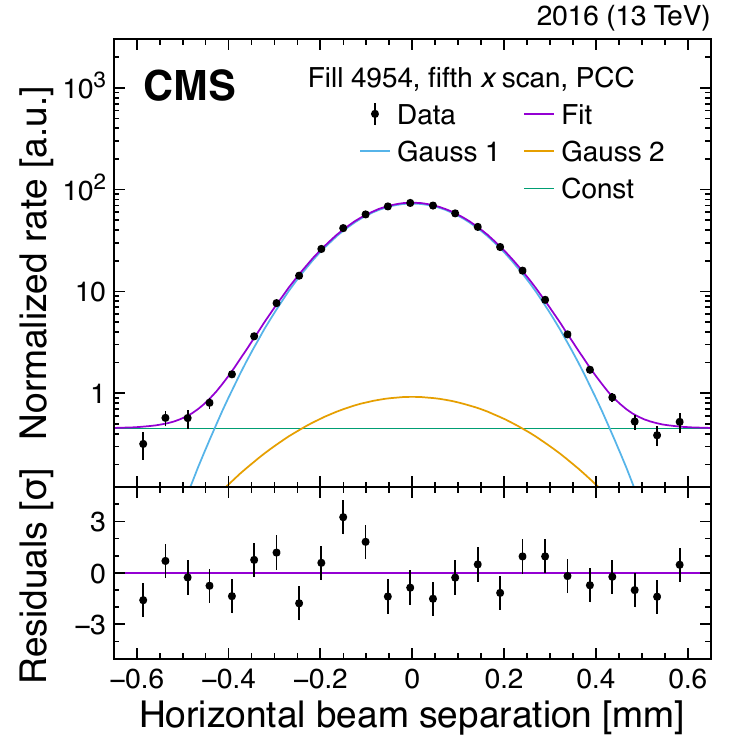}
  \includegraphics[width=.495\textwidth]{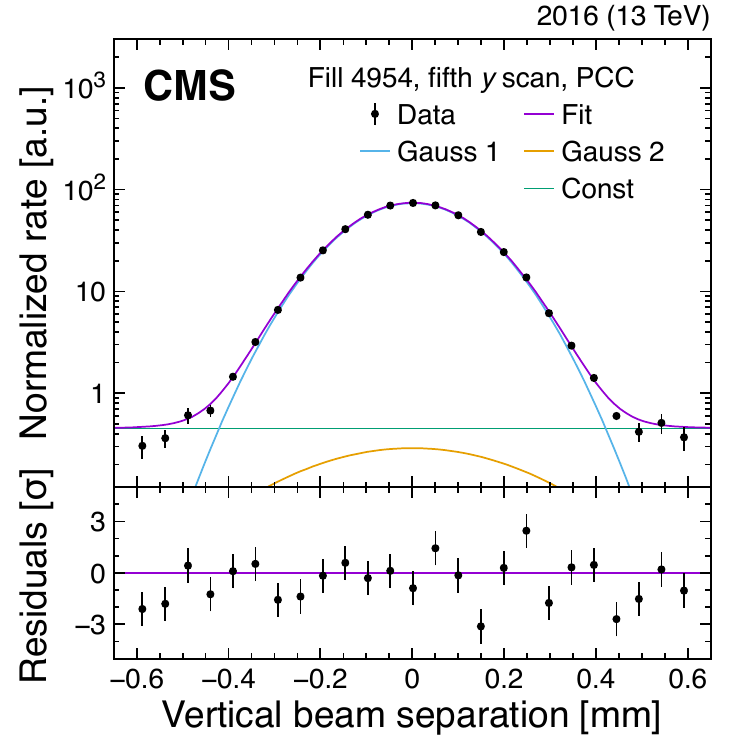}
  \caption{
    Example \vdM scans for PCC for BCID 41, from the last scan pair in fill \FillNumberII,
    showing the rate normalized by the product of beam currents and its statistical uncertainty as a function of the beam separation in the $x$ (left) and $y$ (right) direction,
    and the fitted curves. The purple curve shows the overall double-Gaussian fit, while the blue, yellow, and green curves show the first and second Gaussian components
    and the constant component, respectively.
    All corrections described in Section~\ref{sec:ScanCorrections} are applied.
    The lower panels display the difference between the measured and fitted values divided by the statistical uncertainty.
  \label{fig:Fill4954}}
\end{figure*}

\begin{figure*}[htbp]
  \centering
  \includegraphics[width=.95\textwidth]{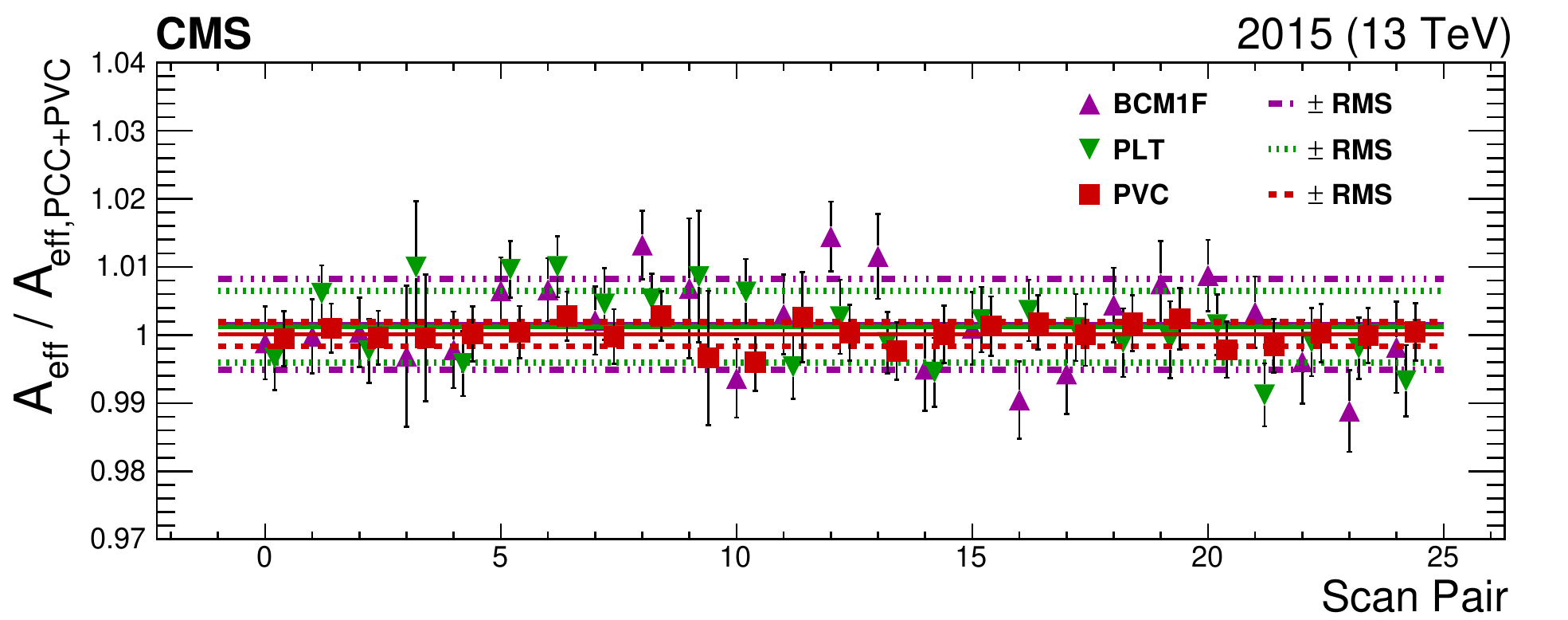}  \\
  \includegraphics[width=.95\textwidth]{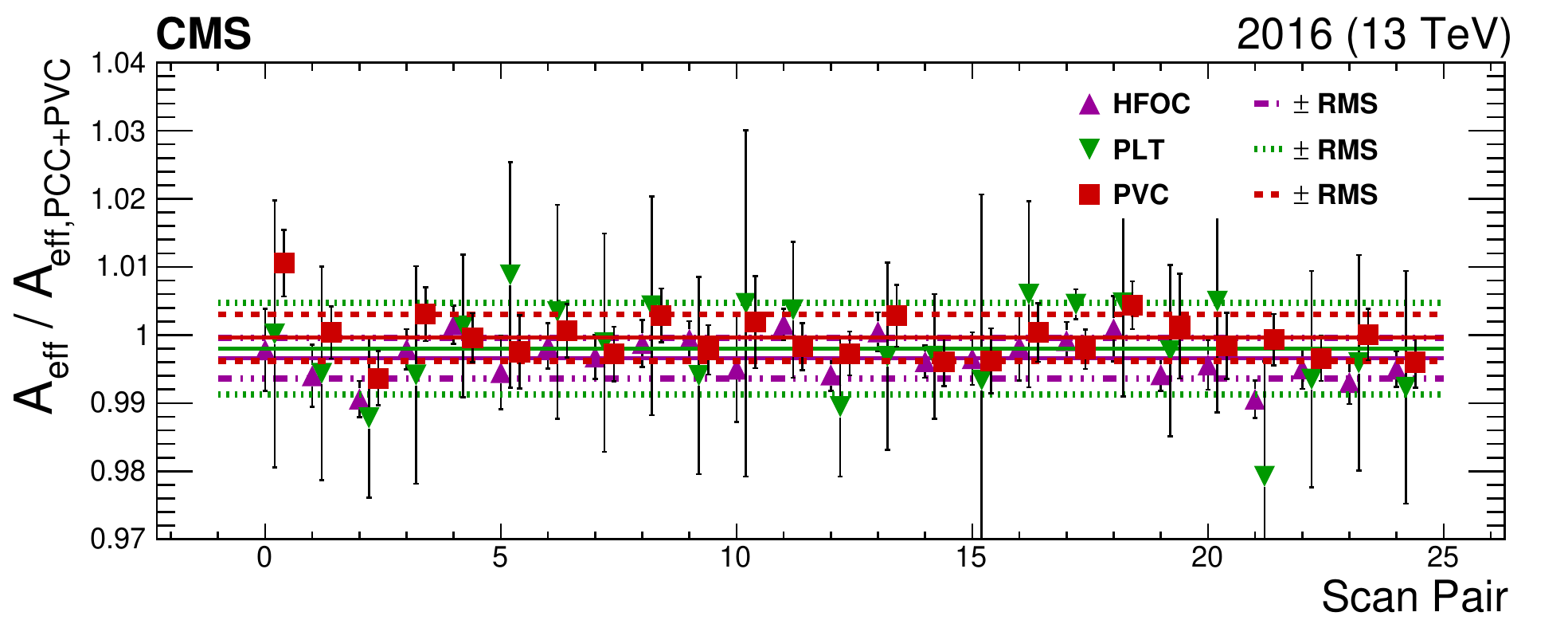}
  \caption{The two figures above show comparisons of effective area (\Aeff) of cross-check luminometers with 
    respect to the nominal PCC+PVC for fills 4266 (upper) and 4954 (lower).  The points are the ratio of 
   the \Aeff of the labeled luminometer to PCC+PVC.  There are 25 \Aeff values because there are five scan pairs with
    five BCIDs analyzed for each scan pair.  The solid lines are the average of all the \Aeff while the bands 
    are the standard deviations.  In both sets of data the  average comparison is compatible with unity within 
    or near the standard deviation.
    \label{fig:Aeff}}
\end{figure*}

Pixel data are collected for PCC and for methods involving collision vertices using a zero-bias trigger, which collects data from five BCIDs with
a total rate of approximately 20\unit{kHz}.
Figure~\ref{fig:Fill4954} shows \vdM scan data from PCC recorded in the fifth scan pair of the session 
in fill 4954. The fit function corresponds to the double-Gaussian formalism of Eq.~(\ref{eq:DG}), 
and the parameters are estimated by simultaneously fitting the PCC and PVC rate measurements. 
An additional constant term is included to estimate the background originating from noncollision sources. 
This function provides a good description of the data in a range that extends over nearly three orders 
of magnitude in rate ($\chi^2/\text{dof}\approx1$ in Fig.~\ref{fig:Fill4954}). 
For other luminometers, background rates are either negligible (PLT and PVC) or estimated and subtracted (BCM1F and HFOC) prior to the beam parameter fit. 
Since the instantaneous luminosity is relatively low, any nonlinear effect has a negligibly small impact in any method. 
The beam-width parameters (Eq.~(\ref{eq:EffectiveSigma})) measured using different luminometers are in excellent agreement, 
which is shown in Fig.~\ref{fig:Aeff} with comparisons of \Aeff with the nominal PCC+PVC results.

\begin{figure}[!htb]
\centering
\includegraphics[width=0.495\textwidth]{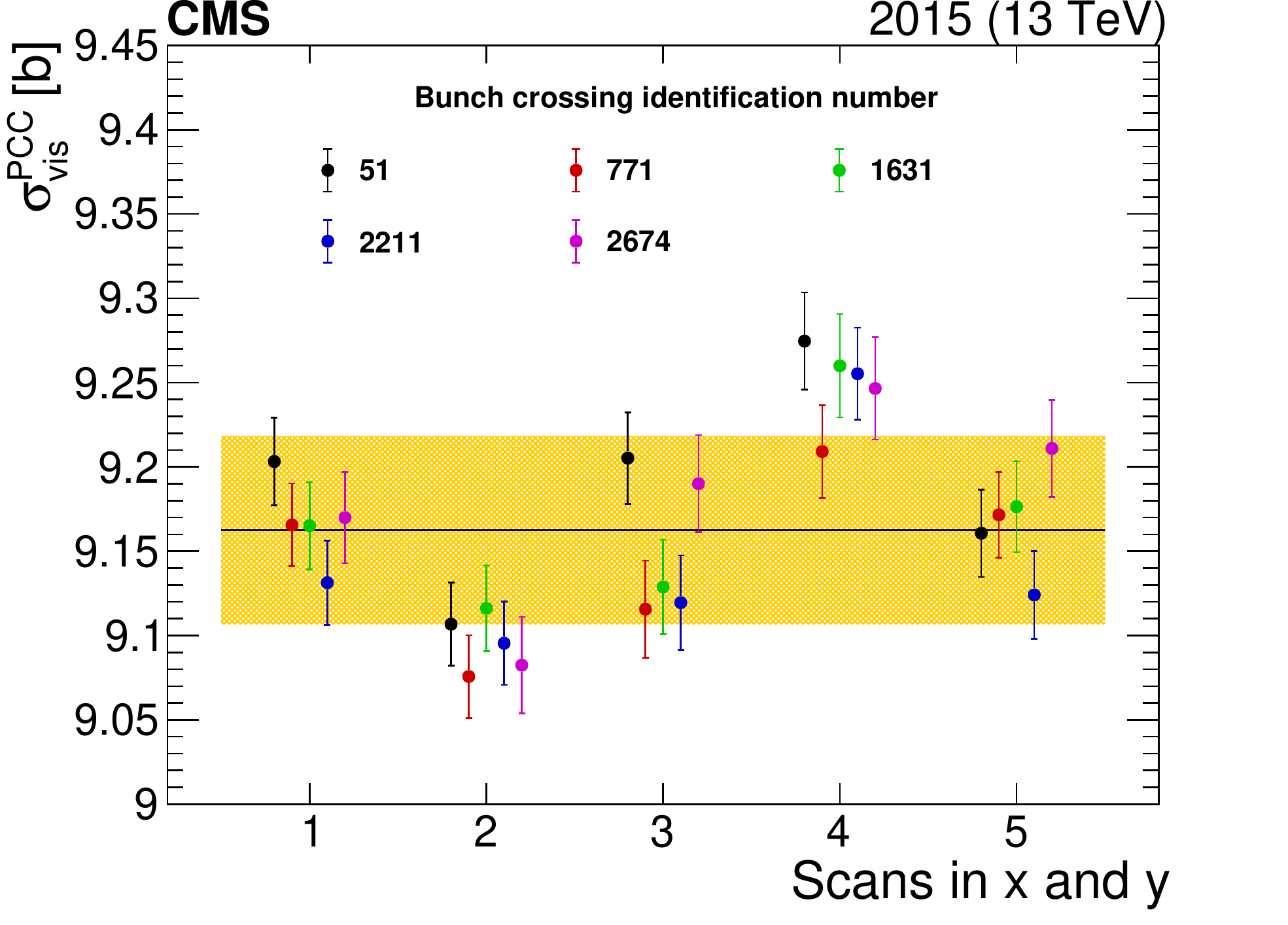}
\includegraphics[width=0.495\textwidth]{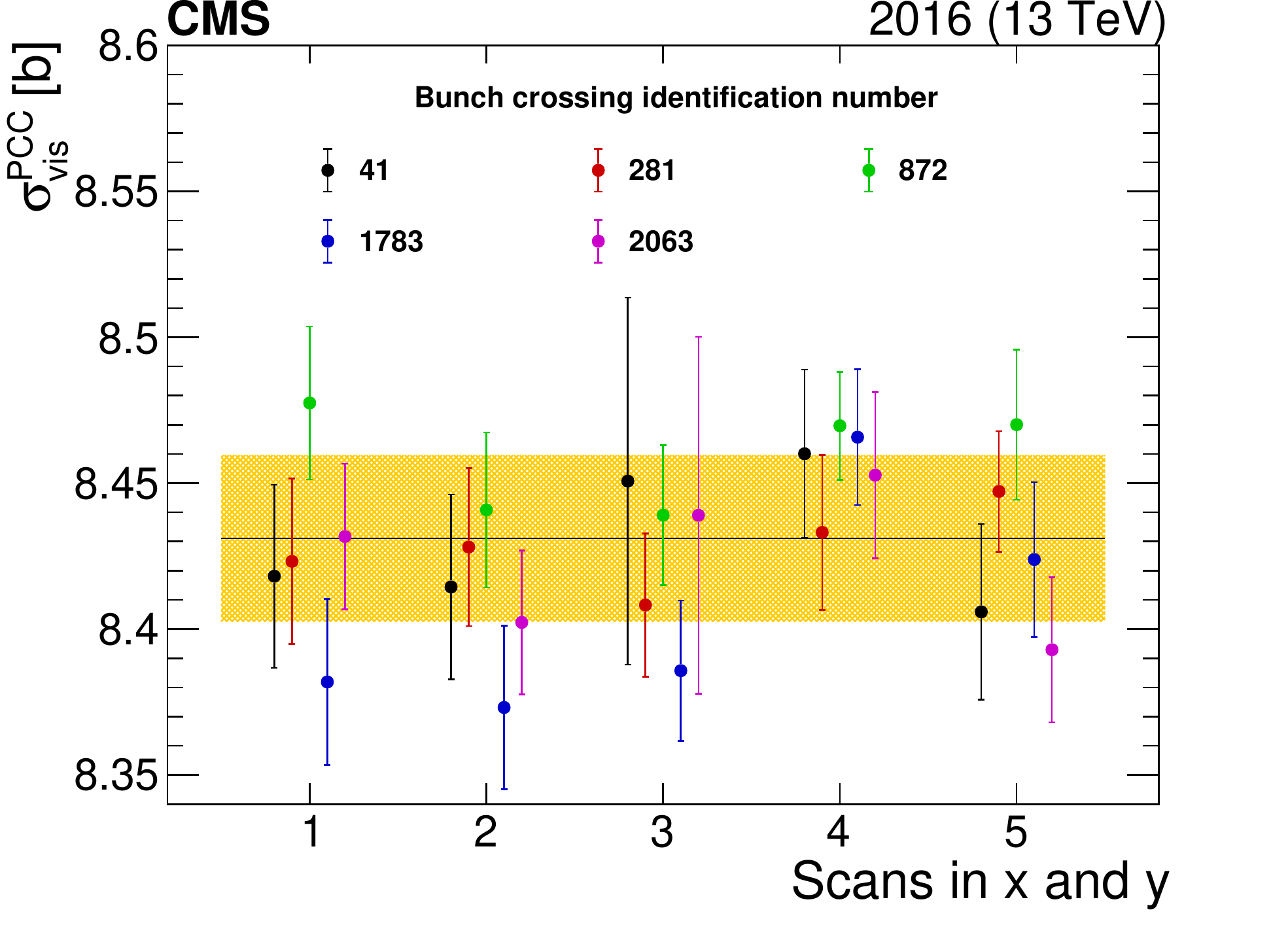}
\caption{
  The measured \sigmaVisPCC, corrected for all the effects described in Section~\ref{sec:ScanCorrections},
  shown chronologically for all \vdM scan pairs (where 3 and 4 are BI scans) taken in fills 4266 (\cmsLeft) 
  and 4954 (\cmsRight), respectively.  Each of the five colliding bunch pairs is marked with a different color.  
  The error bars correspond to the statistical uncertainty propagated from the \vdM fit to \sigmaVisPCC.  
  The band is the standard deviation of all fitted \sigmaVisPCC values.
  \label{fig:ScanVariation}}
\end{figure}

Although the accelerator parameters, such as bunch transverse sizes or intensities, vary during the course of a fill, such changes cancel in the calculation of 
\sigmaVis, which should remain invariant. This is shown in Fig.~\ref{fig:ScanVariation} for the measured \sigmaVisPCC as a function of time for \vdM scans taken 
in fills 4266 and 4954. After including all the effects described in Section~\ref{sec:ScanCorrections}, $\sigmaVisPCC=\PCCINoUnit$ and \PCCII in 2015 and 2016, respectively,
where the bunch-by-bunch fit uncertainty in $\Sigma_x$, $\Sigma_y$, and \muVis is propagated to the measured \sigmaVisPCC per scan.
Since these uncertainties are statistical in nature, they contribute to the scan-to-scan combination in an uncorrelated way.
The assumption of factorizable proton bunch densities limits the level of accuracy in the luminosity scale inferred from Eq.~(\ref{EQ:basiclumi2}). 
A common approach is thus adopted at the LHC that includes a dedicated tailoring of the proton bunch injection chain to minimize 
the emergence of non-Gaussian bunch density distributions~\cite{CERN-ACC-NOTE-2013-0008}.
Since the factorizability between the $x$ and $y$ distributions could impact the \vdM scan result of the different IPs differently, 
CMS reconstructs the individual proton bunch densities during the BI and \vdM scans, as described in Section~\ref{sec:xycorr}.

\subsection{Corrections to \texorpdfstring{\vdM}{vdM} scan data}
\label{sec:ScanCorrections}

Several systematic effects can change the measurement of \sigmaVis, and the following sections describe the
measurement of these effects, the corrections used, and the resulting systematic uncertainty in \sigmaVis.

Adjustments to the bunch-by-bunch charge measurement are made to correct for spurious charge that is present 
outside the nominally filled part of the slot (Section~\ref{sec:current}). 
Then, we correct for potential sources of bias associated with the beam position monitoring at the scale of $\mu$m.
We distinguish between ``orbit drifts'', which we model with smooth, linear functions, 
and residual differences relative to the nominal beam positions, where corrections per scan step are assessed.
Since both effects are time dependent, thereby biasing \sigmaVis incoherently, they are monitored continuously during each scan (Section~\ref{sec:od}).

Another source of correction originates from the electromagnetic interaction between charged particles in the colliding bunches (beam-beam effects); when the beams are displaced, rather than being
head-on, a beam deflection and change in \betastar may be induced. The former causes the beams to be more separated than the nominal value from LHC beam position estimates,
whereas the latter influences the spatial distributions of proton bunches and thus the observed rate.
The resulting corrections to \sigmaVis are evaluated at IP 5~\cite{Balagura:2020fuo,beambeamLLCMWG}, and depend on the LHC optics,
beam parameters, and filling scheme (as discussed in Section~\ref{sec:beambeam}).

The \vdM method requires an accurate knowledge of the beam separation. Possible differences in the absolute scale between the
nominal beam separation produced by the steering of the LHC magnets, as used in Eqs.~(\ref{EQ:Weff})
and~(\ref{EQ:Heff}), and the actual separation are determined by using the LSC procedure (Section~\ref{sec:LS}).

\subsubsection{Beam current calibration and spurious charge}
\label{sec:current}

The LHC beam currents are measured by dedicated devices. The FBCT system is used to measure the current of
individual bunches in 25\unit{ns} bunch slots.  The DCCT system provides a precise (\BeamCurUnc\%) measurement
of the total current for each of the two beams; since it is more precise than the FBCT sum, its scale is used to
normalize the sum of the FBCT measurements.

Both the DCCT and FBCT measurements are sensitive to additional charges outside the actual colliding bunch. These
components must be measured and subtracted.  The LHC radio frequency (RF) cavities operate at 400\unit{MHz}, so 
a single 25\unit{ns} wide bunch slot contains ten 2.5\unit{ns} wide ``RF buckets''. Only one RF bucket in a given bunch 
slot is filled with protons, and, in principle, the other nine RF buckets are empty.  Similarly, of the total 3564 bunch slots,
only a predefined subset is filled, according to the filling scheme.  In practice, however, a 
small amount of spurious charge is present in the nominally empty RF buckets and bunch slots, which should be 
subtracted from the $n_1$ and $n_2$ values in Eq.~(\ref{EQ:basiclumi2}).  The amount of ``ghost'' charge in the 
nominally empty bunch slots is included in the DCCT but not in the FBCT measurement, since the latter is 
insensitive to bunch charges below a certain threshold.  
The out-of-time (satellite) charge occupies RF buckets adjacent to the main bunch. As such, it can 
experience long-range interactions with the main bunch in the other beam and is visible in the FBCT measurement.
The corrected value for $n^j$ (where $j$ denotes the BCID) is therefore given by:
\begin{linenomath}
\begin{equation}
n^j =  \frac{n^j_{\text{FBCT}}\left(1 - f^j_{\text{sat}}\right) }{ \sum_j n^j_{\text{FBCT}}}  N_{\text{DCCT}} \left(1 - f_{\text{ghost}}\right),
\label{EQ:BeamCurrent}
\end{equation}
\end{linenomath}
where $f^j_{\text{sat}}$ represents the per-bunch correction due to the satellite bunch population and
$f_{\text{ghost}}$ is the correction for the ghost charge.

The spurious charge is measured by the LHC LDM system, which provides a precise longitudinal distribution of
the beam charge with a time resolution of 90\unit{ps}.  The data from  the LDMs for fills \FillNumberI and \FillNumberII indicate
that both the ghost and satellite charges are small. The latter is estimated to be $<$0.1\% for each of the two
beams and is neglected.  No particular time dependence for either beam is observed, and the resulting overall
spurious-charge correction in \sigmaVis amounts to $+$\SpuriousEffI and $+$\SpuriousEffII\% in 2015 and 2016, respectively. This is applied as a
correction to the beam currents in Eq.~(\ref{EQ:sigmaVis}).

The ghost charge is also measured using the beam-gas imaging method~\cite{Aaij:2014ida,Barschel:1693671,Coombs:2019nmc},
which compares the beam-gas rates in bunch crossings at IP 8 (the location of the LHCb detector) where only one beam contains protons, or where neither beam contains protons, leading to consistent
results with the LDM measurement. The systematic uncertainty of \SpuriousUnc\% is assigned to cover the difference between the two estimates of the
ghost contributions to the beam current.

\subsubsection{Beam position monitoring}
\label{sec:od}

{\tolerance=800 Although the LHC beam orbits are generally stable during a fill, even a small variation (either random or systematic in nature) in the 
beam positions during scans can significantly affect the resulting calibrations.  The beam positions are measured 
primarily using the DOROS BPM system. The LHC arc BPMs, when possible, are 
used to confirm the stability of the orbits during the scan.\par}

\begin{figure*}[!htb]
\centering
\includegraphics[width=.99\textwidth]{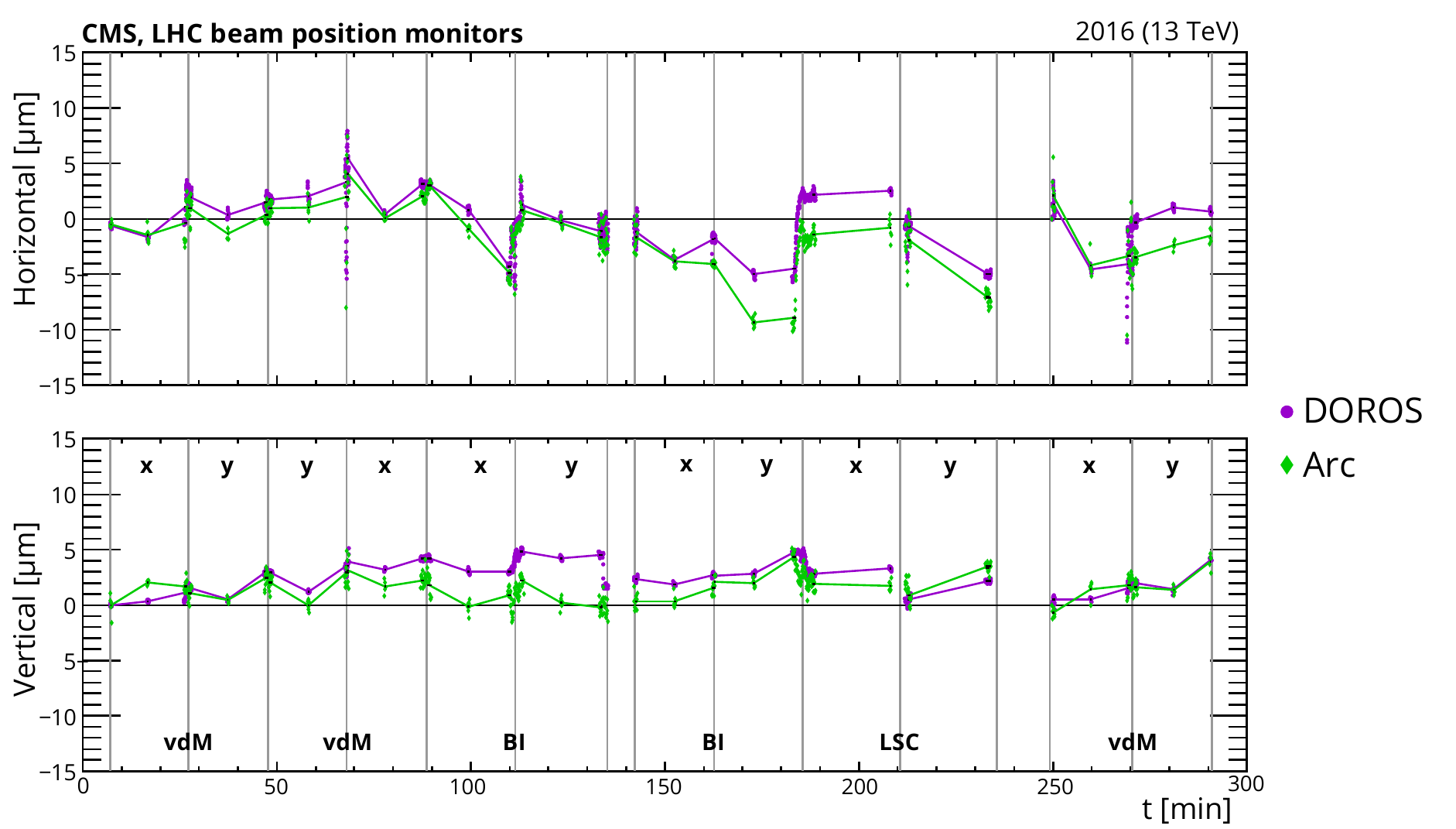}
\caption{
  Effect of orbit drift in the horizontal (upper) and vertical (lower) beam-separation directions during 
  fill \FillNumberII. The dots correspond to the beam positions measured by the DOROS or LHC arc BPMs 
  in $\mu$m at times when the beams nominally collide head-on and in three periods per scan (before, 
  during, and after) represented by the vertical lines.  First-order polynomial fits are subsequently 
  made to the input from BPMs (dots) and are used to estimate the orbit drift at each scan step.  Slow, 
  linear orbit drifts are corrected exactly in this manner, and more discrete discontinuities are 
  corrected on average.  
\label{FIG:OrbitDrift}}
\end{figure*}

To measure the orbit drift, we use the beam position measurements in $x$ and $y$ in three 15\unit{s} periods when the beams are nominally
colliding head-on: immediately before and after each scan,
as well as at the middle point of the scan, where the beams are also head-on. For each scan, a fit using a first-order polynomial is performed from the point before 
the scan to the middle point, and it is used to derive the correction for the first half of the 
scan.  Similarly, a fit from the middle to the point after the scan is used to correct the 
second half of the scan. Figure~\ref{FIG:OrbitDrift} shows the measured positions along with the resulting fits.  In general, 
the orbit drift during the 2015 and 2016 \vdM scans is less than about 5\mum for 
most of the scans.  However, in the third scan of both series, the orbit drift was significant 
enough to shift \sigmaVis by approximately $+$\ODRandomUpEffI\%.  
The corrections are derived using the average of the two BPM systems, and the largest deviation
of the correction from each individual system from the nominal correction is taken as the value of the
systematic uncertainty due to orbit drift. This is typically \ODRandomUncII--\ODRandomUncI\% overall.

\begin{figure*}
  \centering
    \includegraphics[width=0.4\textwidth]{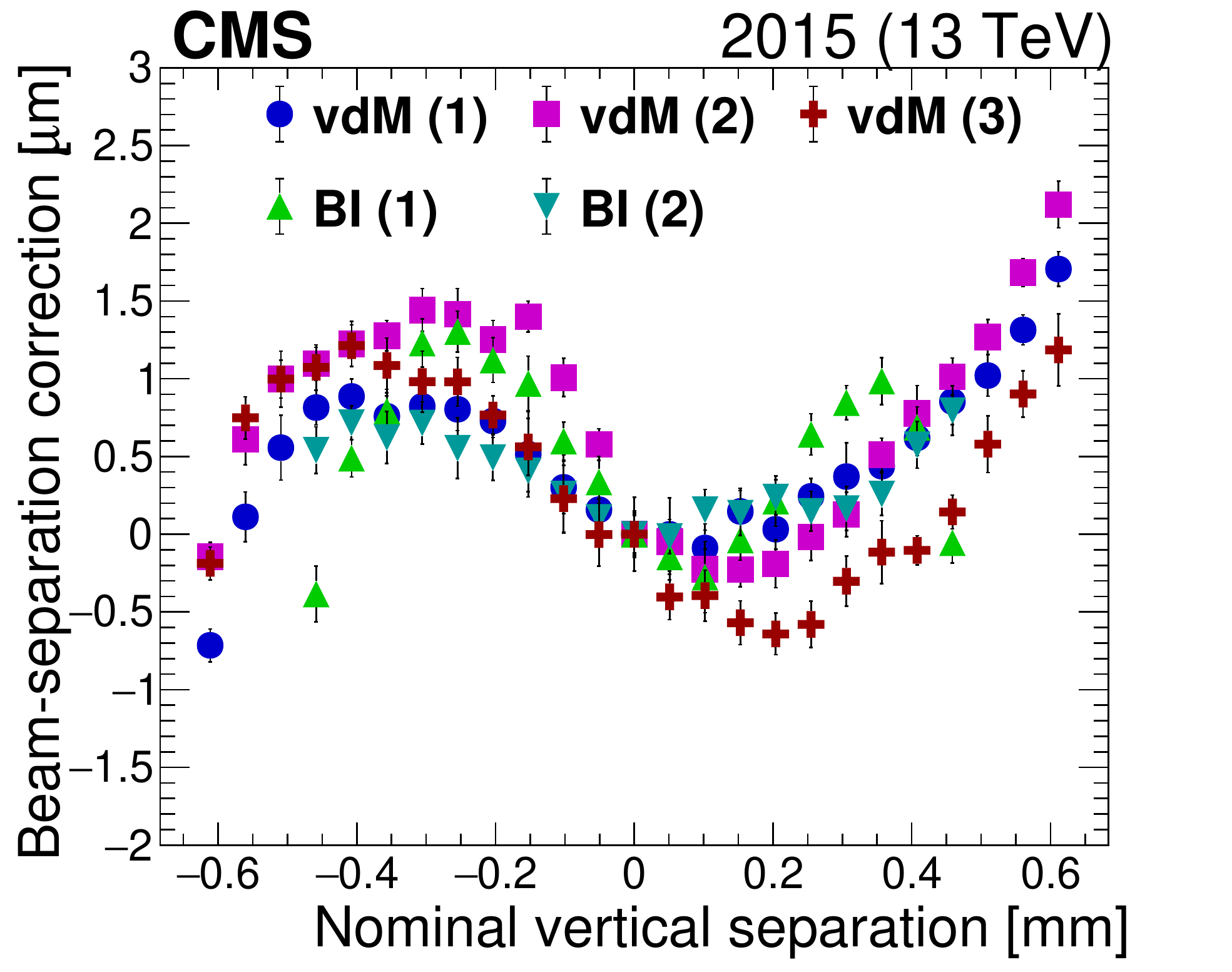}
    \includegraphics[width=0.4\textwidth]{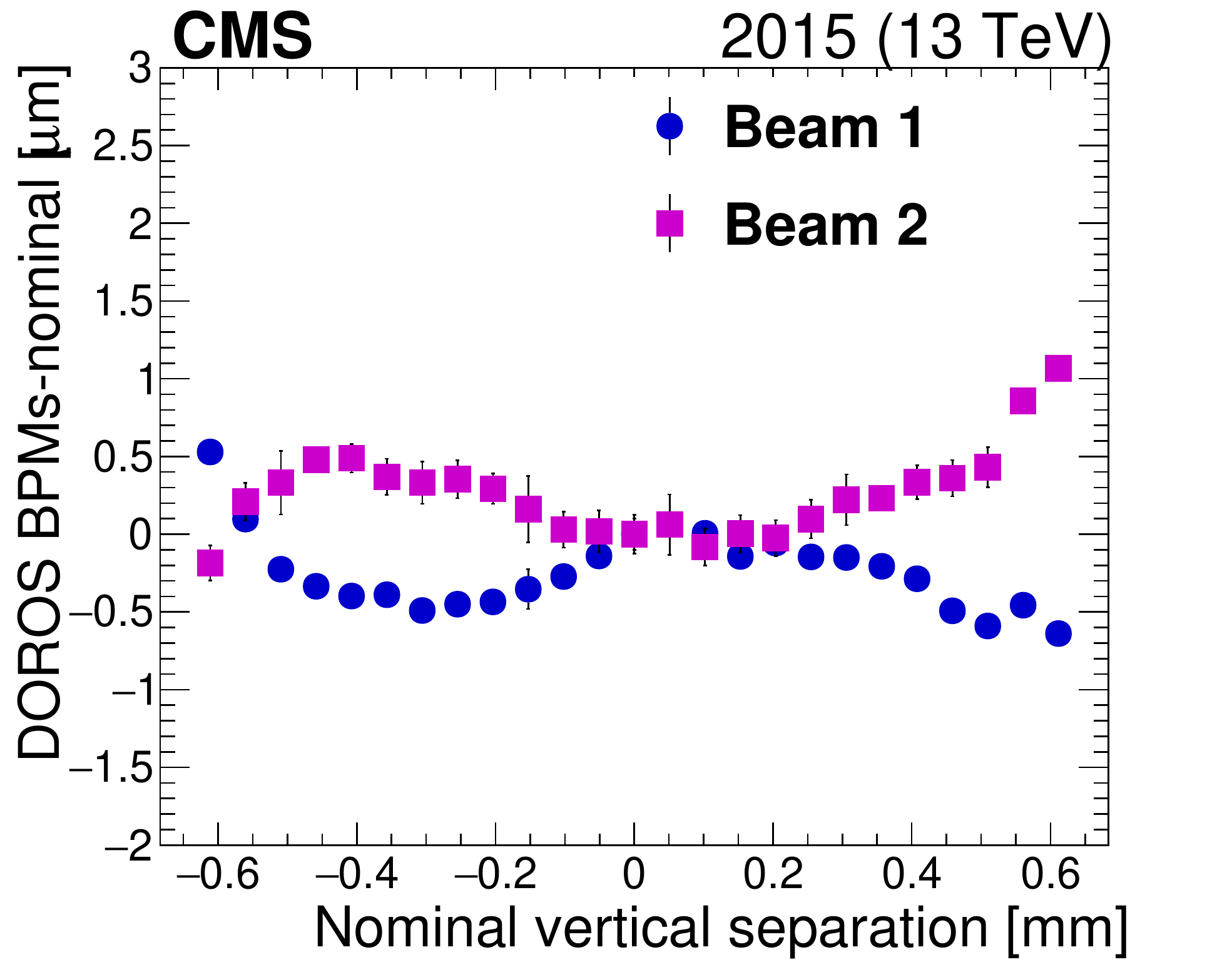} 
    \includegraphics[width=0.4\textwidth]{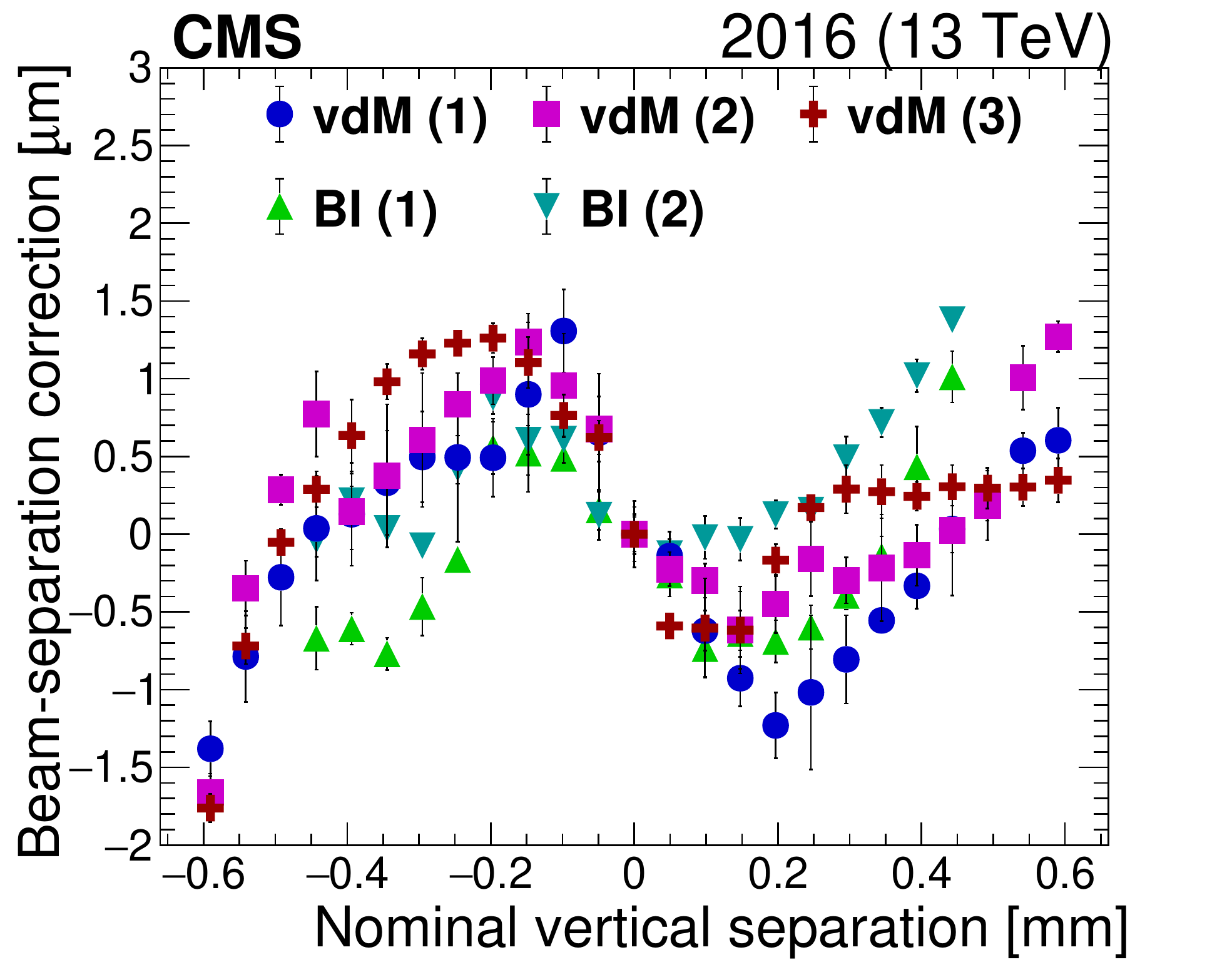}
    \includegraphics[width=0.4\textwidth]{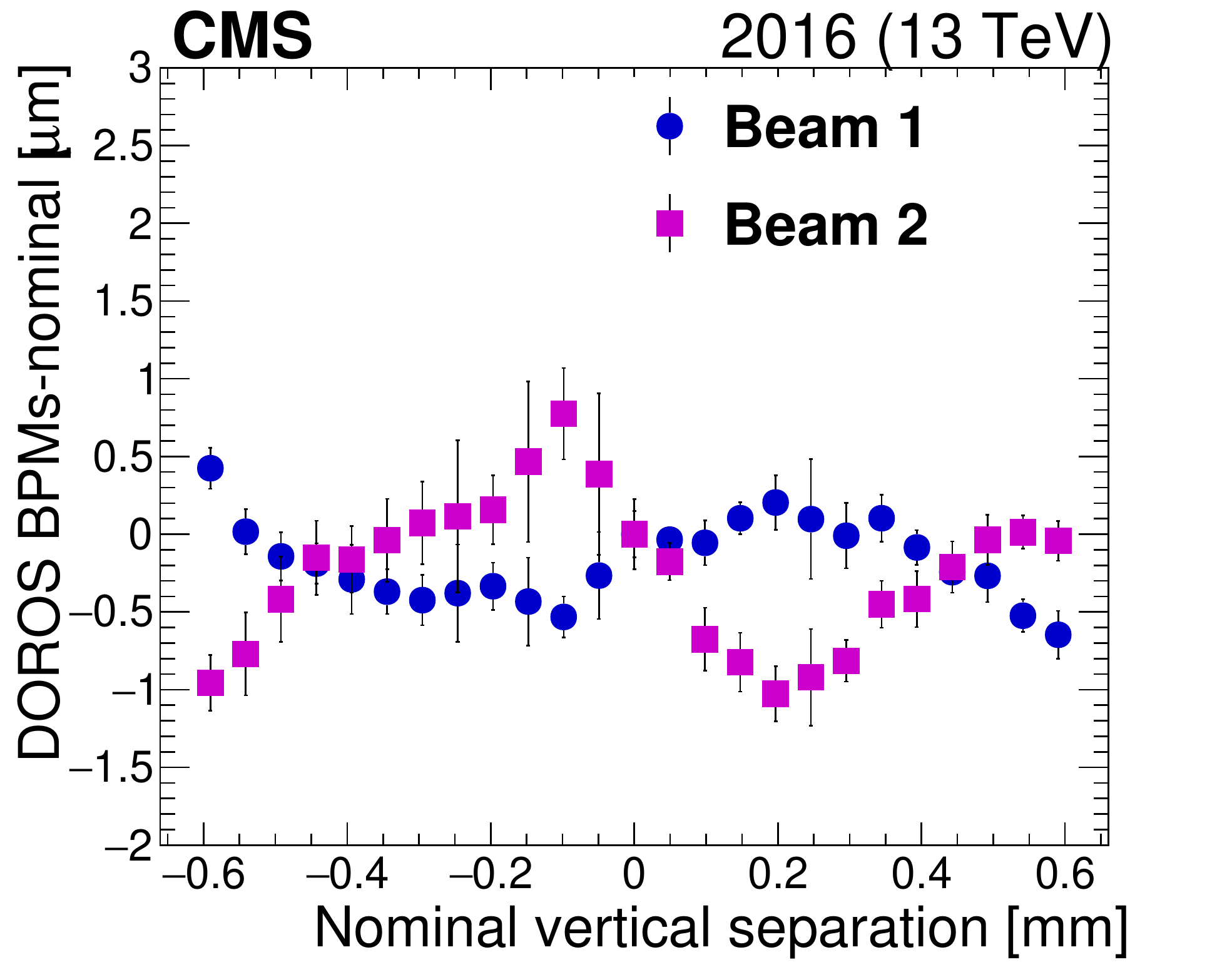} 
    \caption{
        The beam-separation residuals in $y$ during all scans in fills \FillNumberI (upper) and \FillNumberII
        (lower) are shown on the left.  The dots correspond to the difference (in terms of beam separation 
        in $\mu$m) between the corrected beam positions measured by the DOROS BPMs and the beam 
        separation provided by LHC magnets (``nominal''). The error bars denote the standard deviation in 
        the measurements. The figures on the right show the residual position differences per beam between the DOROS BPMs
        and LHC positions for the first vdM scans in $y$ in fills \FillNumberI (upper) and \FillNumberII
        (lower).
    }
\label{FIG:beamres}
\end{figure*}

At each scan step, the actual beam separation can be also affected by systematic or random deviations of the beam positions from their nominal settings, which, in turn,
impact the observed rate at each scan point. The magnitude of this potential bias is evaluated from consecutive single-beam orbit measurements at IP 5,
provided by the DOROS BPMs and with a duration of a few seconds each. They are further corrected for the beam-beam effects (as discussed in Section~\ref{sec:beambeam}) and 
the length scale (as described in Section~\ref{sec:LS}) using the position of reconstructed vertices as the calibration target.
The impact from beam-beam deflection at the location of the DOROS BPMs ($z_{\text{DOROS}}=\pm21.5\unit{m}$ away from IP 5) 
is magnified by a factor of $1+\tan\left(\pi Q_{x/y}\right)z_{\text{DOROS}}/\betastar$, where $Q_{x}$ and $Q_{y}$ are 
the betatron tune values in the $x$ and $y$ directions~\cite{lhcdesign}. Because these values are different, the resulting factors are
2.7 in the $x$ direction and 2.8 in the $y$ direction. The measurements from the DOROS BPMs are integrated over all bunches.
Therefore, the observed beam-beam deflection may be overestimated because of the inclusion of noncolliding, nondeflected bunches. 
In this analysis, a reduction factor of 0.6 is thus applied on top of the geometric factor in both years, which is the approximate 
fraction of the total number of bunches in the vdM fills that collide at IP 5.
The orbit drift, as described above, is also subtracted from the single-beam DOROS measurements before forming the actual beam separation.
Finally, an additional length scale correction is made to DOROS data for each beam and in both of 
the two transverse directions.  The calibration using vertices, both for DOROS and nominal
LHC positions, determines only the average length scale for the two beams. The calibrations of each beam are 
also not necessarily the same for the two sets of data.  Therefore, a final, relative calibration of the
DOROS data is made to align each beam in both transverse directions to the scale of the LHC beams.  
Figure~\ref{FIG:beamres} shows the residual difference in beam separation in all $y$ scans in
2015 and 2016 as well as the residuals per beam in a single scan, which shows symmetric behavior.
The resulting impact on \sigmaVis is in the range \ODSystLowEffI to \ODSystUpEffI and 
\ODSystLowEffII to \ODSystUpEffII\%, with average values of \ODSystEffI and \ODSystEffII\%, 
in 2015 and 2016, respectively. Corrections are applied for each scan, and the uncertainty comes 
from the reduction factor in the beam-beam deflection correction at the location of the DOROS BPMs.

\subsubsection{Beam-beam effects}
\label{sec:beambeam}

We distinguish two types of beam-beam interactions that affect the \vdM and BI scan measurements:
coherent and incoherent beam-beam effects. The total correction originates from the combination of both effects,
which affect the nominal beam separation (coherent) and the detector rate (incoherent) via the change of the beam shapes.

\begin{figure*}[!htb]
\centering
\includegraphics[width=0.475\textwidth]{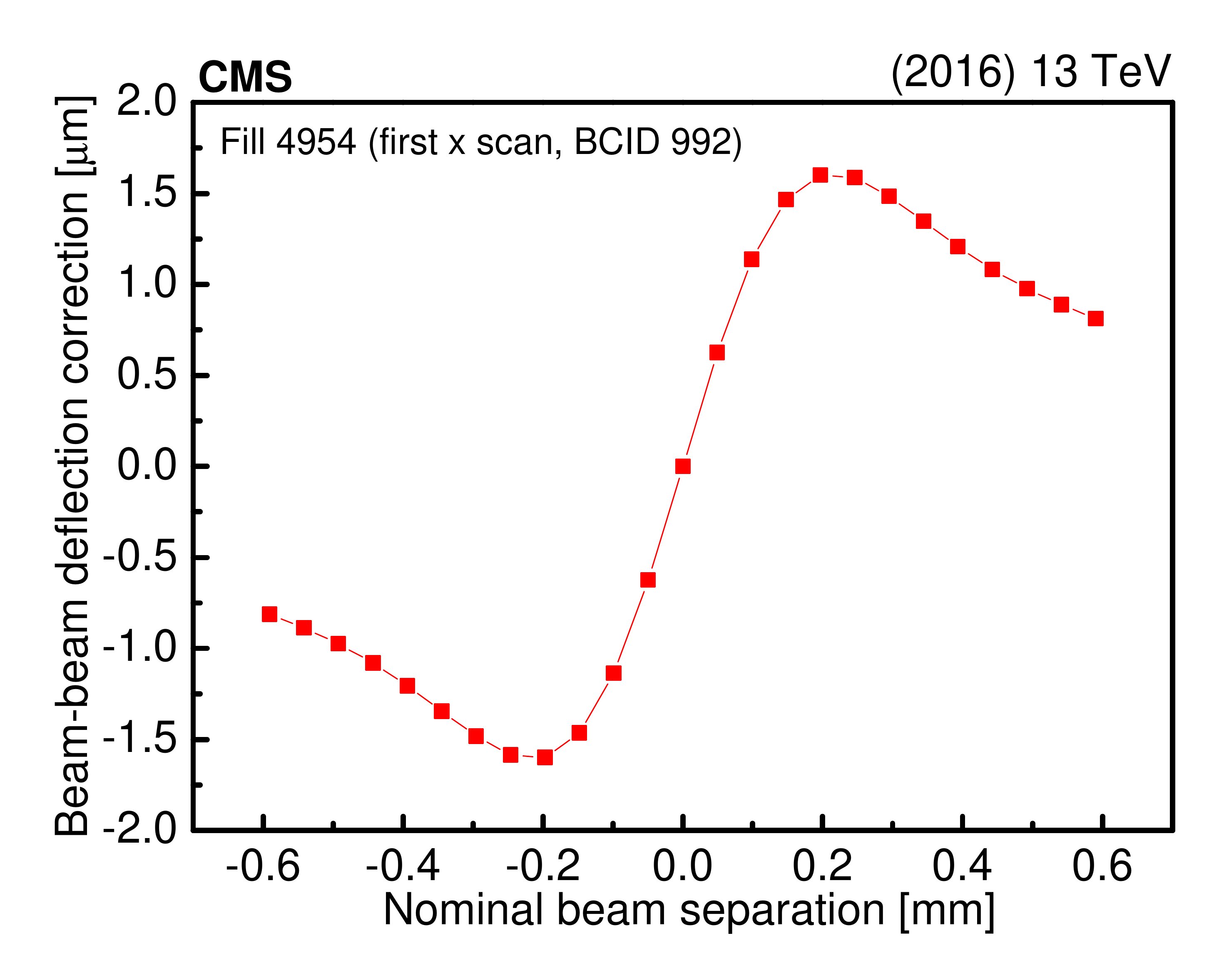}
\includegraphics[width=0.495\textwidth]{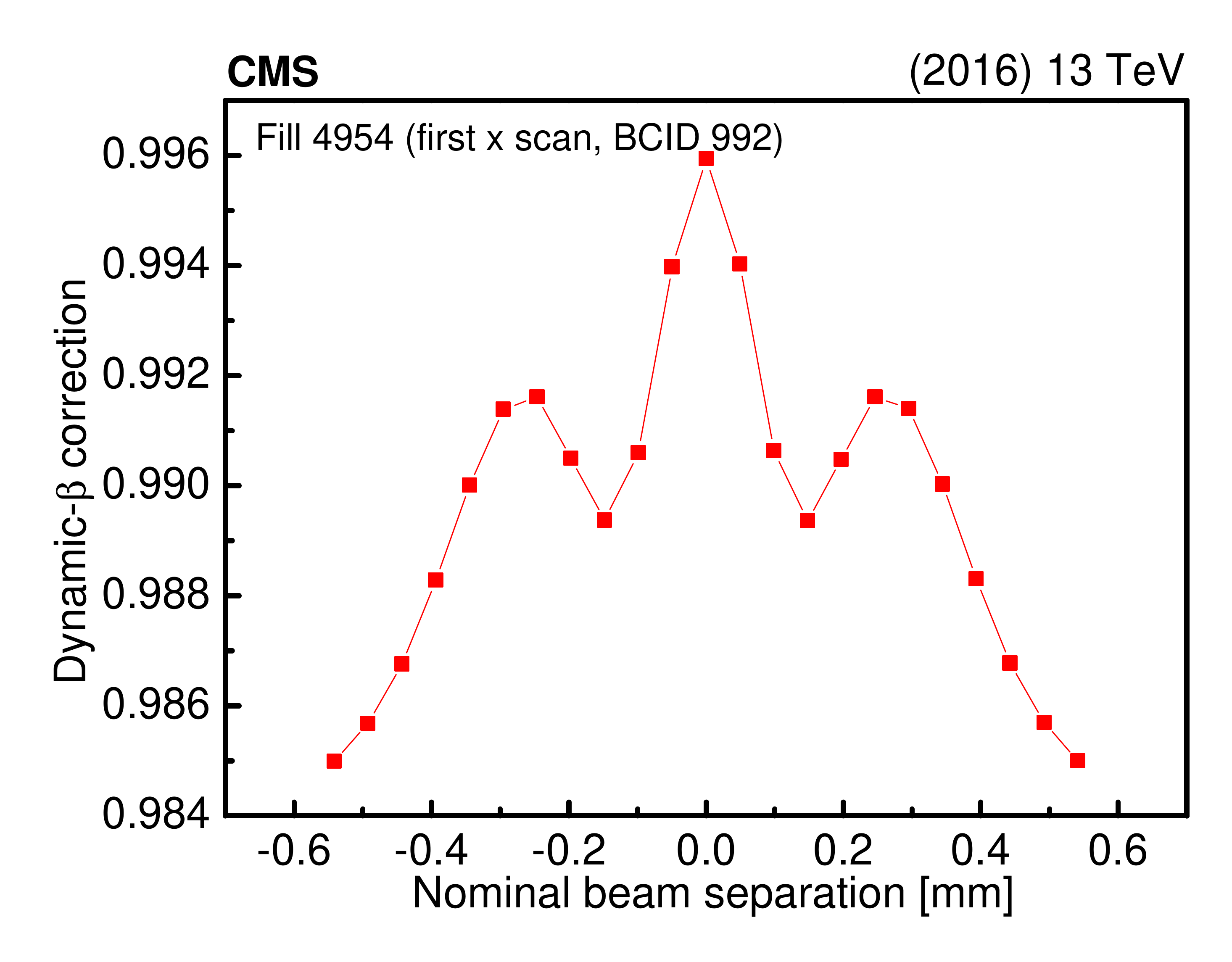}\\
\caption{
  Calculated beam-beam deflection due to closed-orbit shift (left) and the multiplicative rate correction for PLT
  due to the dynamic-$\beta$ effect (right) as a function of the nominal beam separation
  for the beam parameters associated with fill \FillNumberII (first scan, BCID 992). 
  Lines represent first-order polynomial interpolations between any two adjacent values.
\label{FIG:BeamBeam}}
\end{figure*}

The closed orbits of the bunches in the scans are shifted coherently by the angular kick induced by their electromagnetic repulsion, resulting in an increase in the absolute beam separation.
The size of this additional beam-beam deflection depends on the transverse beam size, bunch intensities, collision optics, and
separation between the orbits of colliding bunches.
It is calculated based on the Bassetti--Erskine formalism for the electric field of elliptically distributed bunches, as 
discussed in Ref.~\cite{CERN-ACC-NOTE-2013-0006}.
The orbit shift depends linearly on the separation for small nominal beam separations, reaches a maximum 
near 2\sigmab (${\approx}0.2\unit{mm}$ in fills \FillNumberI and \FillNumberII), and decreases nonlinearly towards zero at larger separations.
Figure~\ref{FIG:BeamBeam} (left)~\cite{Balagura:2020fuo,beambeamLLCMWG} shows the resulting correction as a function of nominal beam separation, for the conditions during the scans in fill \FillNumberII (Table~\ref{Tab:ScanSummary}).
The beam-beam deflection correction increases the $\Sigma_x$ and $\Sigma_y$ values, impacting the \sigmaVis measurement by about $+$\DeflEffI ($+$\DeflEffII)\% in 2015 (2016).

The incoherent effect corresponds to the change of the proton bunch density distribution functions $\rho(x,y)$ at the IP
due to deflection at the per-particle level. It causes a change in the effective \betastar, and thus results in a change in the
measured luminosity. This dynamic evolution of \betastar is usually referred to as the ``dynamic-$\beta$'' effect.
The correction for the dynamic-$\beta$ effect is evaluated numerically by using a dedicated particle tracking program that calculates \Aeff under
different hypotheses~\cite{Balagura:2020fuo,beambeamLLCMWG}. Considering the dynamic-$\beta$ effect independently of the beam-beam deflection,
we obtain the ratio of the detector rate as shown in Fig.~\ref{FIG:BeamBeam} (right). At vdM conditions, the dynamic-$\beta$ correction can be up to about $-$2\% at large values of beam separation.
Figure~\ref{FIG:BeamBeam} shows the effect is typically larger at higher beam separation.
In contrast to the beam-beam deflection, the dynamic-$\beta$ correction thus decreases the original $\Sigma_x$ and $\Sigma_y$ values.
The corresponding impact on the calculated \sigmaVis is about $-$\FocEffI ($-$\FocEffII)\% in 2015 (2016).

The total beam-beam correction (\ie, when both the beam-beam deflection and dynamic-$\beta$ effects are included) results in an increase in the calculated \sigmaVis of about 0.3 (0.2)\%
in 2015 (2016) at IP 5. In addition, when considering further head-on collisions at the IP at the opposite side of the ring (IP 1 at ATLAS), the effect is approximated as a single-IP simulation
but with shifted betatron tune values. The impact on \sigmaVis is enhanced by a factor of about two, leading to a total beam-beam correction of $+$\DefAndFocEffI ($+$\DefAndFocEffII)\% in 2015 (2016).
The uncertainty in this calculation is dominated by the uncertainty in the betatron tune values, which was 
estimated taking into account the symmetric tune spread as well as the full shift due to head-on collisions 
at a second interaction point (in ATLAS at IP 1).  These considerations translate into an uncertainty of 
\DefAndFocUnc\% in the corrected \sigmaVis~\cite{Balagura:2020fuo,beambeamLLCMWG}.

\subsubsection{Length scale calibration}
\label{sec:LS}

In the canonical \vdM formalism described in Section~\ref{SEC:Scan}, it is implicitly assumed that the beam separation is
perfectly known.  Operationally, the nominal displacement of the beams at the IP is achieved based on a local 
distortion (bump) of the orbit using a pair of steering dipoles located on either side of the 
IP~\cite{ref:White}.  The size of the nominal separation is subject to potential uncertainty associated with the 
response of the steering dipoles themselves (\eg, magnet hysteresis) or lattice imperfection~\cite{lhc_optics}, 
\ie, higher multipole components in the quadrupoles located within those orbit bumps.  For a given IP, there 
are four possible bumps, for the two possible displacement directions of the two beams.
 
An accurate calibration for the size of the bumps can be obtained using the CMS tracker.  In 
particular, for small vertex displacements, the uncertainty in the reconstructed vertex position in $x$ or $y$ is 
${\approx}20\mum$ for zero-bias collisions~\cite{TRK-11-001}.  During LSC scans, the data for each separation
distance contains several hundred thousand reconstructed vertices, yielding a position measurement with submicron precision.

The \vdM scans described in Section~\ref{SEC:VdMScanData} are typically done by moving the beams in equal 
steps in opposite directions.  Since the two beams have independent length scales, the full 
separation correction is obtained from the mean of the length scale corrections per beam.  Separate scans, wherein both beams are moved in steps in the 
same direction, are thus required to obtain the LSC. A more detailed description on the relationship between the calibration constant associated with the ``offset''
(\ie, the arithmetic mean between the transverse beam positions) and the observed quantities during LSC scans can be found in Ref.~\cite{Aaij:2014ida}.
Here, for each scan step, the centroid of the luminous region is measured as the mean from a Gaussian fit to the observed vertex positions.
A calibration constant for each transverse direction is extracted with a first-order polynomial fit to the 
difference between the measured mean position and the nominal offset as a function of the latter.  This constant 
corresponds to the average calibration of the bumps of the two beams. It is then applied as a scale factor to correct the nominal beam displacement.

The nominal offset is also affected by the random and systematic beam position deviations described in Section~\ref{sec:od}.
The beam positions at each step are monitored using DOROS BPMs.
We estimate the arithmetic mean of the measured step sizes as a good representation of the nominal settings,
after excluding outlier step sizes based on an iterative procedure.
The difference of the remaining step sizes from the mean is used to correct the nominal offsets, 
and their standard deviation is the uncertainty due to beam position deviations.
The correction improves the quality of the first-order polynomial fits and the forward-to-backward scan agreement.
Consistent results are also found using the LHC arc BPMs to derive the correction for beam position deviations.

\begin{figure}[!htb]
\centering
\includegraphics[width=0.495\textwidth]{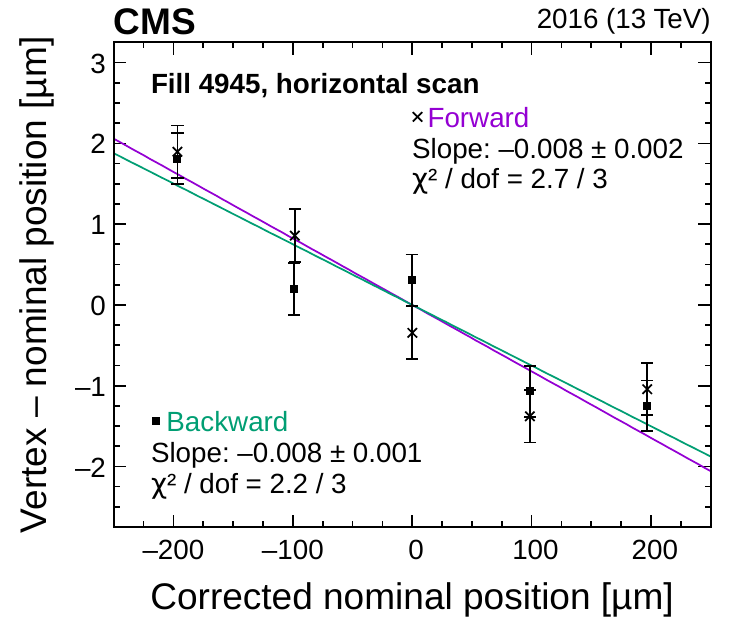}
\includegraphics[width=0.495\textwidth]{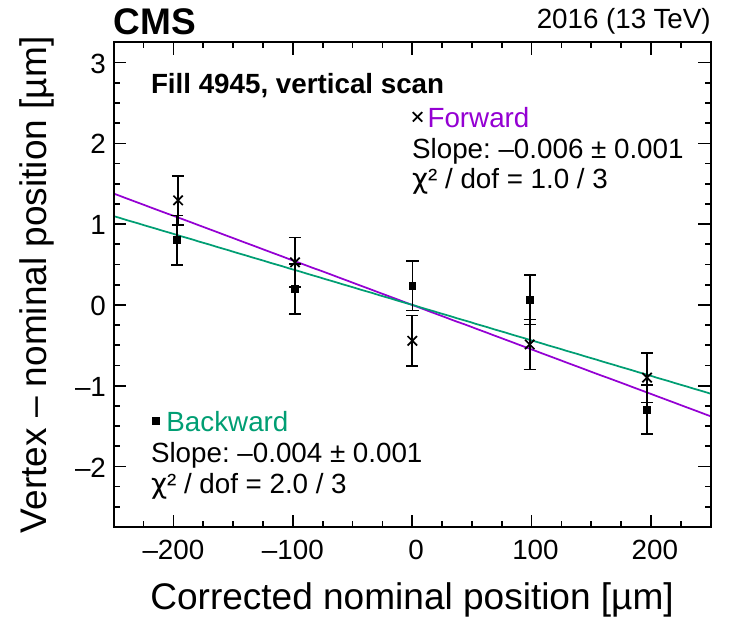}
\caption{
  Fits to LSC forward (purple) and backward (green) scan data for the $x$ (\cmsLeft) and $y$ (\cmsRight) LSC scans in fill \FillNumberIII.
  The error bars denote the statistical uncertainty in the fitted luminous region centroid.
\label{fig:LSscan}}
\end{figure}

The fit results for the $x$ and $y$ LSC scans are shown in Fig.~\ref{fig:LSscan} for fill \FillNumberIII.
The difference between the measured displacement of the beam centroid and the nominal displacement of the beams, corrected 
for the estimated beam position deviation, is plotted as a function of the latter.
In all cases, the data are well described by first-order polynomial fits with calibration constants differing on average
from zero by $-0.3$ and $-0.8$\% in the horizontal plane in 2015 and 2016, respectively,
and by $-0.1$ and $-0.5$\% in the vertical plane.
The combined correction to the visible cross section is $(-\LSCEffI \pm \LSCUncI)$ and $(-\LSCEffII \pm \LSCUncII)$\%.
The total uncertainty, equal to the uncertainty contributions from the $x$ and $y$ planes added in quadrature,
includes the statistical uncertainty in the first-order polynomial fits ($<$0.1\%),
the variation between the two scan directions and the different scans (0.1\%), a tracker alignment
uncertainty ($<$0.1\%), and the uncertainty from the estimated beam position deviations (0.1--0.2\%).

\subsection{Transverse factorizability}
\label{sec:xycorr}

The use of the \vdM scan technique to measure \Aeff relies on the assumption that the proton bunch density functions
are factorizable into $x$- and $y$-dependent components, as described in Section~\ref{SEC:Scan}. If this
condition is not met exactly, the measurements of \Aeff and \sigmaVis will be biased. To correct for this
potential bias, the bunch density distributions are measured independently with two methods, which are used in
a combined way to evaluate \Aeff. In both methods, primary vertices are reconstructed from tracks measured in
the CMS silicon tracker.

\subsubsection{Beam-imaging method}
\label{sec:BI}

In the BI method~\cite{Klute:2016grh,Knolle:2020ifp}, the distributions of reconstructed vertices during BI scans are used to obtain an
image of the transverse bunch profiles integrated over the scanning direction.
A primary vertex resolution comparable to or smaller than the transverse beam sizes is necessary to 
extract the beam profiles from the measured distributions. The two-dimensional distribution
in $x$ and $y$ of the reconstructed vertices
depends on the overlap of the bunch density functions, their transverse separations $\Delta x$ and $\Delta y$, and the vertex resolution $V$ 
of the CMS tracker system as:
\begin{linenomath}
\begin{equation}
    N^{\text{vtx}}(x, y; \Delta x, \Delta y) \propto \rho_1(x,y) \rho_2(x+\Delta x,y+\Delta y) \otimes V.
\end{equation}
\end{linenomath}
The combination of the vertex distributions from all steps of the BI scan in the $x$ direction is approximated as:
\begin{linenomath}
\ifthenelse{\boolean{cms@external}}
{\begin{multline}
    \label{eq:deriveFitModel2}
    \sum_{\Delta x=-4.5\sigma_b}^{+4.5\sigma_b} N^{\text{vtx}}(x, y;\Delta x, \Delta y)
    \approx  \bigg[\int\rho_1(x,y)\rho_2(x+\Delta x,y) \\
    \rd(\Delta x)\bigg]
    \otimes V= \rho_1(x,y) (\mathcal{M}_x \rho_2)(y) \otimes V.
\end{multline}}
{\begin{equation}
    \label{eq:deriveFitModel2}
    \begin{aligned}
    \sum_{\Delta x=-4.5\sigma_b}^{+4.5\sigma_b} N^{\text{vtx}}(x, y;\Delta x, \Delta y)
    & \approx  \bigg[\int\rho_1(x,y)\rho_2(x+\Delta x,y)\rd(\Delta x)\bigg] \otimes V \\
    & = \rho_1(x,y) (\mathcal{M}_x \rho_2)(y) \otimes V.
\end{aligned}
\end{equation}}
\end{linenomath}

Here, $(\mathcal{M}_x \rho_2)(y)=\int\rho_2(x,y)\rd x$ denotes that the proton bunch density of the second 
beam appears marginalized in the direction of the scan.  This results from the assumption that the step size is 
small enough with respect to the width of the bunch densities, so we can replace the sum over discrete scan 
points with a continuous integral over $\Delta x$.  This two-dimensional vertex distribution can be exploited to 
constrain the transverse correlations of the bunch density of the first beam.

Combining four such vertex distributions accumulated during the BI scan set, we reconstruct the two-dimensional proton
bunch densities of the two beams from a simultaneous fit.  This requires knowledge of the 
primary vertex resolution, which is modeled with a two-dimensional Gaussian function.  Convolving with the 
primary vertex resolution is then analytically possible for bunch density models built from Gaussian functions.

Models for the proton bunch density are built from Gaussian distributions 
parameterized with an additional correlation parameter $\varrho$:
\begin{linenomath}
\ifthenelse{\boolean{cms@external}}
{\begin{multline}
    \label{eq:BeamImagingFitModel}
    g_j(x,y) = \frac{1}{2\pi \sigma_{jx} \sigma_{jy} \sqrt{\smash[b]{1-\varrho_j^2}}} \exp{\Bigg(-\frac{1}{2(1-\varrho_j^2)}} \\
      \Bigg[ \frac{x^2}{\sigma_{jx}^2} +\frac{y^2}{\sigma_{jy}^2}-\frac{2\varrho_jxy}{\sigma_{jx}\sigma_{jy}}\Bigg]\Bigg),
\end{multline}}
{\begin{equation}
    \label{eq:BeamImagingFitModel}
    g_j(x,y) = \frac{1}{2\pi \sigma_{jx} \sigma_{jy} \sqrt{\smash[b]{1-\varrho_j^2}}} \exp{\Bigg(-\frac{1}{2(1-\varrho_j^2)}\Bigg[ \frac{x^2}{\sigma_{jx}^2}+\frac{y^2}{\sigma_{jy}^2}-\frac{2\varrho_jxy}{\sigma_{jx}\sigma_{jy}}\Bigg]\Bigg) },
\end{equation}}
\end{linenomath}
where $j$ indicates the beam number ($j=1$ or 2). More complicated models are constructed with sums of
these individual correlated Gaussian distributions. Distributions with a wide tail are better described by
adding a Gaussian component with a small weight and a large width.  Distributions with a flattened central
part can be modeled with an additional component with a small negative weight and a narrow width.
Typically, both nonzero correlation parameters and different widths are required to describe the 
nonfactorizability observed in data.

\begin{figure*}[!htb]
  \centering
  \includegraphics[width=.5\textwidth]{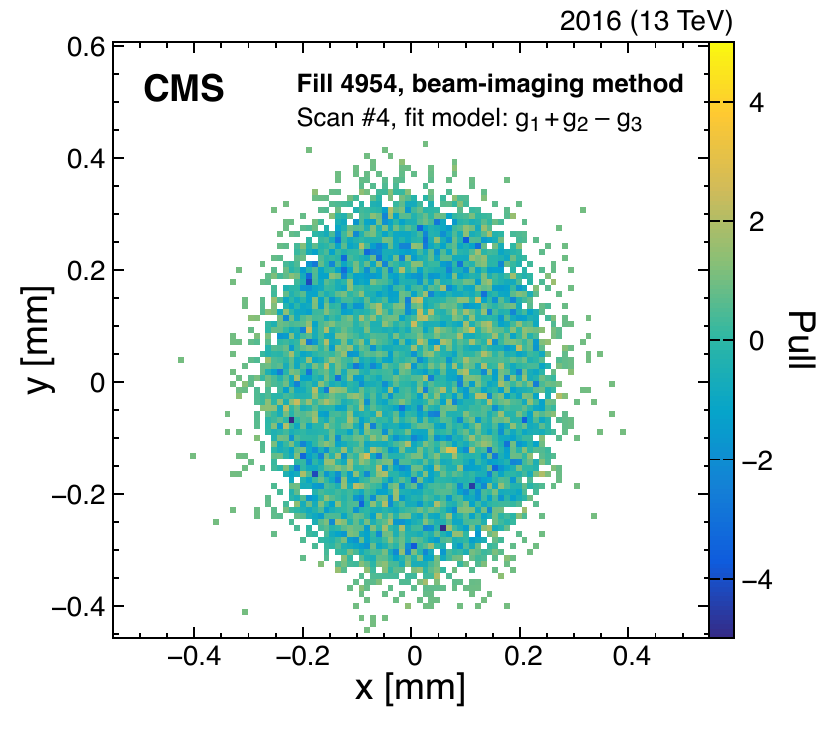} \\
  \includegraphics[width=.45\textwidth]{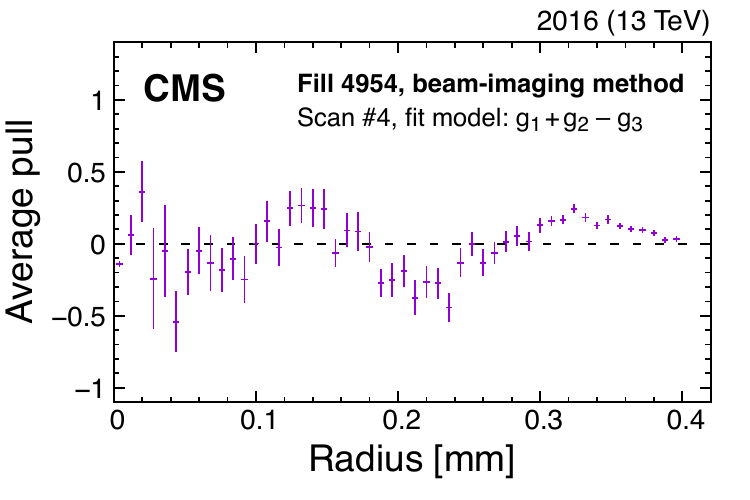}
  \includegraphics[width=.45\textwidth]{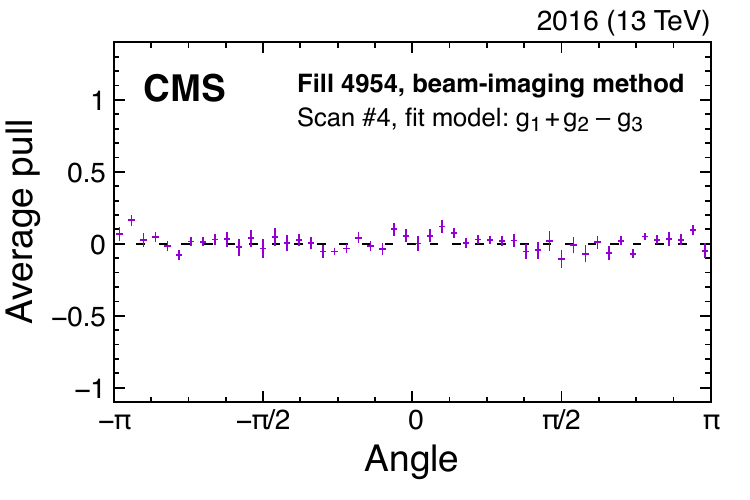}
  \caption{Example of the pull distributions of the fit model of Eq.~(\ref{eq:BeamImagingFitModelComp})
    with respect to the vertex distribution that constrains beam 2 in the $y$ direction recorded in fill \FillNumberII.  The
    upper plot shows the two-dimensional pull distributions, and the lower plots show the per-bin pulls
    averaged over the same radial distance (lower left) or angle (lower right). The error bars in the lower plot denote the standard error in the mean of the pulls in each bin.  The fluctuations observed 
    in the radial projection of the residuals are included in the uncertainty estimation.}
  \label{fig:xycorr:4954pulls}
\end{figure*}

The best description of the BI data collected in 2015 and 2016 for the five bunch crossings used 
is achieved consistently with a sum of three Gaussian distributions, where the narrow component has a 
negative weight:
\begin{linenomath}
\ifthenelse{\boolean{cms@external}}
{\begin{multline}
  \label{eq:BeamImagingFitModelComp}
    \rho_j (x,y) = - w_{j,1} g_{j,1}(x,y) + w_{j,2} g_{j,2}(x,y) \\
+ (1+w_{j,1}-w_{j,2}) g_{j,3}(x,y).
\end{multline}}
{\begin{equation}
  \label{eq:BeamImagingFitModelComp}
    \rho_j (x,y) = - w_{j,1} g_{j,1}(x,y) + w_{j,2} g_{j,2}(x,y) + (1+w_{j,1}-w_{j,2}) g_{j,3}(x,y).
\end{equation}}
\end{linenomath}
Figure~\ref{fig:xycorr:4954pulls} shows the two-dimensional pull distribution, \ie, 
$(N^{\text{vtx}}_{\text{data}}-N^{\text{vtx}}_{\text{fit}})/\sigma_{\text{data}}$, and the one-dimensional projections 
for the vertex distributions collected in the BI scan where the first beam 
is moved vertically for one bunch crossing in fill \FillNumberII.
In these fits, the effects from the beam-beam deflection and dynamic-$\beta$ are included
in the positions of the reconstructed vertices and as per-vertex weights, respectively, whereas the impact of orbit drift is negligibly small.

\begin{figure}[!htp]
    \includegraphics[width=0.45\textwidth]{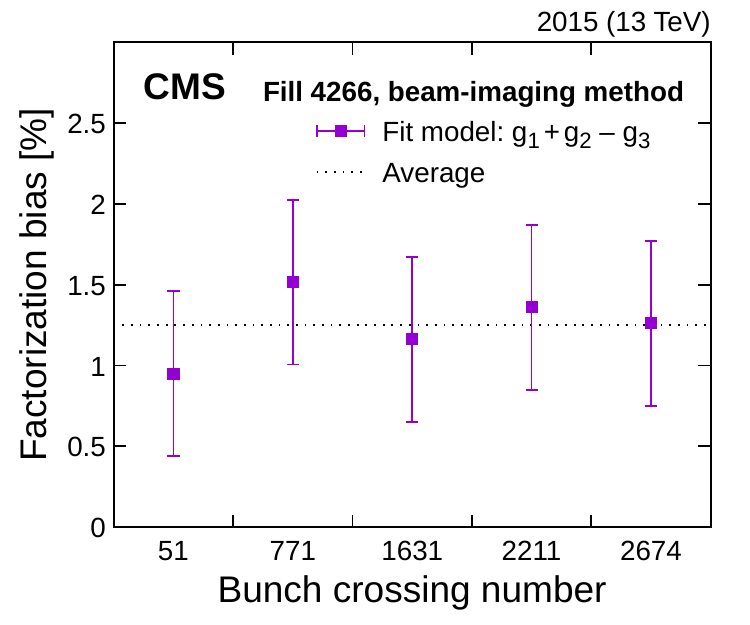}
    \hfill
    \includegraphics[width=0.45\textwidth]{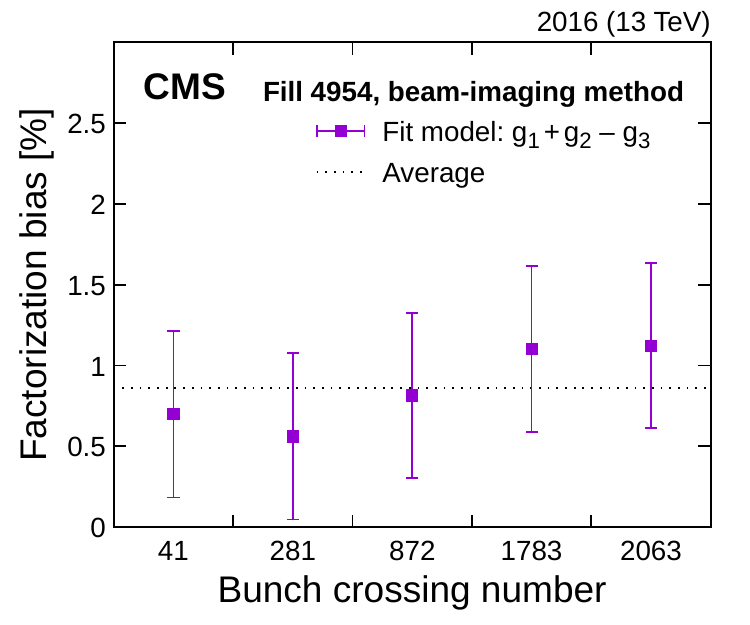}
    \caption{
      Factorization bias estimated from the fits to the BI bunch-by-bunch data in fills \FillNumberI (\cmsLeft) and \FillNumberII (\cmsRight).
      The error bars denote sources of uncertainty (statistical and systematic), added in quadrature, in the factorization bias estimates.
    }
    \label{fig:xycorr:biasperbcid}
\end{figure}

The value of \Aeff can then be calculated from an integration of the overlap of the bunch densities directly (\ie, 
$\Aeff=\iint\rho_1(x,y)\rho_2(x,y)\rd x\rd y$).  This is compared to the value of \Aeff obtained from an
MC simulated \vdM scan pair generated with the reconstructed bunch densities as input, and analyzed with 
the vdM method (\ie, $\Aeff=1/(2\pi\Sigma_x^{\text{MC}}\Sigma_y^{\text{MC}})$).  The difference between the two 
values yields the bias of the \vdM results, and is applied as a correction to $\sigmaVis$ values.  The bias
is computed separately for each bunch crossing, and the results are shown in Fig.~\ref{fig:xycorr:biasperbcid}.
The values for the estimated bias are averaged, resulting in a correction of $+$\BIEffI (\BIffII)\% 
in \sigmaVis for 2015 (2016) because of the assumption of $x$-$y$ factorization.

To estimate the uncertainty in the measured bias, the MC simulation of the \vdM scans is repeated multiple
times and the RMS of the resulting biases is 0.1\% for both years, which is considered as the statistical uncertainty in the \vdM scans.
Additionally, a systematic uncertainty is evaluated with a closure 
test:  simulated models are constructed by randomly drawing parameters of the fit model in Eq.~(\ref{eq:BeamImagingFitModelComp}). 
These are used to simulate MC pseudo-experiments by generating BI scan data, which are then fitted with
the same model and procedure.  Comparing simulated models with fit quality and fitted correction values similar to 
the data fits, the bias obtained from the bunch densities reconstructed from the fit agrees well on 
average with the true bias of the simulated model.  The RMS of the distributions of deviations 
is \BIUnc\% for both years.  We assign this RMS as the systematic uncertainty.

\subsubsection{Luminous region evolution}
\label{sec:LR}

In this method, which was inspired by Ref.~\cite{Aaboud2016}, the luminosity and luminous region geometry are used to 
reconstruct the bunch density distributions in three dimensions and as a function of time.
Using single-beam parameters, described in the following, bunch profiles are then generated for simulated \vdM scans and treated as genuine \vdM scan data.
Similar to the BI method, the impact of factorization is extracted by comparing the ``measured'' luminosity extracted from the one-di\-men\-sional \vdM simulated bunch profiles with
the ``true'' luminosity from the computed four-dimensional ($x$, $y$, $z$, $t$) overlap integral of the single-bunch distributions.
The luminous region is modeled by a three-dimensional ellipsoid whose parameters (nine in total) are extracted from an unbinned maximum likelihood fit of a three-dimensional Gaussian function
to the spatial distribution of the primary vertices~\cite{TRK-11-001}. The vertex resolution is determined from data as part of the fitting procedure.

\begin{figure*}[htp]
  \centering
    \includegraphics[width=0.495\textwidth]{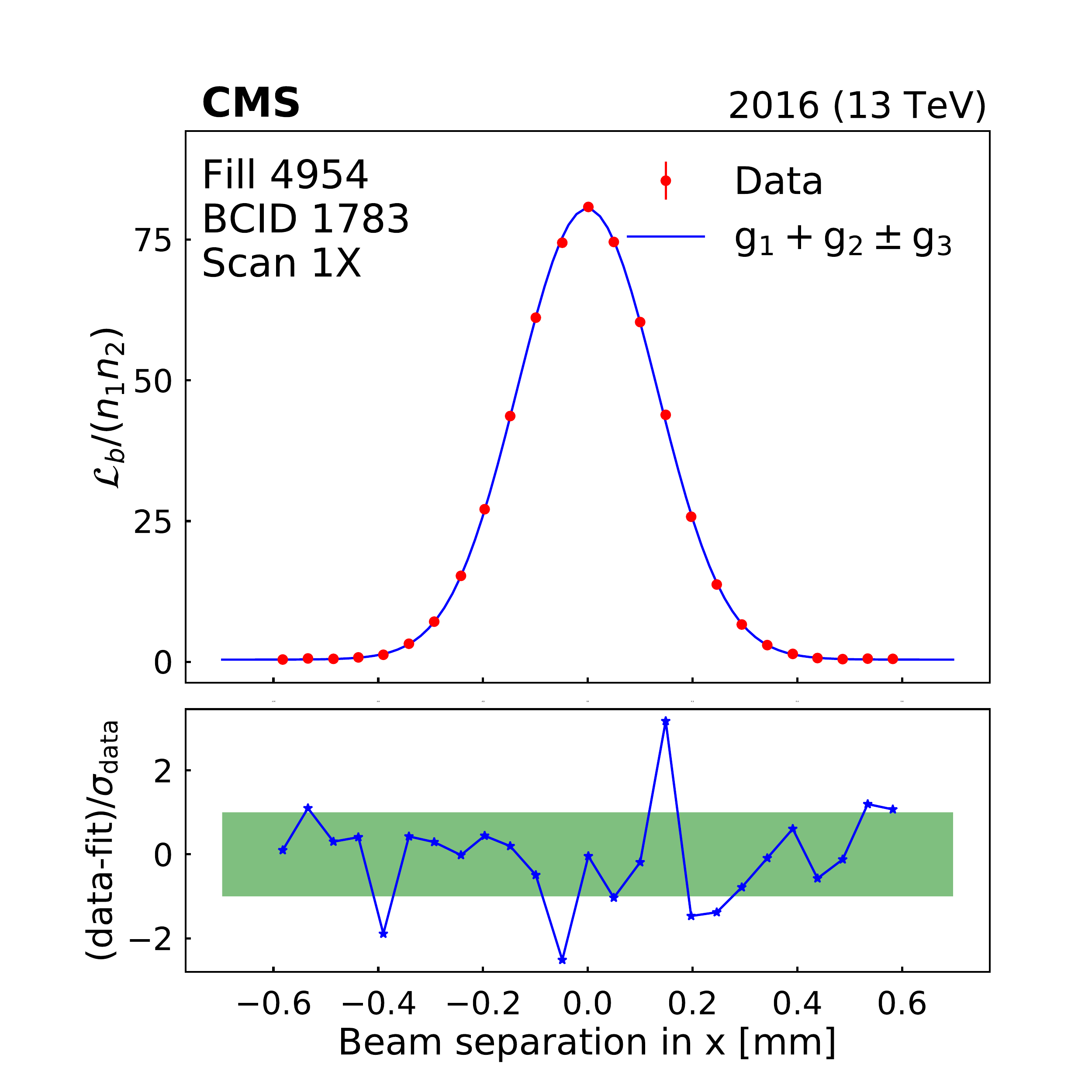}
    \includegraphics[width=0.495\textwidth]{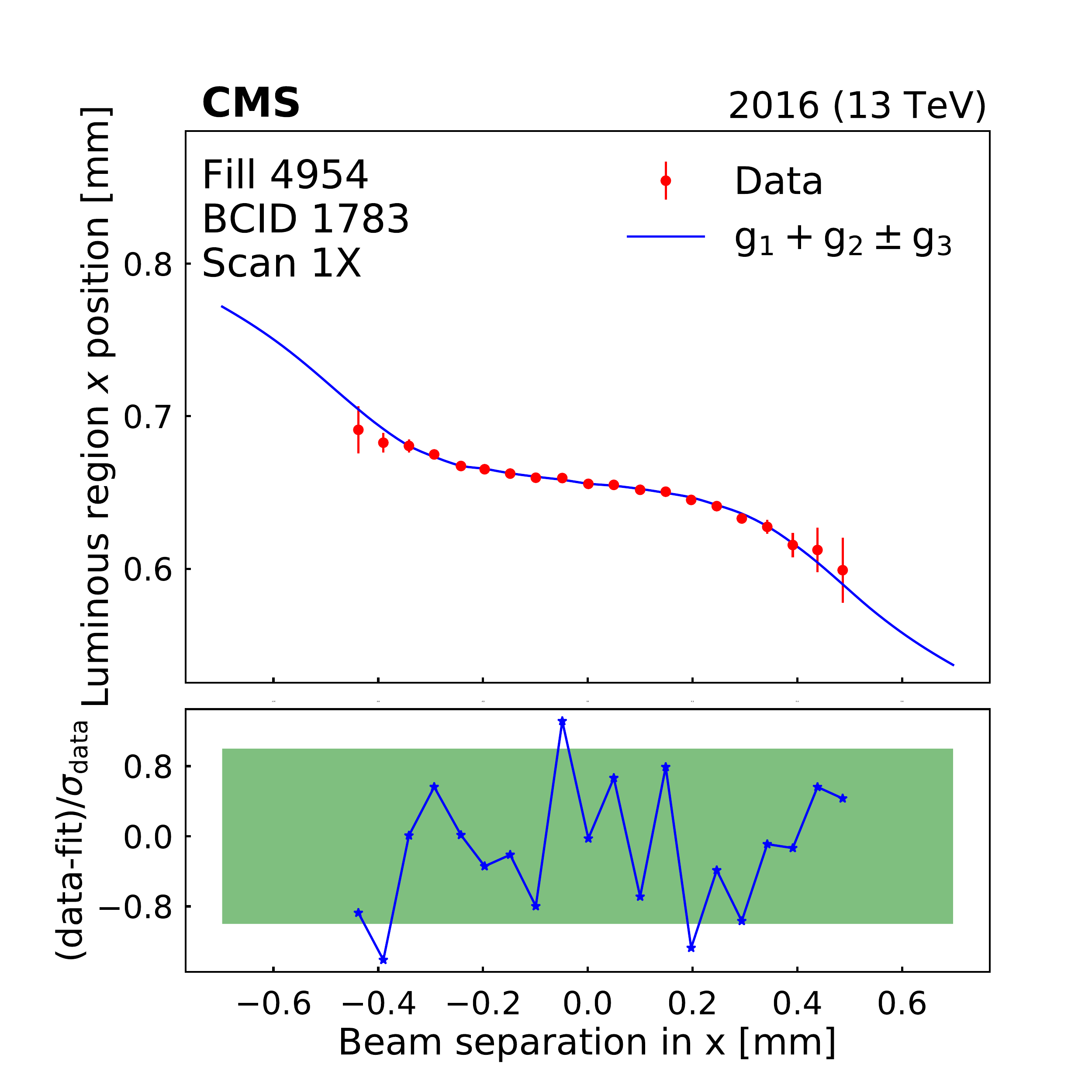}
    \includegraphics[width=0.495\textwidth]{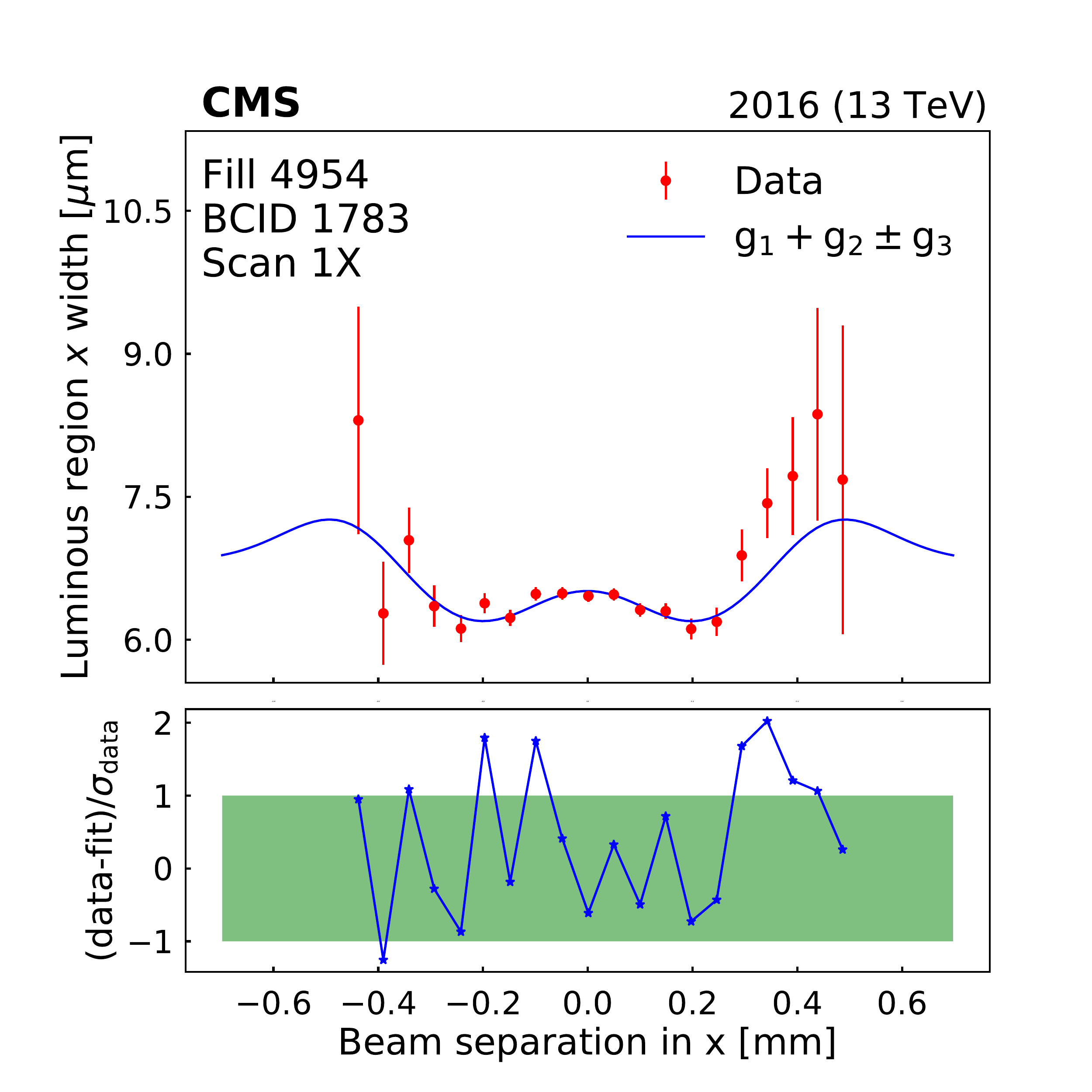}
    \includegraphics[width=0.495\textwidth]{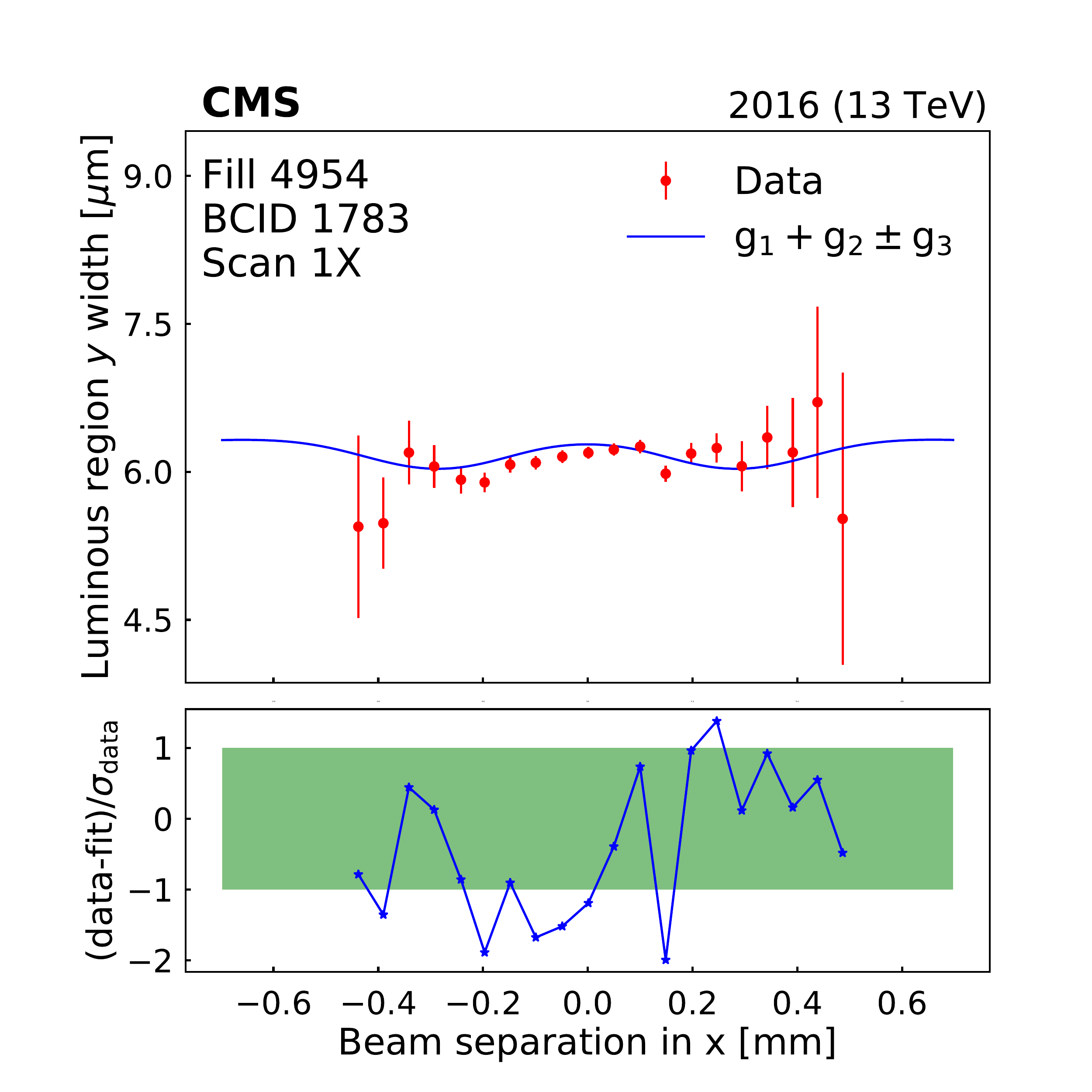}
    \caption{Beam-separation dependence of the luminosity and some luminous region parameters during the first horizontal \vdM scan in fill \FillNumberII.
      The points represent the luminosity normalized by the beam current product (upper left), the horizontal position of the luminous centroid (upper right),
      and the horizontal and vertical luminous region widths (lower left and right).
      The error bars represent the statistical uncertainty in the luminosity, and the fit uncertainty in the luminous region parameters.
      The line is the result of the three-Gaussian ($g_1+g_2\pm g_3$) fit 
      described in the text. In all cases, the lower panels show the one-dimensional pulls.
    }
    \label{fig:lumRegAna}
\end{figure*}

{\tolerance=800 The bunch profiles $\rho_j(x,y,z)$, parameterized per beam $j$, are the sum of three individual Gaussian distributions $g_{j,1\ldots3}(x,y,z)$ with common mean, but arbitrary width and orientation parameters (referred to as ``bunch parameters'' in the following):
\begin{linenomath}
\ifthenelse{\boolean{cms@external}}
{\begin{multline}
    \label{eq:LuminousRegionFitModelComp}
    \rho_j(x,y,z) = w_{j,1} g_{j,1}(x,y,z) + w_{j,2} g_{j,2}(x,y,z) \\
    + (1-w_{j,1}-w_{j,2}) g_{j,3}(x,y,z).
  \end{multline}}
{\begin{equation}
    \label{eq:LuminousRegionFitModelComp}
    \rho_j(x,y,z) = w_{j,1} g_{j,1}(x,y,z) + w_{j,2} g_{j,2}(x,y,z) + (1-w_{j,1}-w_{j,2}) g_{j,3}(x,y,z).
  \end{equation}}
\end{linenomath}
The overlap integral of Eq.~(\ref{eq:LuminousRegionFitModelComp}) is evaluated at each scan step to predict the true luminosity and the geometry of the luminous region for a given set of bunch parameters. In this calculation, we consider the impact of beam-beam effects, LSC, and orbit drifts. The bunch parameters are then adjusted according to a $\chi^2$ minimization procedure to determine the best-fit centroid position, orientation, and the widths (corrected for the primary vertex resolution) of the luminous region measured at each step of a BI or \vdM scan. An example of a fit to the PCC luminosity and luminous region geometry is illustrated in Fig.~\ref{fig:lumRegAna} for one of the horizontal scans in fill \FillNumberII and a subset of the three-dimensional ellipsoid parameters.  One of the four figures shows the variation in the beam width in $y$ during the $x$-separation beam scan, which is indicative of nonfactorization.  The goodness of fit is better than $\chi^2 / \text{dof} = 1.8$ for both years, with some systematic deviations being apparent mainly in the tails of the scan. The fits are repeated by substituting PLT as the luminosity input, but no particular dependence is seen.\par}

\begin{figure*}[htp]
  \centering
    \includegraphics[width=0.495\textwidth]{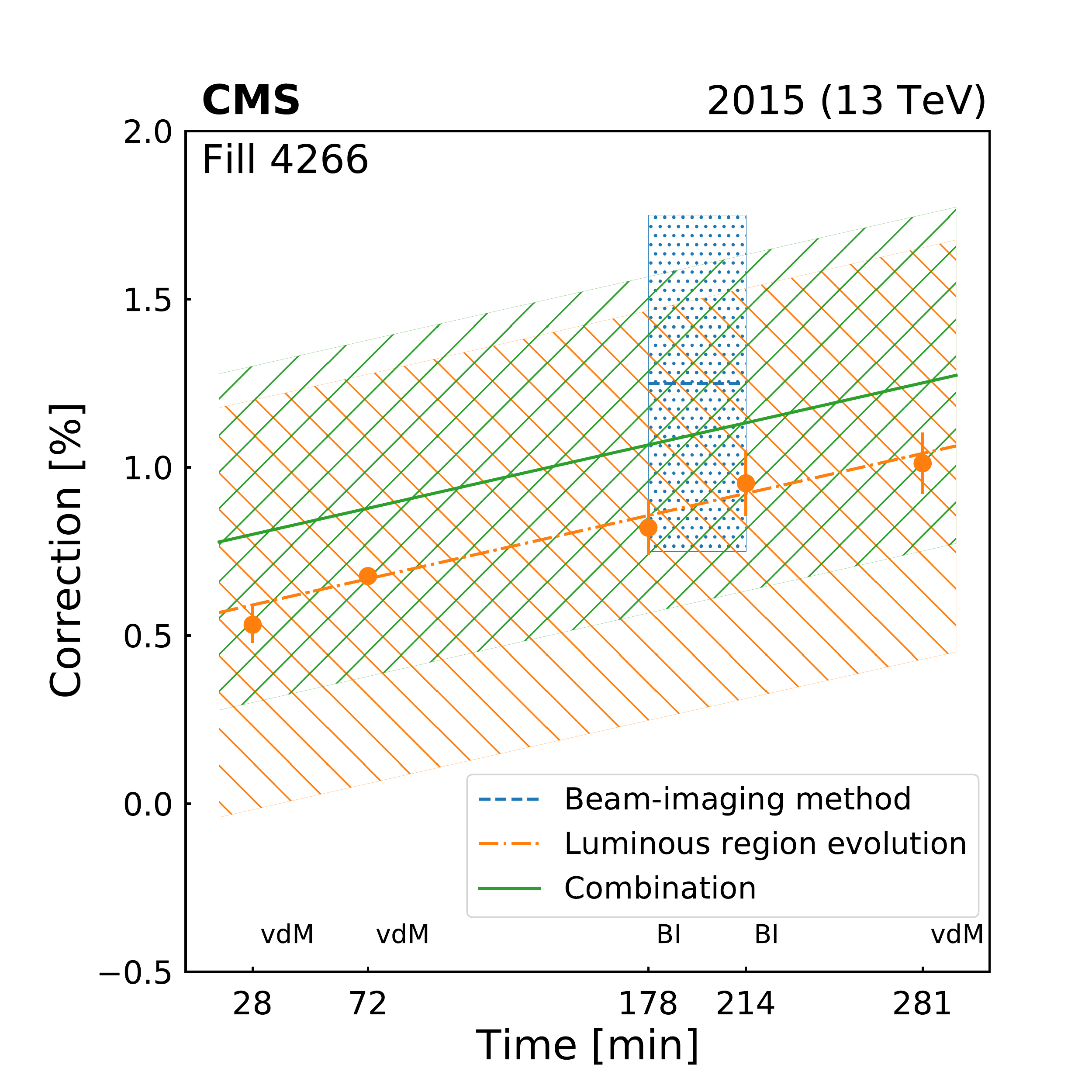}
    \includegraphics[width=0.495\textwidth]{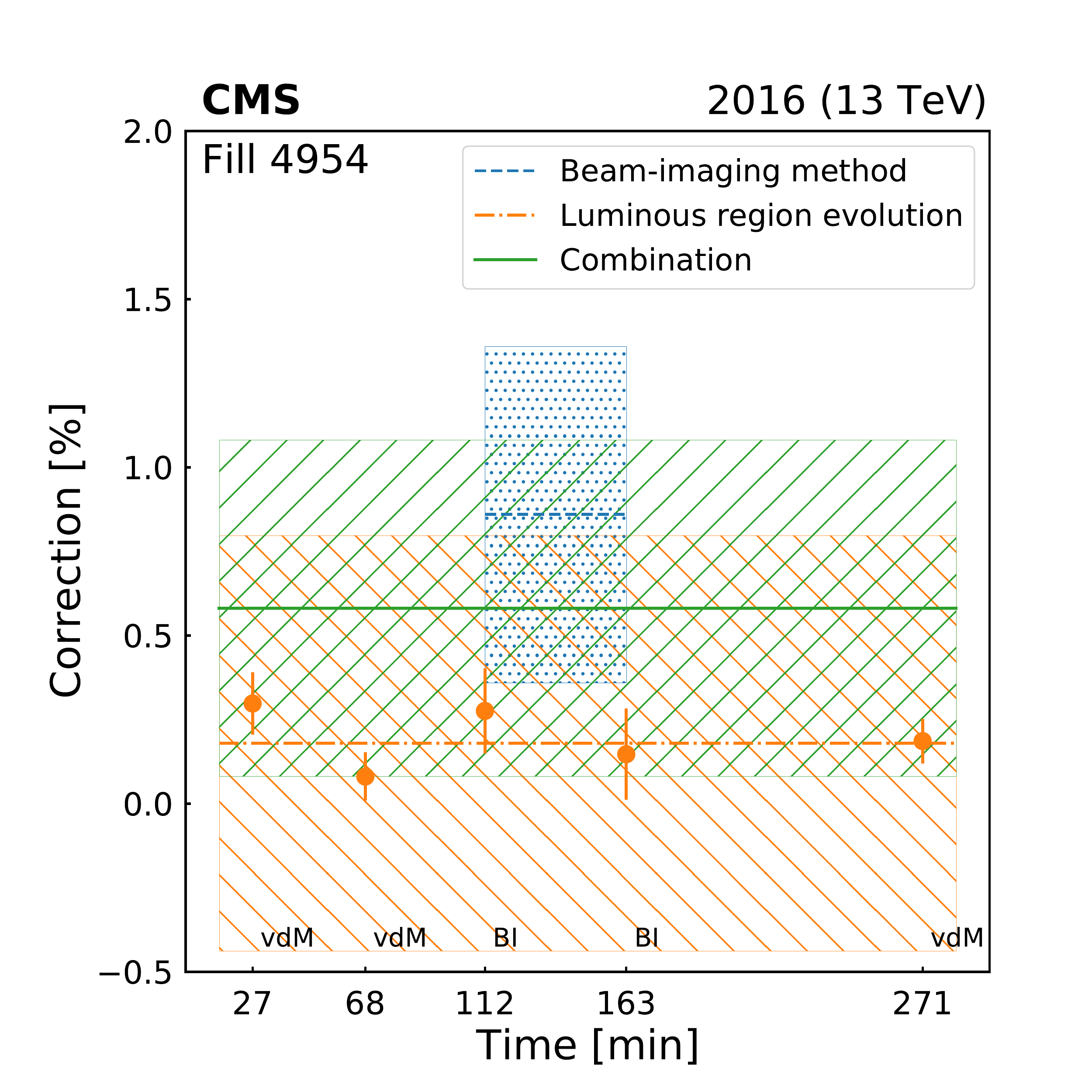}
    \caption{Ratio of the \sigmaVis evaluated from the overlap integral of the reconstructed single-bunch profiles in two (BI method)
      or three (luminous region evolution) spatial dimensions to that determined by the \vdM method, assuming factorization, and their combination.
      The central values are displayed as points or with a line while the corresponding full uncertainties are shown as hatched areas.  
      Different methods (including the combination) are color coded.
      Each point corresponds to one scan pair in fills \FillNumberI (left) and \FillNumberII (right).
      The statistical uncertainty is shown by the error bars.
}
\label{fig:rNF}
\end{figure*}

This procedure is applied to all (\ie, BI and \vdM) scans in fills \FillNumberI and \FillNumberII, and the results are summarized in Fig.~\ref{fig:rNF}.
The \sigmaVis extracted from the standard \vdM analysis with the assumption that factorization is valid
is smaller by \LRLowEffI--\LRUpEffI (\LREffII)\% than that computed from the reconstructed single-bunch parameters in fill \FillNumberI (\FillNumberII).
Similar to the evaluation in the BI method, the uncertainty amounts to \LRUnc\%. This uncertainty is dominated by the standard deviation in simulation-driven
closure tests, and includes the fit uncertainty in data and the contributions from beam-beam effects, length scale, and orbit drift.
These observations are thus consistent with the ones obtained in Section~\ref{sec:BI} in terms of absolute magnitude during the BI scans. 
The two results are combined to produce the final correction in \sigmaVis of 
$+(\BIAndLRLowEffI\text{--}\BIAndLRUpEffI\pm\BIAndLRUnc)$ and $+(\BIAndLREffII\pm\BIAndLRUnc)$\% in 2015 and 2016, respectively.
The final corrections retain the time evolution derived uniquely from the luminous region evolution method.

\section{Rate corrections under physics running conditions}
\label{SEC:RateCorrections}

The calibration scans described in the previous sections are performed with a small number of well-separated 
proton bunches with low bunch intensity.  In contrast, during nominal conditions, the collision rate 
is generally maximized to produce large data sets for physics measurements and searches.  This section describes 
the corrections that are applied to uncalibrated luminometer rates to ensure that the final luminosity values
are accurate. These corrections, summarized in Table~\ref{tab:corrections} for 2016, compensate for out-of-time pileup, efficiency, and nonlinearity effects for each 
individual luminometer.

\begin{table*}[htbp]
  \topcaption{Summary of the rate corrections under physics running conditions in 2016 applied separately to each luminometer.
    For HFOC, two distinct sources of out-of-time pileup corrections are provided.
    In the first and second columns, the \vdM calibration condition and the relative agreement of the luminometers in terms of \Aeff relative to PCC during fill \FillNumberII are given, respectively. The DT luminosity is also corrected for a very small additional muon rate from beam halo and cosmic sources, which is treated as a constant per fill.}
  \label{tab:corrections}
  \centering
  \cmsTable{
\begin{tabular}{cccccc}
\hline
                & \vdM calibrated & \vdM calibration         & Out-of-time pileup  & Efficiency       &  Nonlinear      \\
                &                 & agreement to PCC [\%]    & corrections [\%]    & corrections [\%] &  response [\%]      \\
\hline
PCC            & Yes              & \NA                      & 0--4                & 1                & \NA             \\
DT             & No               & \NA                      & \NA                & \NA              & \NA               \\
HFOC           & Yes              & 0.2                      & 0--15, 1--5         & 1                & 0--10         \\
PLT            & Yes              & 0.1                      & \NA                 & 0--10            & $-$0.2 to $+$1.4/(Hz/$\mu$b) \\
PVC            & Yes              & \hspace{-3.7mm}$<$0.1    & \NA                 & \NA              & \NA             \\
RAMSES         & No               & \NA                      & \NA                 & \NA              & \NA             \\
\hline
\end{tabular}
}
\end{table*}

\subsection{Out-of-time pileup corrections}
\label{SUBSEC:afterglow}
{\tolerance=800 The measurements in most detectors have out-of-time pileup contributions that do not arise from the in-time \pp collision
within the 25 ns window of the bunch crossing. Ideally, these contributions should be subtracted from all
bunch crossings before the total instantaneous luminosity is computed.  There are generally two types of
effects that are considered: spillover of electronic signals and real additional response from material 
activation.  These are denoted as type 1 (T1) and 2 (T2) afterglow, respectively.\par}

The T1 afterglow generally only impacts the following bunch crossing because electronic signals tend to decline
exponentially and hence two bunches later (50\unit{ns}) the signal is again below threshold.  The T1
contribution in bunch $n+1$ from bunch $n$ is proportional to $\Lb(n)$.  Thus, the model for the correction
is:
\begin{linenomath}
\begin{equation}
\lumi_{\text{b,corr}}(n+1)=\lumi_{\text{b,uncorr}}(n+1)-\alpha_{\text{T1}} \lumi_{\text{b,corr}}(n),
\label{eqn:type1model}
\end{equation}
\end{linenomath}
where $\alpha_{\text{T1}}$ is detector dependent and sometimes time dependent; $\alpha_{\text{T1}}$ 
ranges from 0.005 for BCM1F to 0.02 for HFOC to as large as 0.09 for PCC.

In contrast, T2 afterglow tends to impact all bunch crossings, because the half-life of the activated material
can be longer than several bunch crossings. The response can be modeled with a single- or double-ex\-po\-nen\-tial
distribution. The impact of T2 afterglow varies by filling scheme and by detector. In fills where \nb is low
and where the bunches are well separated, the T2 corrections are very small and often
completely negligible, as is the case by design in the \vdM calibration fills.  When LHC fills contain several
hundred bunches, the corrections start to contribute at the percent level in most bunches. With maximally full
filling schemes, the corrections can be up to about 4 (15)\% for PCC (HFOC).

Although there are clearly two distinct components, a combined (T1 and T2) model can be constructed that gives
the response for a specific bunch crossing, accounting for contributions from all other 3563 bunch crossing slots.
This model is referred to as the single-bunch response (SBR).  The
SBR for HFOC luminosity is taken directly from data in a reference fill with $\nb=2$ for
approximately the first half of the bunch crossings, and the bunches in the second half are smoothly
extrapolated using an exponential model.  The SBR is normalized to $\Lb(n)$ and it is then subtracted from all
other bunch slots. This procedure is repeated for all bunch crossings.

\begin{figure}[htp]
\centering
\ifthenelse{\boolean{cms@external}}{
\includegraphics[width=0.45\textwidth]{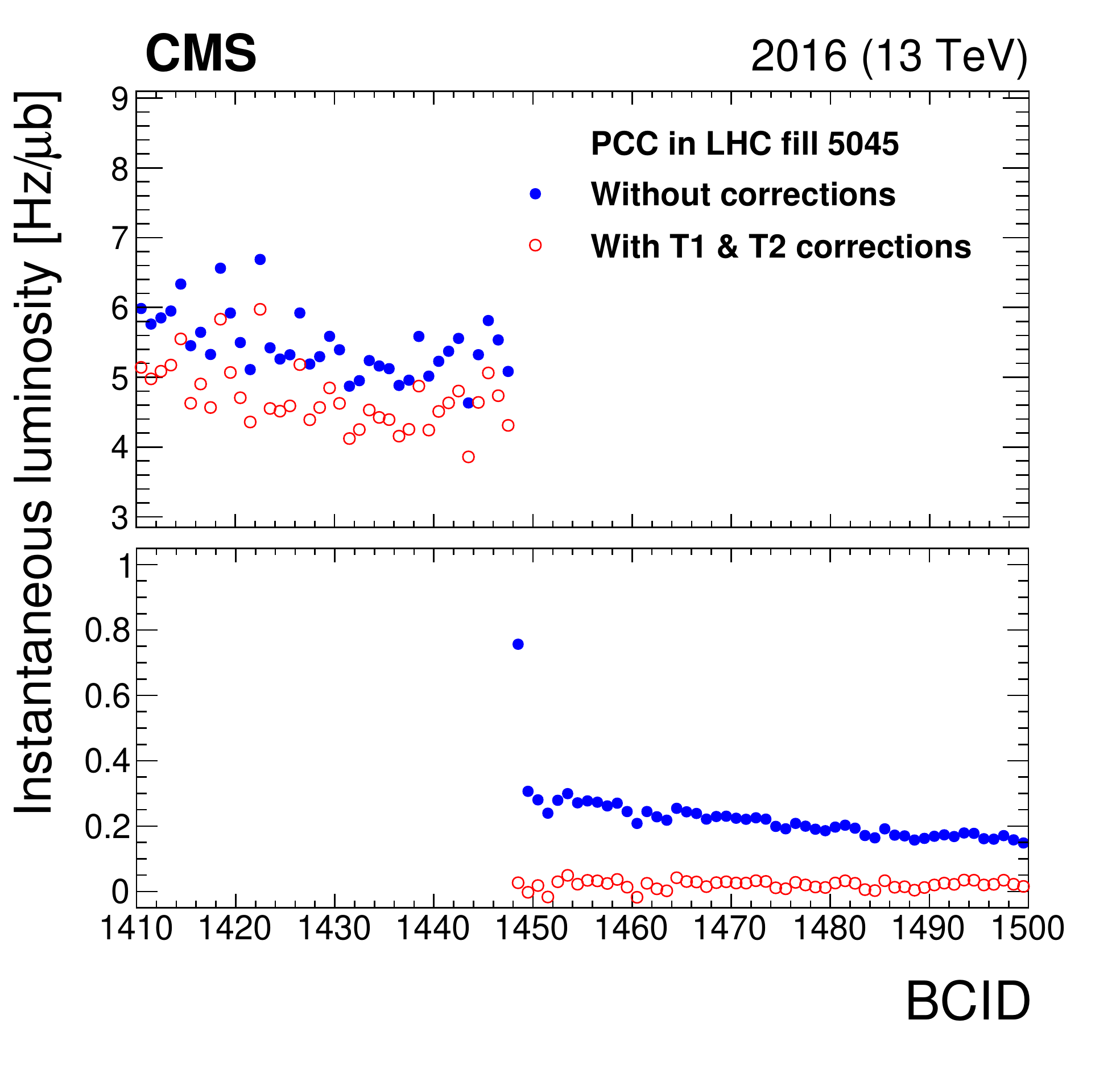}
\includegraphics[width=0.45\textwidth]{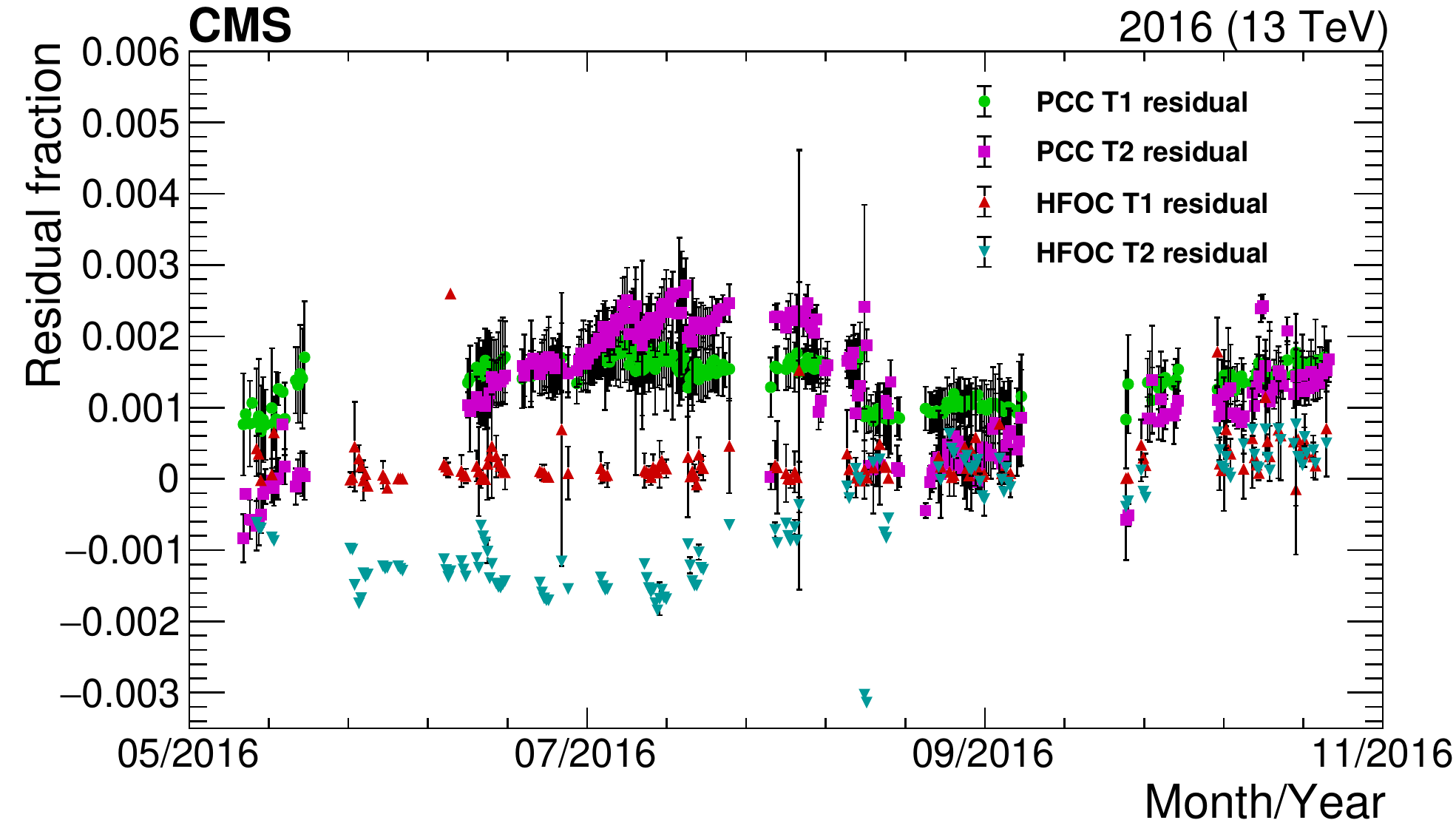}
}{
\includegraphics[width=0.4\textwidth]{Figure_015-a.pdf}
\raisebox{0.7cm}{\includegraphics[width=0.55\textwidth]{Figure_015-b.pdf}}
}
\caption{The \cmsLeft plot shows the instantaneous luminosity measured from PCC as a function of BCID before (filled blue points) and after (open red
  points) afterglow corrections are applied for each colliding bunch. The upper panel shows a subset of bunch crossings colliding at
  IP 5, and the lower panel shows empty bunch crossings (the scale is different in the two panels to show differences more clearly).  The
  open red points in the lower panel lie close to 0, indicating that any residual PCC response is small in
  empty bunch slots. The \cmsRight plot shows the estimated residual T1 and T2 afterglow as a function of 
  time during the full range of 2016 data for both PCC and HFOC, which use the same afterglow subtraction 
  methodology.}
\label{fig:pcc_afterglow}
\end{figure}

After the corrections from the SBR are applied, empty bunch slots, where there are no collisions, should have a
rate of zero.  For PCC, the SBR is determined by optimizing $\alpha_\text{T1}$, which is time dependent and 
measured in intervals of about 20\unit{min}, and the parameters of the 
exponential used for T2 corrections, such that there is minimal residual rate in the noncolliding bunch slots.  
Figure~\ref{fig:pcc_afterglow} shows per-bunch data in a fill from 2016 before and after the afterglow 
corrections for PCC are applied.

The empty bunch slots are also used to estimate the residual afterglow after the full set of corrections
is applied.  The corrected rate in the first empty bunch slot after a colliding bunch slot is used to estimate
the residual T1 response.  Likewise, the 2nd to 30th empty bunch slots are used to estimate the residual T2
effect.  This procedure is performed for the entire 2015 and 2016 data sets for PCC and HFOC luminosity measurements.  A
window covering all residuals over the course of each data set is used as the systematic uncertainty in the
final corrections.  The resulting uncertainty for PCC in the two corrections is $\TypeIUncI\bigoplus\TypeIIUncI$ 
($\TypeIUncII\bigoplus\TypeIIUncII$)\% in 2015 (2016).

These types of per-bunch luminosity corrections are applied for PCC, HFOC, and BCM1F, whereas PLT is almost
completely background free and no such correction is needed. Since the DT and RAMSES measurements integrate over all
bunch crossings, out-of-time pileup corrections can only be applied on average to the integrated rates.
For DT these amount to 0--1\%, while no corrections are applied to RAMSES.

A second type of T1 afterglow affects the HFOC luminosity. This is the case where the afterglow from a preceding
bunch and the signal from the current bunch are both under the threshold to be counted as a hit, but their sum
exceeds the threshold. This effect is referred to as the ``bunch train effect'', because it affects only
active bunches preceded by other active bunches (that is, bunches within a train, as opposed to ``leading''
bunches at the beginning of a train).  The method previously described for estimating T1 afterglow does not
include this contribution.  This effect is measured in a dedicated study comparing the double ratio of the
leading bunch in a train relative to the second bunch for HFOC divided by the same ratio for PCC.  A single 
correction model with magnitude 1--5\%, linearly increasing with instantaneous luminosity, is determined 
utilizing most valid data from 2016.

\subsection{Efficiency corrections}
Radiation damage can affect the detector response by reducing efficiency, increasing
noise, or both. Noise is typically a small effect for most luminometers, but reduced response in detectors due
to radiation damage can have significant (percent-level) effects, and so corrections are required.
Corrections are measured against a stable benchmark relative to the performance at or near the \vdM scans,
and are applied to \sigmaVis.  Shifts of 0--10\% in detector response in the PLT in 2016
are corrected using RAMSES as a benchmark, whereas the impact of radiation damage on the HF efficiency is 
corrected using a parameterization derived from a model of aging.  An HF efficiency correction of 1\% is derived
by measuring the average energy deposits in the HF in events characterized by the presences of \PZ bosons
that decay to two muons with large transverse momentum. 

A further efficiency correction is necessary for the PCC measurement.
The pixel detector has a static internal memory buffer for data storage before the
trigger decision is taken.  When the buffer is filled, the oldest data overflows and are lost.
This effect is proportional to the total instantaneous luminosity, and it can be estimated by studying the frequency 
of missing pixel clusters in otherwise well-reconstructed tracks~\cite{TRK-11-001}.  In 2016, the effect was 1.0\%
at 1.4\ten{34}\percms.  A correction proportional to total instantaneous luminosity is applied, and the 
total impact on integrated luminosity is 0.2\%.  Since the total luminosity in 2015 is substantially lower, no 
correction is applied.  The PCC also has very small noise corrections.

\subsection{Nonlinear response}
\label{sec:linearity}
In the absence of out-of-time pileup, the PCC luminosity is expected 
to be linear, according to simulations, so no corrections are applied.  Moreover, the ratios of PCC to both DT and RAMSES 
luminosity measurements are highly compatible as a function of the instantaneous luminosity without any corrections.
The HFOC response in 2015 and 2016, on the other hand, exhibits significant nonlinearity compared to the
other luminometers. The main source of nonlinearity is the uncalibrated ADC-to-charge conversion
applied at the time of data taking. Data from fill 5416, which exhibit a wide range of instantaneous luminosity, 
are used to model the correction for HFOC with a fourth-order polynomial.  This smooth function extrapolates to the 
\sigmaVis calibration at low pileup within uncertainty.  This single model is used to correct the nonlinear behavior of 
HFOC (0--10\% higher response when compared to PCC) throughout 2016.

As described in Section~\ref{SEC:lumi_algos}, nonlinearity corrections are also needed for PLT. The 
corrections are modeled with a first-order polynomial.  The parameters are time dependent, because of changes 
in the PLT operating conditions during the course of 2015 and 2016.  These corrections, amounting to $-$0.2 to $+$1.4/(Hz/$\mu$b),
are derived by comparing with RAMSES data in five different periods.

\section{Detector stability and linearity}
\label{SEC:Performance}

After the rate corrections are applied (as discussed in Section~\ref{SEC:RateCorrections}), 
comparisons between different luminometers are performed to assess remaining systematic effects 
impacting the luminosity measurement.  Since PCC is expected to suffer the least from nonlinearities, as described 
in Section~\ref{sec:PCC}, once out-of-time pileup effects are corrected, PCC is the preferred luminometer 
for these data sets in the following estimates for stability and potential nonlinearity.

\subsection{Upper bounds on stability}
\begin{figure*}[htp]
\centering
\includegraphics[width=0.895\textwidth]{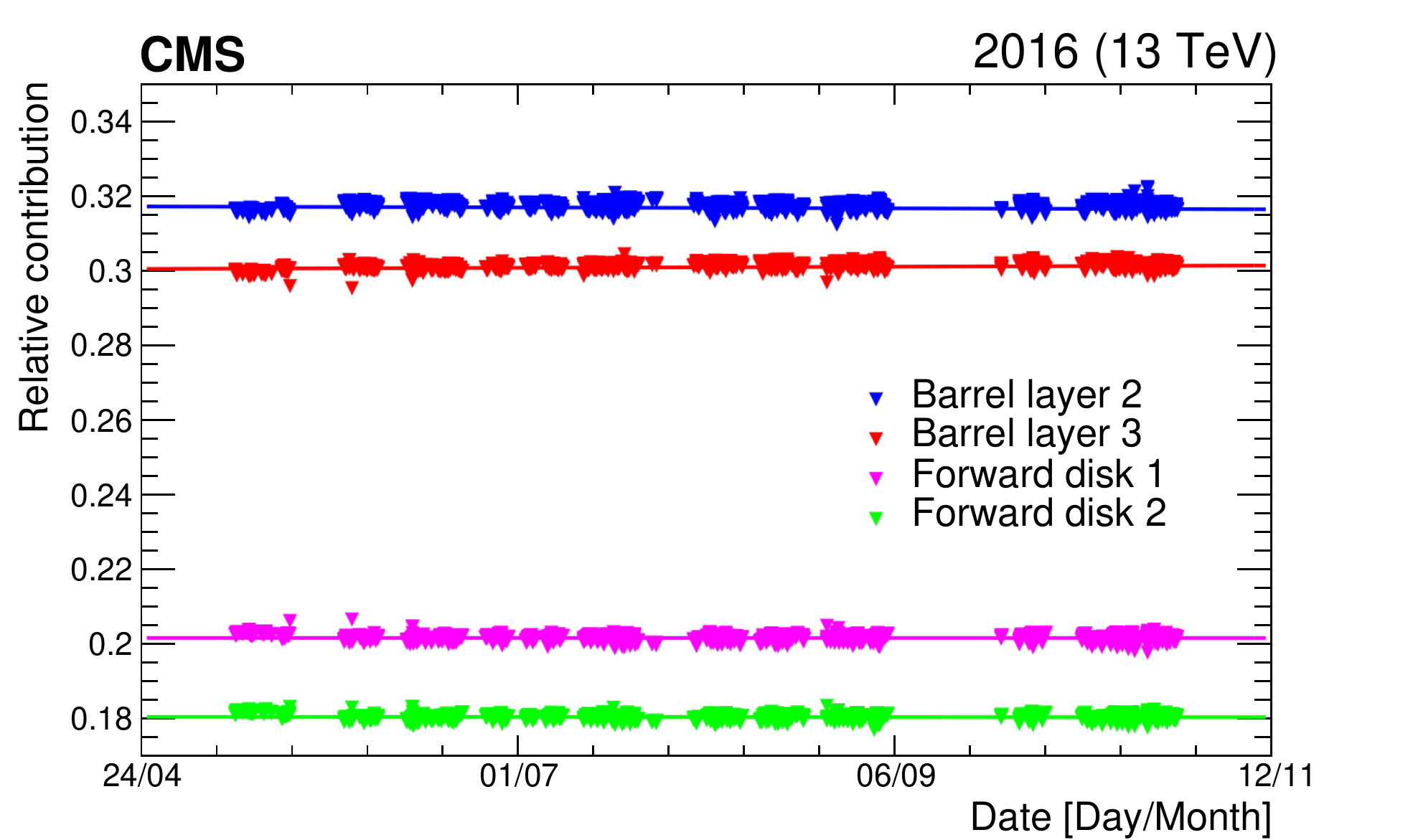}
\caption{
The relative contribution to the total number of observed pixel clusters from the four regions of the pixel
detector used in the luminosity measurement (barrel layers 2 and 3, and inner and outer forward pixel disks),
as a function of time throughout 2016. The lines represent first-order polynomial fits to the relative contributions from each region.
 \label{fig:stabilityPCC}}
\end{figure*}

One measurement of potential instability in the PCC luminosity comes from intrinsic monitoring 
(\ie, comparing rates from different sections/parts of the subdetector over time). In a perfectly 
stable system, the fractional rates among different subcomponents would exhibit no variation with 
time.  Figure~\ref{fig:stabilityPCC} shows the result of applying out-of-time pileup corrections (as discussed in 
Section~\ref{SUBSEC:afterglow}) separately to each pixel layer or disk.
The sum of the per-region corrections matches the total, nominal correction made for all PCC regions 
to better than 0.1\%.  After the corrections are applied, the relative rates are quite stable over 
the course of 2015 and 2016. This is also shown in Fig.~\ref{fig:stabilityPCC}, where the relative PCC rates over time are simultaneously fit to a first-order polynomial.

\begin{figure*}[htp]
\centering
\includegraphics[width=0.895\textwidth]{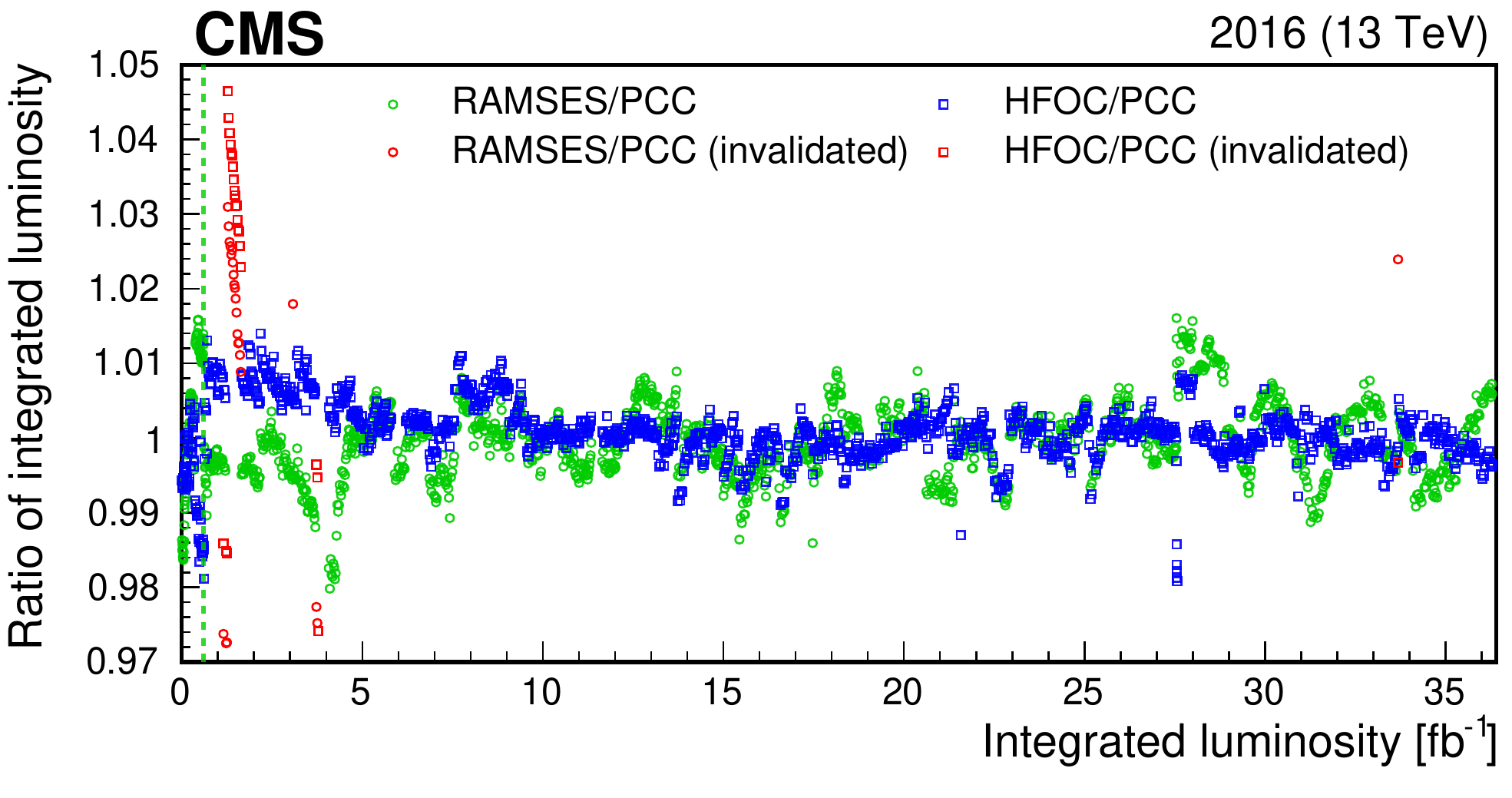}
\caption{The luminosity measurements from PCC, HFOC, and RAMSES are compared as a function of the 
integrated luminosity in 2016. Comparison among three luminometers facilitates the
identification of periods where a single luminometer suffers from transient stability issues. 
The ratios that are plotted in red contain invalidated data. 
The dashed line delineates the \vdM calibration (fill \FillNumberII).}
\label{fig:timestability}
\end{figure*}

However, this method cannot detect global shifts in \sigmaVis, and so it is crucial to make comparisons with
completely independent systems. With multiple independent systems available for comparison, luminometers
displaying brief periods of instability can be clearly identified.  The cross-detector comparison is repeated 
for the entire data set for each year to detect periods where a single luminometer experiences transient effects 
(\eg, data quality issues, some detector components off, anomalous signals, etc.).  
Figure~\ref{fig:timestability} shows the ratio of the luminosity measurements for different pairs of detectors 
throughout 2016, highlighting (in red) periods where the ratios significantly deviate from unity and so the associated data are invalidated.

\begin{figure*}[htp]
\centering
\includegraphics[width=0.495\textwidth]{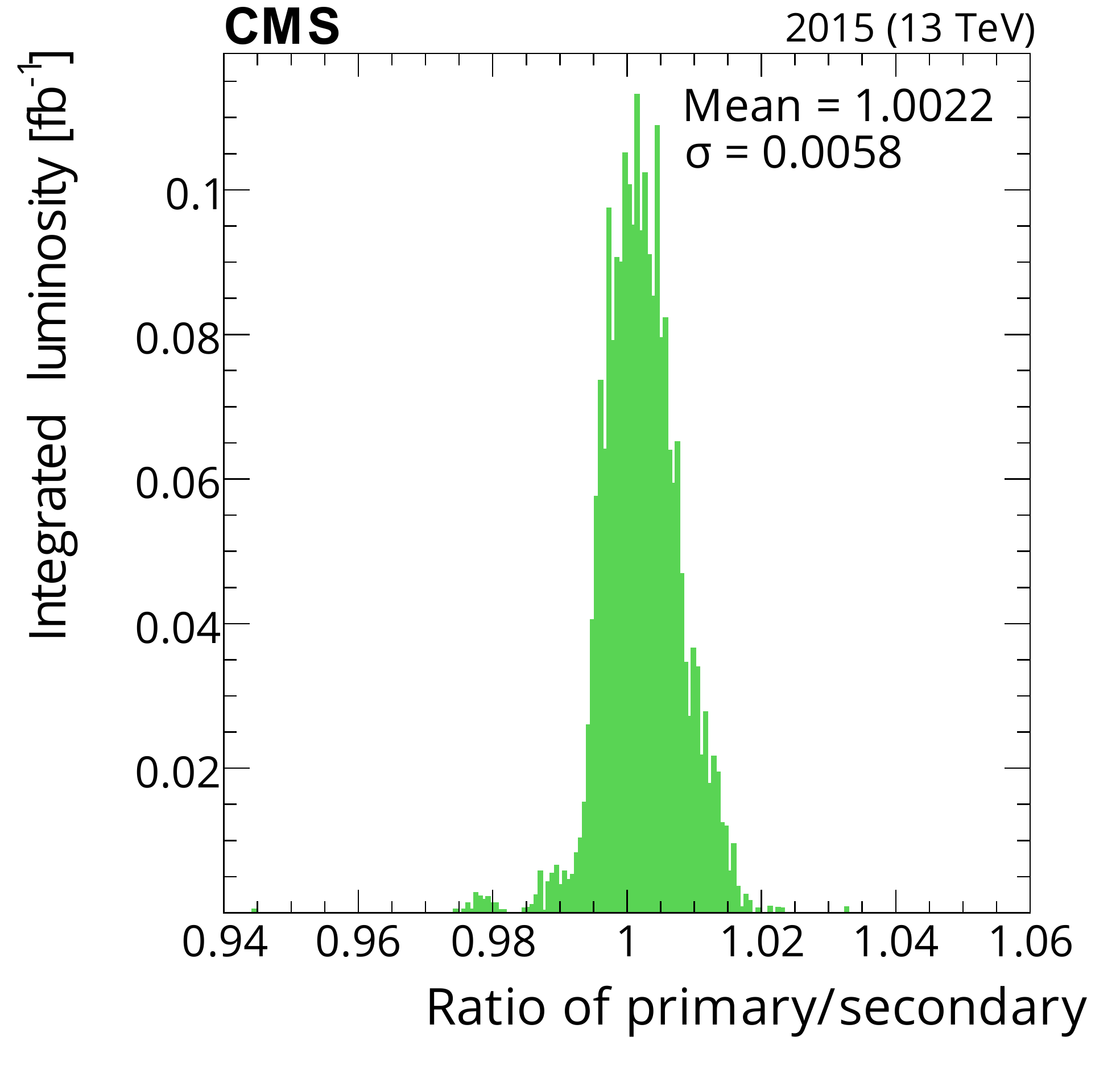}
\includegraphics[width=0.495\textwidth]{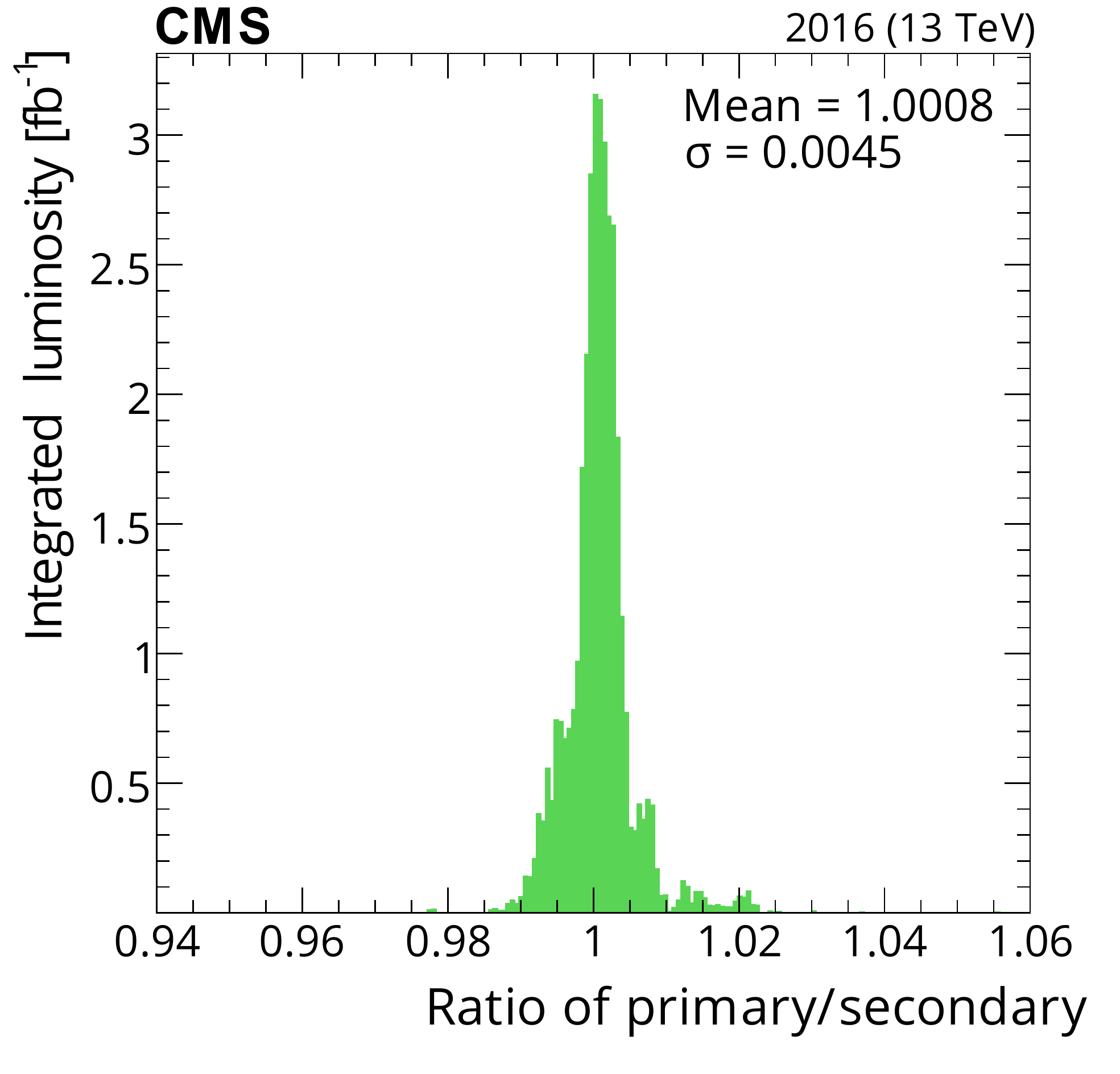}
\caption{
  The ratio of the primary (best available) to secondary (next-best available) luminosity as computed in time windows of approximately 20\unit{min} each.  
  The left plot shows the 2015 results (principally PCC/RAMSES), and the right plot shows the 2016 results (principally PCC/HFOC).  Each entry is weighted by the integrated
  luminosity for the time period.}
\label{fig:stability}
\end{figure*}

After the exclusion of invalidated data, which amount to $\lesssim$5\% for each luminometer, the remaining 
input from different luminometers is used to assess an upper limit on the stability of the luminosity. 
PCC measurements are valid for 98.3 (94.3)\% of the data set in 2015 (2016).  The rest of the luminosity
is provided by the next most stable luminometer, which is RAMSES (HFOC) for 2015 (2016).  
The primary luminosity, which is PCC or luminosity from the next most stable detector when PCC is unavailable, is 
compared with the next-best available luminometer (secondary). 
In Fig.~\ref{fig:stability}, the latter is selected using the lowest standard deviation in the ratio 
relative to PCC over fixed time intervals of approximately 20\unit{min} each.  
The position of the mean shows the agreement between the luminometers on the integrated luminosity.  The width 
reflects stability effects, as well as residual statistical uncertainty in the luminosity measurement in each 
interval. From the distribution over the course of each year, the width is an upper limit on the 
uncertainty due to time dependencies in the luminometers.  For 2015 (2016) a systematic uncertainty due to 
detector stability of \StabUncI (\StabUncII)\% is derived.

\subsection{Time dependence of linearity}
We make use of two methods for assessing the detector linearity. The primary method compares the ratio of
the instantaneous luminosity from two luminometers per fill as a function of the instantaneous luminosity, 
which is estimated from the numerator.  A first-order polynomial fit is performed and the slope is extracted. 
The slopes per fill are then studied as a function of time. No significant deviation over time is observed between 
DT/PCC or RAMSES/PCC and HFOC/PCC, DT/PCC, or RAMSES/PCC in 2015 and 2016, respectively.

\begin{figure}[htp]
\centering
\includegraphics[width=0.495\textwidth]{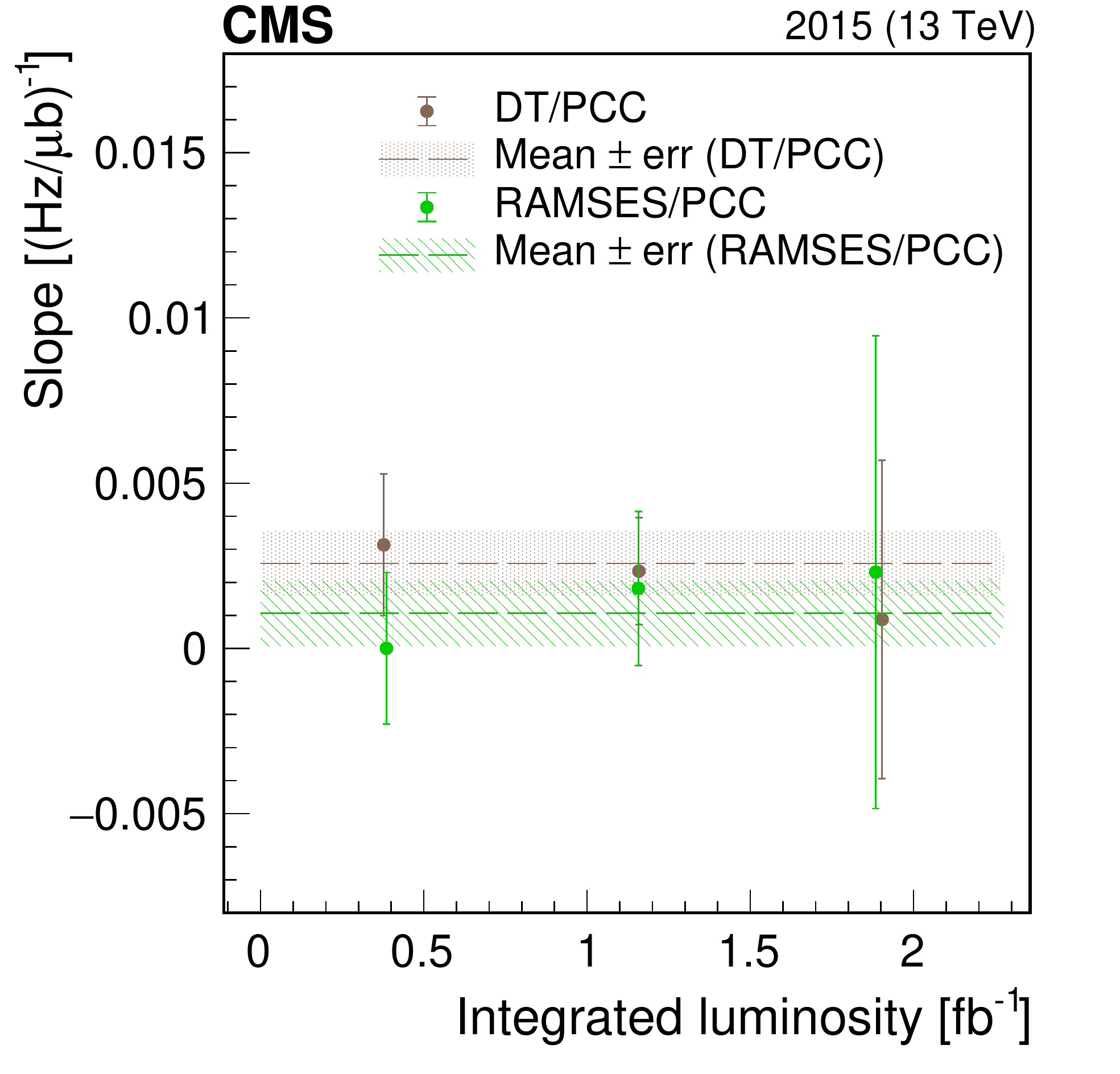}
\includegraphics[width=0.495\textwidth]{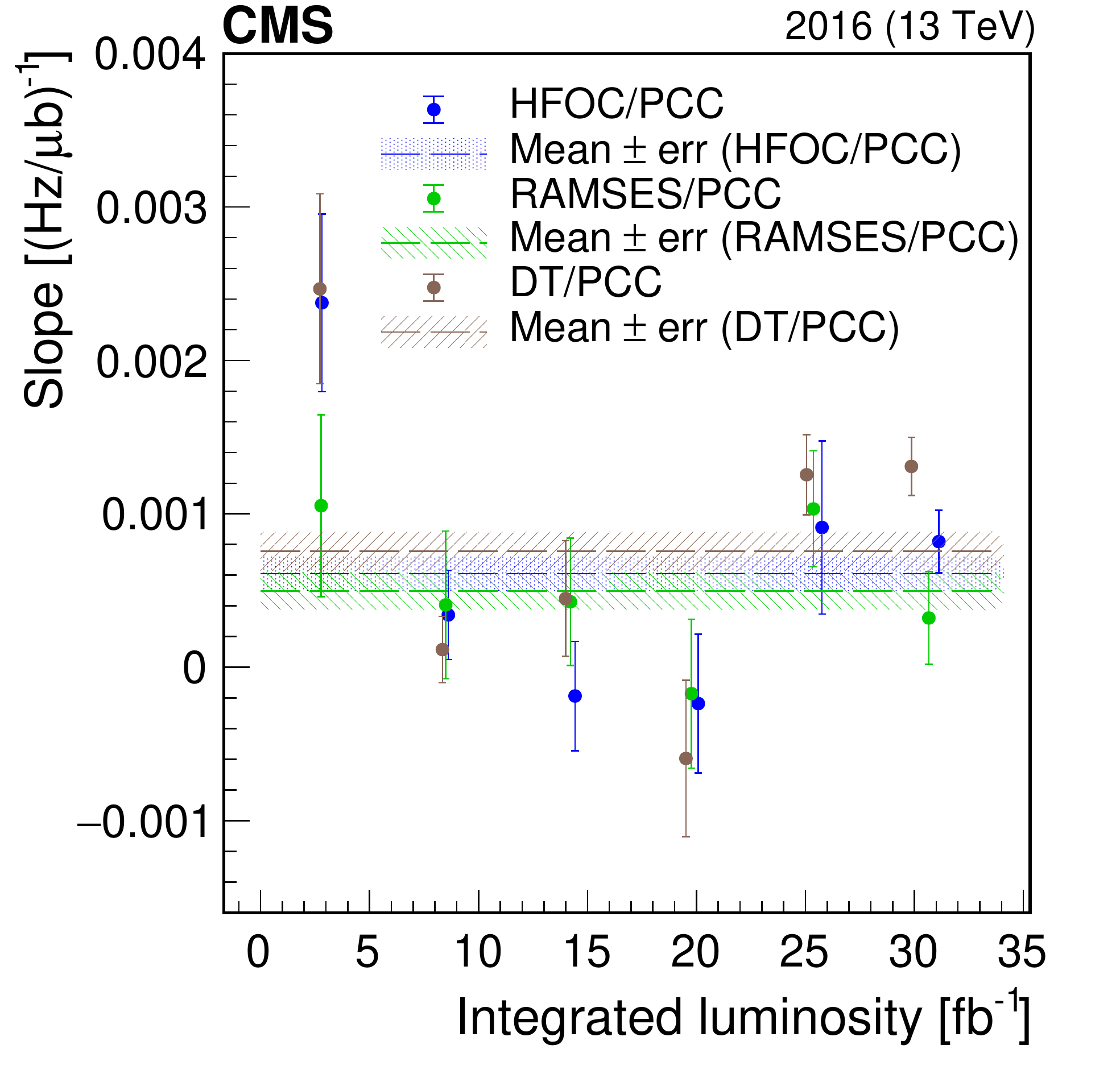}
\caption{
  Linearity summary for 2015 (\cmsLeft) and 2016 (\cmsRight) at \highEnergy.
  The slopes are plotted for each detector relative to PCC.  The markers are averages of fill-by-fill slopes from fits 
  binned in roughly equal fractions of the total integrated luminosity through the year.  The error bars on the markers are the propagated statistical uncertainty from fitted slope parameters in each fill, which are weighted by integrated luminosities of each fill.  The dashed lines and corresponding hatched
  areas show the average from the entire data set and its uncertainty.}
\label{fig:linearityOverall}
\end{figure}

{\tolerance=800 To estimate the uncertainty, the fitted slopes are weighted according to the per-fill integrated luminosity.
The mean values deviate slightly from 0, and the largest deviation is the systematic uncertainty 
in the linearity of PCC luminosity.  Figure~\ref{fig:linearityOverall} shows the summary of these slopes  
for 2015 and 2016 at \highEnergy both for the whole year, and for subsets of each data set with equal
luminosity. The largest average slope is 0.26 (0.08)\%/(Hz/$\mu$b) in 2015 (2016), which translates into a 
\LinearityUncI (\LinearityUncII)\% uncertainty in the integrated luminosity of the 2015 (2016) data set, where the 
average \Lb is approximately 2.0 (3.3) Hz/$\mu$b.\par}

The alternative method makes use of the entire data set throughout the year, and extracts a single relative
slope with a first-order polynomial fit. To remove effects from variations in the absolute luminosity scale
over time, the per-fill ratios are shifted such that their extrapolation at zero luminosity is unity.  The
results are consistent with the primary method described above.

\section{Total luminosity correction and uncertainty}
\label{SEC:SysErr}

For each data set, final rate corrections and final calibrations are applied to data in small time windows of 
$2^{18}$ LHC orbits, approximately 23 seconds.  All the measurements are summed to derive a total integrated luminosity measurement.
The contributions to the systematic uncertainty in the integrated luminosity are divided into two general categories:  
\begin{itemize}
  \item ``normalization'' uncertainty in the absolute luminosity scale, \sigmaVis, determined from the \vdM scan procedure
  \item ``integration'' uncertainty associated with \sigmaVis variations over time (stability) and pileup 
    (linearity and out-of-time pileup corrections)
\end{itemize}
The magnitudes of the corrections applied to the absolute normalization from the \vdM calibration are listed in
Table~\ref{TAB:SystematicEffect}, and Table~\ref{TAB:SystematicError} summarizes the sources of uncertainty.  
The dominant sources of normalization uncertainty are associated with the beam position monitoring (as discussed 
in Section~\ref{sec:od}), transverse factorizability (as explained in Section~\ref{sec:xycorr}), and beam-beam effects 
(as described in Section~\ref{sec:beambeam}).

The dominant sources of integration uncertainty arise from the linearity and stability of the primary relative 
to secondary luminosity measurements over the course of each year (as discussed in Section~\ref{SEC:Performance}).  In addition, 
the subleading systematic uncertainty due to out-of-time pileup corrections is considered for the PCC method since 
it is primarily used for the luminosity estimate.

Several sources of normalization uncertainty are considered to be correlated for the years studied 
because the scan procedures and analysis methodology are identical between the two \vdM calibrations.  
The sources of the normalization uncertainty that are not correlated between the two \vdM programs, and are partly statistical in nature, are
the orbit drift, along with the scan-to-scan and bunch-to-bunch variations in the measured \sigmaVis. 
The latter are collectively referred to as ``other variations in \sigmaVis'' in Table~\ref{TAB:SystematicError}.

Among the sources of integration uncertainty, the afterglow corrections are treated identically in 
the two data sets, and so this source of systematic uncertainty is correlated.  The estimate of the 
uncertainty due to linearity is considered to be correlated, since it is derived from the PCC linearity in both years.  On the
other hand, the stability assessment is based on cross-detector comparisons.  Although PCC is the 
primary luminometer in each data set, the secondary luminometer is different for each year.
Since the source of instability cannot be assessed and contains time-dependent features, the uncertainty is not correlated.

The tool used for providing luminosity values to physics analyses applies the corrections to the raw
luminosity values using the average per-bunch luminosity, rather than the individual bunch-by-bunch
values. This potentially introduces an error in the case where these corrections include a nonlinear term and
the bunch-by-bunch luminosity varies significantly among bunches. We evaluated the effect of this approximation on 2016
data, and found that the overall impact on the integrated luminosity was $<$0.1\%.

Finally, the quantity measured by the luminometers is the luminosity delivered to CMS; however, the quantity
of interest to most physics analyses is the luminosity corresponding to the data actually recorded by the CMS
DAQ system. These are related by the deadtime, as obtained from the trigger and clock system of
CMS~\cite{Khachatryan:2016bia}. In 2015 this measurement was affected by an algorithm issue in the trigger
system and has an uncertainty of \DeadTimeUncI\%, but this problem was resolved before data taking began in 2016,
so in 2016 the impact is negligible ($<$\DeadTimeUncII\%) and uncorrelated with 2015.

When applying the vdM calibration to the entire periods, the total integrated luminosity is \LumiI\fbinv with a relative precision of
\TotUncI\% in 2015, and \LumiII\fbinv with a relative precision of \TotUncII\% in 2016.  The combined $2015+2016$
luminosity measurement has a precision of \TotUncII\%, which is the same as the 2016 precision since it is the 
significantly larger data set and the precision in 2015 is similar.

\begin{table*}
  \centering
  \topcaption{
    Summary of the BCID-averaged corrections to \sigmaVis (in \%) obtained with the \vdM scan calibrations at \highEnergy in 2015 and 2016.
    When a range is shown, it is because of possible scan-to-scan variations.
    To obtain the impact on \sigmaVis, each correction is consecutively included, the fits are redone following the order below, and the result is compared with the baseline.
    The impact from transverse factorizability is obtained separately (as discussed in Section~\ref{sec:xycorr}).}
  \label{TAB:SystematicEffect}
  \begin{tabular}{lcc}
    \hline
    \multirow{2}{*}{Source} & \multicolumn{2}{c}{Impact on \sigmaVis [\%]} \\
    & 2015 & 2016  \\
    \hline
    Ghost and satellite charge               & $+$\SpuriousEffI                        & $+$\SpuriousEffII \\
    Orbit drift                              & $+$\ODRandomLowEffI to $+$\ODRandomUpEffI & $+$\ODRandomLowEffII to $+$\ODRandomUpEffII \\
    Residual beam position corrections       & \ODSystLowEffI to \ODSystUpEffI         & \ODSystLowEffII to \ODSystUpEffII \\
    Beam-beam effects                        & $+$\DefAndFocEffI                       & $+$\DefAndFocEffII  \\
    Length scale calibration                 & $-$\LSCEffI                             & $-$\LSCEffII   \\
    Transverse factorizability               & $+$\BIAndLRLowEffI to $+$\BIAndLRUpEffI & $+$\BIAndLREffII  \\
    \hline
  \end{tabular}
\end{table*}

\begin{table*}
  \centering
  \topcaption{
    Summary of contributions to the relative systematic uncertainty in \sigmaVis (in \%) at \highEnergy
    in 2015 and 2016.  The systematic uncertainty is divided into groups affecting the description of the 
    \vdM profile and the bunch population product measurement (normalization), and the measurement of 
    the rate in physics running conditions (integration).  The fourth column indicates whether the sources of uncertainty are 
    correlated between the two calibrations at \highEnergy.}
  \label{TAB:SystematicError}
  \begin{tabular}{lccc}
    \hline
    Source                                        & 2015 [\%] & 2016 [\%] & Corr  \\
    \hline
    \multicolumn{4}{c}{Normalization uncertainty} \\
    \multicolumn{1}{l}{\textit{Bunch population}}\\
    \hspace{+2mm} Ghost and satellite charge      & \SpuriousUnc & \SpuriousUnc & Yes  \\
    \hspace{+2mm} Beam current normalization      & \BeamCurUnc & \BeamCurUnc & Yes  \\
    \multicolumn{1}{l}{\textit{Beam position monitoring}}\\
    \hspace{+2mm} Orbit drift                     & \ODRandomUncI & \ODRandomUncII & No  \\
    \hspace{+2mm} Residual differences            & \ODSystUncI & \ODSystUncII & Yes  \\
    \multicolumn{1}{l}{\textit{Beam overlap description}}\\
    \hspace{+2mm} Beam-beam effects               & \DefAndFocUnc & \DefAndFocUnc & Yes \\
    \hspace{+2mm} Length scale calibration        & \LSCUncI & \LSCUncII & Yes   \\
    \hspace{+2mm} Transverse factorizability      & \BIAndLRUnc & \BIAndLRUnc & Yes   \\
    \multicolumn{1}{l}{\textit{Result consistency}}\\
    \hspace{+2mm} Other variations in \sigmaVis   & \OtherVarUncI & \OtherVarUncII & No  \\[\cmsTabSkip]
    \multicolumn{4}{c}{Integration uncertainty} \\
    \multicolumn{1}{l}{\textit{Out-of-time pileup corrections}}\\
    \hspace{+2mm} Type 1 corrections              & \TypeIUncI & \TypeIUncII & Yes \\
    \hspace{+2mm} Type 2 corrections              & \TypeIIUncI & \TypeIIUncII & Yes \\
    \multicolumn{1}{l}{\textit{Detector performance}}\\
    \hspace{+2mm} Cross-detector stability        & \StabUncI & \StabUncII & No   \\
    \hspace{+2mm} Linearity                       & \LinearityUncI & \LinearityUncII & Yes   \\
    \multicolumn{1}{l}{\textit{Data acquisition}}\\
    \hspace{+2mm} CMS deadtime                    & \DeadTimeUncI & \hspace{-3.9mm}$<$\DeadTimeUncII & No   \\[\cmsTabSkip]
    Total normalization uncertainty               & \VdmUncI & \VdmUncII & \NA  \\
    Total integration uncertainty                 & \IntUncI & \IntUncII & \NA \\
    Total uncertainty                             & \TotUncI & \TotUncII & \NA  \\
    \hline
  \end{tabular}
\end{table*}

\section{Summary}
\label{SEC:Summary}

The luminosity calibration using beam-separation (van der Meer, \vdM) scans has 
been presented for data from proton-proton collisions recorded by the CMS experiment in 2015 and 2016
when all subdetectors were fully operational.  
The main sources of systematic uncertainty are related to
residual differences between the measured beam positions and the ones provided by the operational settings of the LHC magnets,
the factorizability of the transverse spatial distributions of proton bunches,
and the modeling of effects on the proton distributions due to electromagnetic interactions among protons in the colliding bunches.
When applying the \vdM calibration to the entire data-taking period, 
the relative stability and linearity of luminosity subdetectors (luminometers) are included
in the uncertainty in the integrated luminosity measurement as well.

The resulting relative precision in the calibration from the \vdM scans is \VdmUncI (\VdmUncII)\% in 2015 
(2016) at \highEnergy; the integration uncertainty due to luminometer-specific effects contributes \IntUncI (\IntUncII)\%, resulting in a total uncertainty
of \TotUncI (\TotUncII)\%; when applying the \vdM calibration to the entire periods, the total integrated luminosity is \LumiI (\LumiII)\fbinv.

{\tolerance=800 The final precision is among the best achieved at bunched-beam hadron colliders.
Advanced techniques are used to estimate and correct for the bias associated with
the beam position monitoring at the scale of $\mu$m, the factorizability of the transverse beam distribution, and beam-beam effects.  
In addition, detailed luminometer rate corrections and the inclusion of novel measurements (such as the data 
from the Radiation Monitoring System for the Environment and Safety) lead to
precise estimates of the stability and linearity over time.\par}

In the coming years, a similarly precise calibration of the real-time luminosity delivered to the
LHC will become increasingly important for standard operations.  Under those conditions, the impact of 
out-of-time pileup effects is expected to be larger, but in principle they can be mitigated using techniques described in this paper.

\begin{acknowledgments}
\hyphenation{Bundes-ministerium Forschungs-gemeinschaft Forschungs-zentren Rachada-pisek}
We congratulate our colleagues in the CERN accelerator departments for the excellent performance of the LHC machine. We thank all contributors to the LHC Luminosity Calibration and Monitoring Working Group and, in particular, its co-coordinator Witold Kozanecki for the discussions on beam instrumentation and beam physics.  We also thank the technical and administrative staffs at CERN and at other CMS institutes for their contributions to the success of the CMS effort. In addition, we gratefully acknowledge the computing c enters and personnel of the Worldwide LHC Computing Grid and other centers for delivering so effectively the computing infrastructure essential to our analyses. Finally, we acknowledge the enduring support for the construction and operation of the LHC, the CMS detector, and the supporting computing infrastructure provided by the following funding agencies: the Austrian Federal Ministry of Education, Science and Research and the Austrian Science Fund; the Belgian Fonds de la Recherche Scientifique, and Fonds voor Wetenschappelijk Onderzoek; the Brazilian Funding Agencies (CNPq, CAPES, FAPERJ, FAPERGS, and FAPESP); the Bulgarian Ministry of Education and Science; CERN; the Chinese Academy of Sciences, Ministry of Science and Technology, and National Natural Science Foundation of China; the Ministerio de Ciencia Tecnolog\'ia e Innovaci\'on (MINCIENCIAS), Colombia; the Croatian Ministry of Science, Education and Sport, and the Croatian Science Foundation; the Research and Innovation Foundation, Cyprus; the Secretariat for Higher Education, Science, Technology and Innovation, Ecuador; the Ministry of Education and Research, Estonian Research Council via PRG780, PRG803 and PRG445 and European Regional Development Fund, Estonia; the Academy of Finland, Finnish Ministry of Education and Culture, and Helsinki Institute of Physics; the Institut National de Physique Nucl\'eaire et de Physique des Particules~/~CNRS, and Commissariat \`a l'\'Energie Atomique et aux \'Energies Alternatives~/~CEA, France; the Bundesministerium f\"ur Bildung und Forschung, the Deutsche Forschungsgemeinschaft (DFG), under Germany's Excellence Strategy -- EXC 2121 ``Quantum Universe" -- 390833306, and under project number 400140256 - GRK2497, and Helmholtz-Gemeinschaft Deutscher Forschungszentren, Germany; the General Secretariat for Research and Technology, Greece; the National Research, Development and Innovation Fund, Hungary; the Department of Atomic Energy and the Department of Science and Technology, India; the Institute for Studies in Theoretical Physics and Mathematics, Iran; the Science Foundation, Ireland; the Istituto Nazionale di Fisica Nucleare, Italy; the Ministry of Science, ICT and Future Planning, and National Research Foundation (NRF), Republic of Korea; the Ministry of Education and Science of the Republic of Latvia; the Lithuanian Academy of Sciences; the Ministry of Education, and University of Malaya (Malaysia); the Ministry of Science of Montenegro; the Mexican Funding Agencies (BUAP, CINVESTAV, CONACYT, LNS, SEP, and UASLP-FAI); the Ministry of Business, Innovation and Employment, New Zealand; the Pakistan Atomic Energy Commission; the Ministry of Science and Higher Education and the National Science Center, Poland; the Funda\c{c}\~ao para a Ci\^encia e a Tecnologia, Portugal; JINR, Dubna; the Ministry of Education and Science of the Russian Federation, the Federal Agency of Atomic Energy of the Russian Federation, Russian Academy of Sciences, the Russian Foundation for Basic Research, and the National Research Center ``Kurchatov Institute"; the Ministry of Education, Science and Technological Development of Serbia; the Secretar\'{\i}a de Estado de Investigaci\'on, Desarrollo e Innovaci\'on, Programa Consolider-Ingenio 2010, Plan Estatal de Investigaci\'on Cient\'{\i}fica y T\'ecnica y de Innovaci\'on 2017--2020, research project IDI-2018-000174 del Principado de Asturias, and Fondo Europeo de Desarrollo Regional, Spain; the Ministry of Science, Technology and Research, Sri Lanka; the Swiss Funding Agencies (ETH Board, ETH Zurich, PSI, SNF, UniZH, Canton Zurich, and SER); the Ministry of Science and Technology, Taipei; the Thailand Center of Excellence in Physics, the Institute for the Promotion of Teaching Science and Technology of Thailand, Special Task Force for Activating Research and the National Science and Technology Development Agency of Thailand; the Scientific and Technical Research Council of Turkey, and Turkish Atomic Energy Authority; the National Academy of Sciences of Ukraine; the Science and Technology Facilities Council, UK; the US Department of Energy, and the US National Science Foundation.

Individuals have received support from the Marie-Curie program and the European Research Council and Horizon 2020 Grant, contract Nos.\ 675440, 724704, 752730, 765710, and 824093 (European Union); the Leventis Foundation; the Alfred P.\ Sloan Foundation; the Alexander von Humboldt Foundation; the Belgian Federal Science Policy Office; the Fonds pour la Formation \`a la Recherche dans l'Industrie et dans l'Agriculture (FRIA-Belgium); the Agentschap voor Innovatie door Wetenschap en Technologie (IWT-Belgium); the F.R.S.-FNRS and FWO (Belgium) under the ``Excellence of Science -- EOS" -- be.h project n.\ 30820817; the Beijing Municipal Science \& Technology Commission, No. Z191100007219010; the Ministry of Education, Youth and Sports (MEYS) of the Czech Republic; the Lend\"ulet (``Momentum") Program and the J\'anos Bolyai Research Scholarship of the Hungarian Academy of Sciences, the New National Excellence Program \'UNKP, the NKFIA research grants 123842, 123959, 124845, 124850, 125105, 128713, 128786, and 129058 (Hungary); the Council of Scientific and Industrial Research, India; the National Science Center (Poland), contracts Opus 2014/15/B/ST2/03998 and 2015/19/B/ST2/02861; the National Priorities Research Program by Qatar National Research Fund; the Ministry of Science and Higher Education, project no. 0723-2020-0041 (Russia); the Programa de Excelencia Mar\'{i}a de Maeztu, and the Programa Severo Ochoa del Principado de Asturias; the Thalis and Aristeia programs cofinanced by EU-ESF, and the Greek NSRF; the Rachadapisek Sompot Fund for Postdoctoral Fellowship, Chulalongkorn University, and the Chulalongkorn Academic into Its 2nd Century Project Advancement Project (Thailand); the Kavli Foundation; the Nvidia Corporation; the SuperMicro Corporation; the Welch Foundation, contract C-1845; and the Weston Havens Foundation (USA).
\end{acknowledgments}

\bibliography{auto_generated}
\cleardoublepage \appendix\section{The CMS Collaboration \label{app:collab}}\begin{sloppypar}\hyphenpenalty=5000\widowpenalty=500\clubpenalty=5000\vskip\cmsinstskip
\textbf{Yerevan Physics Institute, Yerevan, Armenia}\\*[0pt]
A.M.~Sirunyan$^{\textrm{\dag}}$, A.~Tumasyan
\vskip\cmsinstskip
\textbf{Institut f\"{u}r Hochenergiephysik, Wien, Austria}\\*[0pt]
W.~Adam, J.W.~Andrejkovic, T.~Bergauer, S.~Chatterjee, M.~Dragicevic, A.~Escalante~Del~Valle, R.~Fr\"{u}hwirth\cmsAuthorMark{1}, M.~Jeitler\cmsAuthorMark{1}, N.~Krammer, L.~Lechner, D.~Liko, I.~Mikulec, F.M.~Pitters, J.~Schieck\cmsAuthorMark{1}, R.~Sch\"{o}fbeck, M.~Spanring, S.~Templ, W.~Waltenberger, C.-E.~Wulz\cmsAuthorMark{1}
% ...
\vskip\cmsinstskip
\dag: Deceased\\
1:  Also at Vienna University of Technology, Vienna, Austria\\
% ...
\end{sloppypar}
%%% END EDITABLE REGION %%%
% skeleton_end
\end{document}